\documentclass[iop,numberedappendix]{emulateapj}

\usepackage[]{graphicx}
\usepackage[active]{srcltx}
\usepackage{aas_macros}
\usepackage{amsmath}
\usepackage[caption=false]{subfig}
\usepackage{hyperref}

\renewcommand{\d}{\mbox{d}}
\renewcommand{\P}{\mathcal{P}}

\newcommand{\D}{\mathcal{D}}
\newcommand{\M}{\mathcal{M}}

\newcommand{\N}{\mathcal{N}}
\newcommand{\K}{\mathcal{K}}
\newcommand{\R}{\mathcal{R}}
\newcommand{\A}{\mathcal{A}}
\newcommand{\Ph}{\mathrm{Ph}}
\newcommand{\Cm}{\mathrm{Cm}}
\newcommand{\Pol}{\mathrm{Pol}}
\newcommand{\UnP}{\mathrm{UnP}}
\newcommand{\tot}{\mathrm{tot}}
\newcommand{\DPh}{\D_{\Ph}}
\newcommand{\DCm}{\D_{\Cm}}
\newcommand{\DPhPol}{\D_{\Ph}^{\Pol}(\phi,\theta)}
\newcommand{\DPhUnP}{\D_{\Ph}^{\UnP}(\phi,\theta)}
\newcommand{\DCmPol}{\D_{\Cm}^{\Pol}(\phi,\theta)}
\newcommand{\DCmUnP}{\D_{\Cm}^{\UnP}(\phi,\theta)}
\newcommand{\vDPhPol}{\D_{\Ph}^{\Pol}(\varphi,\vartheta)}
\newcommand{\vDPhUnP}{\D_{\Ph}^{\UnP}(\varphi,\vartheta)}

\begin{document}

\title{On the operation of X-ray polarimeters with a large field of view}

\author{Fabio Muleri}
\affil{INAF-IAPS, Via del Fosso del Cavaliere 100, I-00133 Roma, Italy}
\email{fabio.muleri@iaps.inaf.it}

\begin{abstract}
The measurement of the linear polarization is one of the hot topics of High Energy Astrophysics. Gas
detectors based on photoelectric effect have paved the way for the design of sensitive instruments
and mission proposals based on them have been presented in the last few years in the energy range
from about 2 keV to a few tens of keV. As well, a number of polarimeters based on Compton scattering
are approved or discussed for launch on-board balloons or space satellites at higher energies. These
instruments are typically dedicated to pointed observations with narrow field of view telescopes or
collimators, but there are also projects aimed at the polarimetry of bright transient sources, like
Soft Gamma Repeaters or the prompt emission of Gamma Ray Bursts. Given the erratic appearance of
such events in the sky, these polarimeters have a large field of view to catch a reasonable number
of them and, as a result, photons may impinge on the detector off-axis. This changes dramatically
the response of the instrument to polarization, regardless if photoabsorption or Compton scattering
is involved. Instead of the simple cosine square dependency expected for polarized photons which are
incident on-axis, the response is never purely cosinusoidal and a systematic modulation appears also
for unpolarized radiation. We investigate the origin of these differences and present an analytic
treatment which proves that actually such systematic effects are the natural consequence of how
current instruments operate. Our analysis provides the expected response of photoelectric or Compton
polarimeters to photons impinging with any inclination and state of polarization.
\end{abstract}


\section{Introduction}

Polarimetry of astrophysical sources is the only probe in X-rays which still today is substantially
unexplored. After the first pioneering result in the '70s with the positive detection for the
Crab Nebula \citep{Novick1972,Weisskopf1978}, only very recently instrumental improvements have
allowed to reach a sufficient sensitivity to justify renewed efforts. While above $\sim30$~keV the
most sensitive technique remains that of Compton scattering, at lower energies gas detectors able to
image the track of the photoelectron have been developed during the last ten years
\citep{Costa2001,Bellazzini2007,Black2007,Bellazzini2010c} and provide thanks to the larger
efficiency a valuable alternative to Bragg diffraction at 45$^\circ$ and Thomson scattering around
90$^\circ$ \citep{Novick1975}. The use of photoelectric polarimeters is particulary effective in the
soft X-ray range, between about 2 and 10 keV \citep{Black2007,Muleri2008,Muleri2010}, but efforts
are carried out to extend this energy range up to a few tens of keV
\citep{Muleri2006,Hill2007,Soffitta2010,Fabiani2012}.

Almost all astrophysical sources should emit partially polarized photons \citep[for a review,
see][]{Meszaros1988,Weisskopf2009,xraypol2010}, but there are only a few classes for which a
polarization higher than 10\% is expected. The possibility to measure polarization at and below this
level requires to collect hundreds of thousands of photons and therefore X-ray polarimeters are
usually conceived to point the source for a long exposure time. Such a condition makes
convenient to have a narrow field of view to reduce the background and minimize the confusion, and
actually this has been a common approach for missions dedicated to X-ray polarimetry, e.g. GEMS
\citep{Jahoda2010}, POLARIX \citep{Costa2010c}, PoGOLite \citep{Kamae2008} or XIPE
\citep{Soffitta2013}. Nonetheless, the narrow field of view precludes to these missions the
observation of fast transient sources, that is, short-lived sources whose flux decreases so rapidly
to not give enough time to re-point the instrument and perform the measurement, and this has
motivated an high interest in instruments with a large field of view. The primary scientific
objective, but not the only, is the prompt emission of Gamma Ray Bursts (GRBs) which lasts less than
a few minutes. These sources are very appealing for X-ray polarimetry because some of them are very
bright and because there are indications that the prompt phase may be highly polarized.
\citet{Coburn2003} claimed a very high polarization for GRB021206, $\P\sim(80\pm20)\%$,
although this result was subsequently questioned \citep{Rutledge2004,Wigger2004}, and other authors
reported similar high polarization for GRB930131 and GRB 960924 \citep{Willis2005}, GRB041219a
\citep{Kalemci2007,McGlynn2007,Gotz2009} and GRB061122 \citep{McGlynn2009}. More recently, the Gamma
Ray Burst Polarimeter (GAP), the first instrument specifically designed for GRBs polarimetry,
reported the detection of polarized radiation at a considerable level in case of a few other GRBs
\citep{Yonetoku2011b, Yonetoku2012}. Such results, although limited to very bright events and often
with a low significance, have raised a widespread interest in X/soft $\gamma$-ray polarimetry of
GRBs and eventually have led many authors to recognize polarimetry as a tool of great importance to
probe the magnetic field in the jet and its evolution during the afterglow
\citep{Waxman2003,Lazzati2006,Toma2009}. 

The significance of the scientific case has stimulated the design of new instruments, both
photoelectric and Compton, specifically conceived for GRBs, such as POLAR \citep{Produit2005,
Produit2010, Orsi2011}, GRAPE \citep{Bloser2009} or POET \citep{Hill2008}. The fact that these
polarimeters have a large field of view, often covering a significant fraction of the sky, has the
practical consequence that the instrument can not be rotated around the incident direction to
average possible nonuniformities. Moreover, the photons impinge on the detector with an inclination
which can arrive at several tens of degrees and in this condition the measured response of
the instrument is very different from the simple cosine square dependency which is found when
photons are incident on-axis \citep{Yonetoku2006,Xiong2009}. Attempts have been put forward to
remove the systematic effects observed off-axis and reconduct the response to the well-know
cosine square behavior, so that the polarization of the incident photons could be derived with the
standard analysis already developed for on-axis radiation. However, their effectiveness is
questionable, especially when the polarization is comparable to the systematic signal. An
alternative approach is to use Monte Carlo simulations for deriving the response of the
instrument to beams with different states of polarization and find out by comparison the curve
``closer'' to that measured. This approach allows to derive the polarization of the incident
photons, but it does no give any insight into the origin of the off-axis systematic effects.

In this paper we study the off-axis response of both photoelectric and Compton polarimeters by a
novel point of view. We start from the physics of the interaction to demonstrate that the
differences with respect to the response obtained on-axis are simply the natural consequence of how
current instruments work. Both photoelectric and Compton polarimeters are similar at this regard
and therefore we will treat these techniques in parallel throughout the paper. From this premise, we
present an analytic method to calculate the expected response for any inclination and state of
polarization. This work is organized as follows: we summarize the relevant physics of photoelectric
absorption and Compton scattering and the operation of the polarimeters based on these interactions
in Section~\ref{sec:Basic}; in Section~\ref{sec:Method} we describe our method to calculate the
expected modulation curve, applying it both to on-axis and off-axis photons; in
Section~\ref{sec:ModInclined} we show how the expected modulation curve depends on the incident
direction and on the polarization of the impinging photons; we discuss how this poses some
additional requirements on the instrument design and compare our results with those by
previous authors in Section~\ref{sec:Discussion} and eventually draw our conclusions in
Section~\ref{sec:Conclusions}.

\section{Basic principles} \label{sec:Basic}

\subsection{Photoelectric absorption and Compton scattering} \label{sec:IntProc}

\begin{figure}[tbp]
\begin{center}
\subfloat[\label{fig:PhAngles}]{\includegraphics[angle=0,totalheight=6cm]{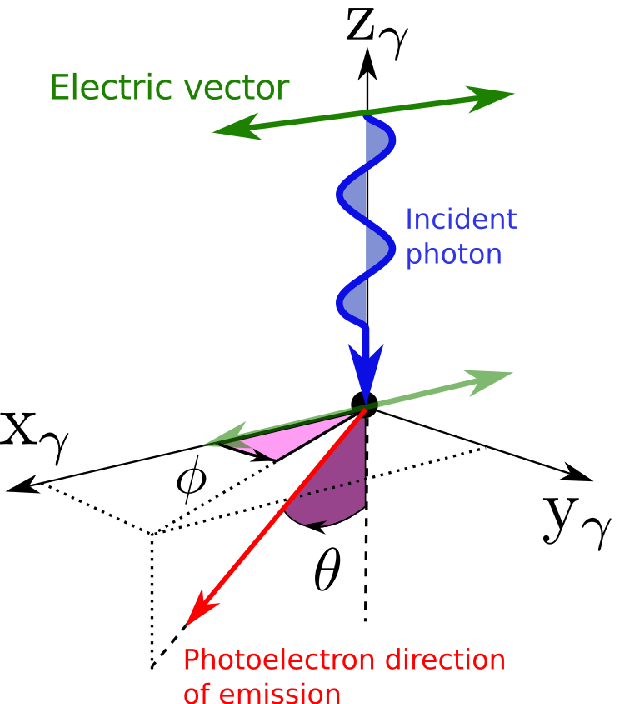}}
\hspace{5mm}
\subfloat[\label{fig:ComptonAngles}]{\includegraphics[angle=0,totalheight=6cm]{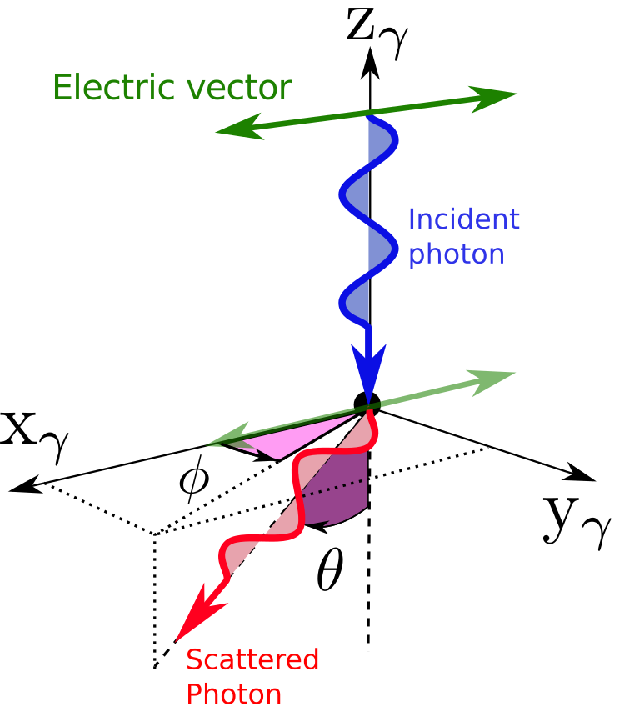}}
\end{center}
\caption{Definition of the relevant angles for photoelectric absorption (a) and Compton
scattering (b). The axis labels have the subscript $\gamma$ to indicate that the frame of
reference is constructed from the photon characteristics, that is the $z$-axis coincides with the
incident direction and the $x$-axis with the polarization vector.}
\label{fig:Angles}
\end{figure}

The photoelectric effect is sensitive to the polarization of the absorbed photon because the
emission direction of the ejected photoelectron is affected by the photon electric field. The
dependency on the polarization in case of a photon with energy $E$ which is absorbed by an atom
of atomic number $Z$ is described by the differential cross section of the process
\citep{Heitler1954}
\begin{equation}
\frac{\d\sigma_{\Ph}}{\d\Omega}=r_0^2\alpha^4 Z^5\left(\frac{m_e c^2}{E}\right)^{\frac{7}{2}}
\frac{4\sqrt{2} \sin^2\theta\,\cos^2\phi}{\left(1-\beta\cos\theta\right)^4} \; ,
\label{eq:dsdO_Ph}
\end{equation} 
where $\beta$ is the photoelectron velocity in units of the speed of light $c$, $\alpha$ is the
fine-structure constant, $m_e$ is the rest mass of the electron and $r_0$ is its classical
radius. The polar angle $\theta$ is that formed by the direction of the incident photon and of the
photoelectron, while $\phi$ is the azimuthal angle of the photoelectron with respect to the
direction of polarization (see Figure~\ref{fig:PhAngles}). Equation~(\ref{eq:dsdO_Ph}) shows that
the probability of emission in a certain azimuthal direction $\phi$ is modulated as $\cos^2\phi$ and
therefore a photoelectron is more probably produced along the direction of the electric field of the
absorbed photon ($\phi=0$). On the contrary the ejection orthogonally to it is suppressed
(${d\sigma}/{d\Omega}=0$ if $\phi=\pi/2$). 

Strictly speaking, Equation~(\ref{eq:dsdO_Ph}) is valid only in the case of absorption by spherical
symmetric shells. The photoelectrons emitted as a consequence of the absorption from other shells
are not completely modulated and, as a matter of fact, they can be emitted also orthogonally to
polarization \citep{Ghosh1983}. For the sake of simplicity, we consider in the following only the
absorption of the K-shell, which is spherical symmetric and therefore Equation~(\ref{eq:dsdO_Ph})
holds. This assumption is justified because the working energy band of a photoelectric polarimeter
is usually chosen to make largely dominant the absorption from the K-shell, of at least an order
of magnitude, to fully exploit the complete modulation with polarization.

The differential cross section expressed in Equation~(\ref{eq:dsdO_Ph}) allows by definition to
derive the probability that a photoelectron is emitted within the elementary solid angle $\d\Omega$
in the $(\phi;\theta)$ direction. In the following we will be interested only in the the
\emph{angular} distribution of the photoelectrons which we name $\D_\Ph^\Pol$ and $\D_\Ph^\UnP$ in
case of linearly polarized and unpolarized incident photons. Then, we define
\begin{subequations}
\label{eq:DPh}
\begin{align}
\DPhPol &=\frac{\sin^2\theta\,\cos^2\phi}{\left(1-\beta\cos\theta\right)^4} \; ; \label{eq:DPhPol}\\
\DPhUnP &=\frac{1}{2}\frac{\sin^2\theta}{\left(1-\beta\cos\theta\right)^4} \; . \label{eq:DPhUnP}
\end{align}
\end{subequations}
In Equation~(\ref{eq:DPhUnP}) we substituted the $\cos^2\phi$ term with its average value over all
azimuthal angles because in case of unpolarized radiation there is no preferred azimuthal direction
and
\begin{equation}
\frac{\int_0^\pi\cos^2(\phi)\d\phi}{2\pi} = \frac{1}{2} \; . \notag
\end{equation}

The angular distribution $\DPh$, intending both $\D_\Ph^\Pol$ and $\D_\Ph^\UnP$, comprises three
contributions: the azimuthal dependency expressed by the $\cos^2\phi$ factor for polarized photons
or by a constant contribution for unpolarized radiation; the polar dependency given by
$\sin^2\theta$ and the factor $1/\left(1-\beta\cos\theta\right)^4$ which can be considered as an
energy correction to it.  If we assume $\beta=0$ and then neglect the last contribution, the
polar distribution of the events peaks for $\theta=\pi/2$ and therefore the photoelectron  is
emitted with more probability in the plane orthogonal to the direction of the incident photon. In
this assumption, there is complete symmetry between the emission above and below this plane and any
direction $\theta=\bar{\theta}$ is equivalent to that $\theta=-\bar{\theta}$. The energy
correction breaks this symmetry and makes more probable the emission in the semi-space opposite to
the incident direction of photon, basically because of the need to conserve the initial momentum. In
the following, we will refer to this effect as \emph{forward bending}.

The scattering of a photon is, as well as photoabsorption, an interaction which is sensitive to
polarization. The information is contained in the scattering direction and also in this case a
cosine square modulation is obtained for polarized radiation. In fact, the Klein-Nishina formula
which gives the differential cross section of the process in the simple hypothesis of scattering on
a free electron at rest is \citep{Heitler1954}
\begin{equation}
\frac{d\sigma_{\Cm}}{d\Omega}=\frac{1}{2}r_0^2\frac{E'^2}{E^2}\left[\frac{E}{E'}
+\frac{E'} {E}-2\sin^2\theta\,\cos^2\phi\right] \; , \label{eq:dsdO_Cm}
\end{equation}
where $E$ and $E'$ are the energy of the photon before and after the interaction, $\theta$ is the
angle of scattering and $\phi$ is the angle which the plane of scattering form with that containing
the polarization vector and the direction of the incident photon (see
Figure~\ref{fig:ComptonAngles}). $E$ and $E'$ are related by \citep{Heitler1954}
\begin{equation}
\frac{E'}{E}=\frac{1}{1+\varepsilon(1-\cos\theta)}
\label{eq:EnCmp}
\end{equation}
where $\varepsilon=E/m_e\,c^2$.

The direction of scattering is modulated as $\cos^2\phi$ with a peak in the direction orthogonal to
polarization, that is ${d\sigma}/{d\Omega}$ is maximum if $\phi=\pi/2$. The modulation is never
complete, namely ${d\sigma}/{d\Omega}\neq0$ if $\phi=0$, unless we restrict ourselves to the case
$\theta=\pi/2$ and to the low energy limit for which $E'=E$ (Thomson scattering). Unfortunately
these assumptions are not suitable for what we are going to deal with in the following. Compton
polarimeters are being used above a few tens of keV and in this energy range the ratio $E'/E$
differs significantly from one, $E/E'\gtrsim0.1$. Moreover, it is not practically convenient to
constrain the scattering angle strictly around $\theta=\pi/2$ because the detection efficiency would
be reduced accordingly.

Analogously to photoelectric absorption, we define the angular distribution of the scattered photons
in case of polarized and unpolarized radiation as
\begin{subequations}
\label{eq:DCm}
\begin{align}
\DCmPol &=\frac{1}{1+\varepsilon\left(1-\cos\theta\right)}+\frac{1}{
\left[1+\varepsilon\left(1-\cos\theta\right)\right]^3}+ \notag \\
&-\frac{2\sin^2\theta\cos^2\phi}{\left[1+\varepsilon\left(1-\cos\theta\right)\right]^2} \; ; \\
\DCmUnP &=\frac{1}{1+\varepsilon\left(1-\cos\theta\right)}+\frac{1}{
\left[1+\varepsilon\left(1-\cos\theta\right)\right]^3}+ \notag \\
&-\frac{\sin^2\theta}{\left[1+\varepsilon\left(1-\cos\theta\right)\right]^2} \; ,
\end{align}
\end{subequations}
where we made use of Equation~(\ref{eq:EnCmp}).

It is worth stressing that the two functions representing the angular distribution of
the photoelectrons and of the scattered photons, $\DPh$ and $\DCm$, shares many similarities. In
addition to the similar azimuthal dependency, the variable $\varepsilon$  plays a role in the latter
similar to $\beta$ in $\DPh$. As a matter of fact, if we restrict ourselves to the Thomson
scattering regime, that is $E'=E$ and then $\varepsilon=0$, the angular dependence reduces to
$\mathcal{D}_\Cm^\Pol=2\left(1-\sin^2\theta\cos^2\phi\right)$. This function is symmetric for the
scattering below and above the plane orthogonal to the incident direction, although in this case the
probability of emission peaks at $\theta=0$ and $\theta=\pi$ and not at $\theta=\pi/2$ as for
photoelectric absorption. Dropping the Thomson scattering assumption, the forward/back symmetry of
the scattering is broken and an effect of forward bending similar to that of photoelectric
absorption occurs. 

\subsection{Operation of polarimeters} \label{sec:Polarimeters}

A polarimeter, either photoelectric or Compton, measures the polarization by reconstructing the
geometry of the interaction which occurs in the instrument. In the case of photoelectric
instruments, this implies to derive the direction of emission of the photoelectron while for
Compton polarimeters the scattering direction has to be inferred. 

\begin{figure*}
\begin{center}
\subfloat[\label{fig:PhSketch}]{\includegraphics[angle=0,totalheight=7.8cm]{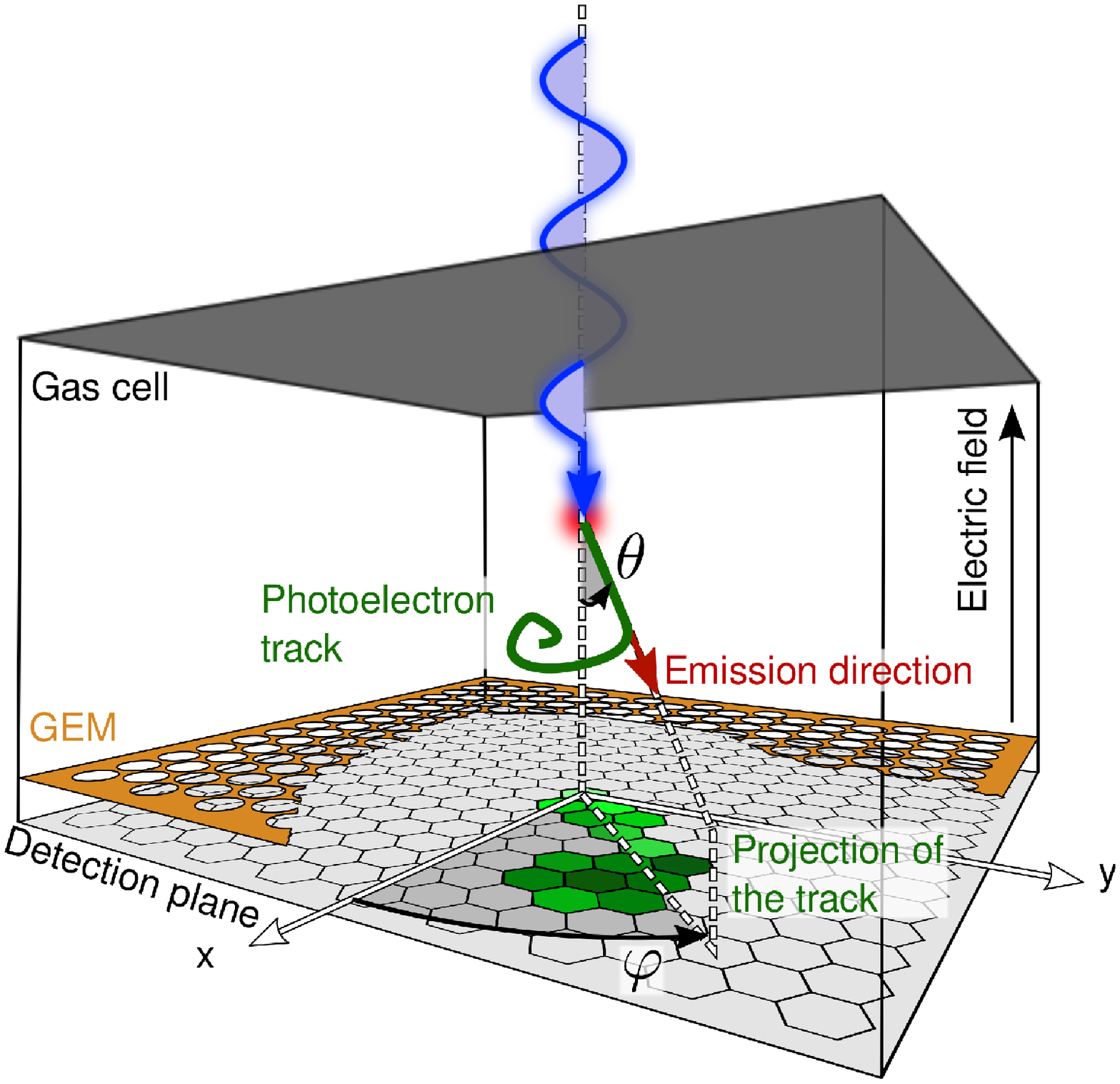}}
\hspace{5mm}
\subfloat[\label{fig:ComptonSketch}]{\includegraphics[angle=0,totalheight=7.8cm]{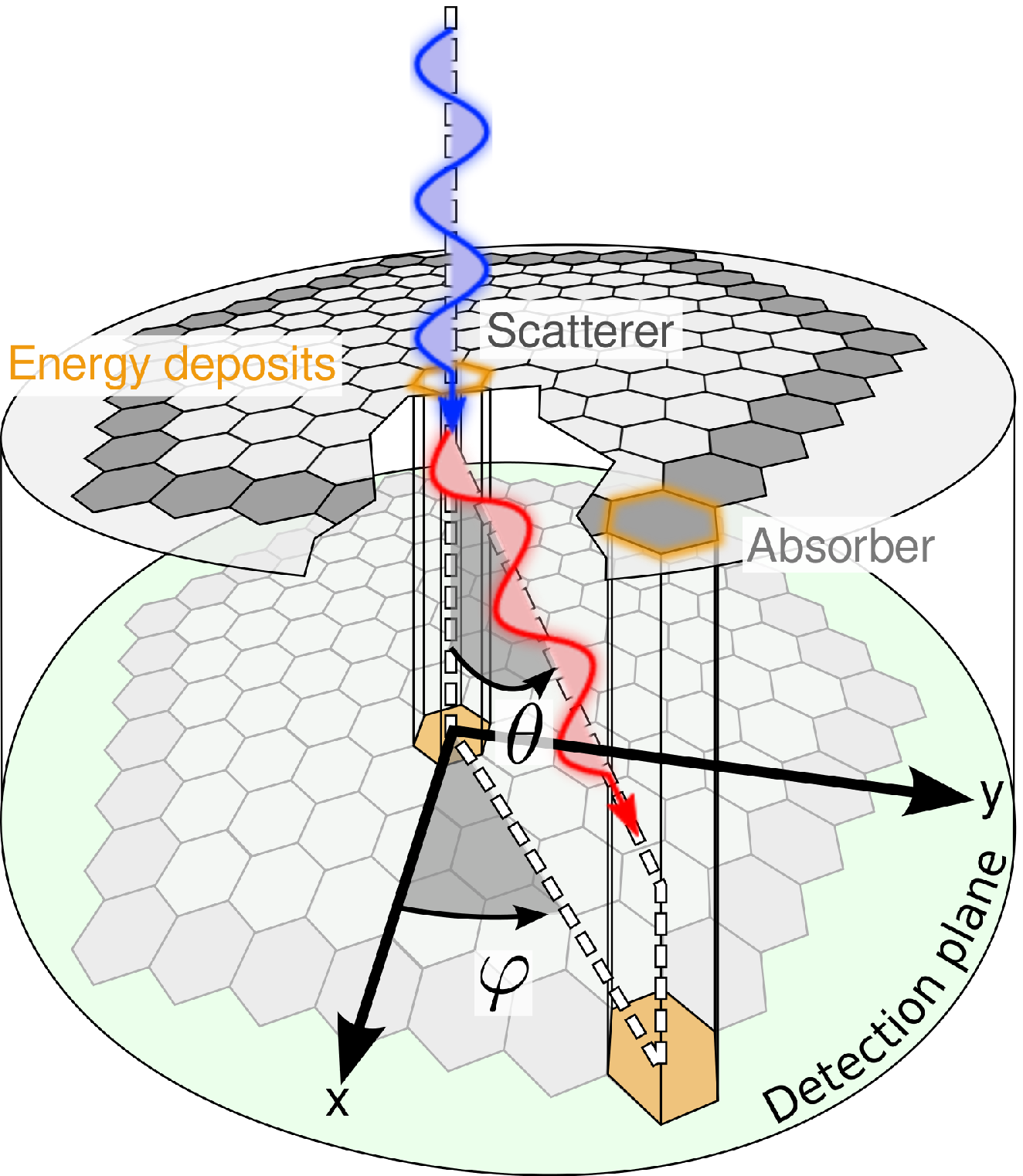}}
\end{center}
\caption{Sketch of a photoelectric (a) and of a Compton (b) polarimeter. The angle of
emission/scattering as defined throughout this paper are also indicated.}
\label{fig:Sketch}
\end{figure*}

The operation of a real photoelectric polarimeter, the Gas Pixel Detector
\citep{Costa2001,Bellazzini2010c}, is sketched in Figure~\ref{fig:PhSketch}. When a photon is
absorbed in the gas cell, the path of the photoelectron emitted is distinguished thanks to the
electron-ion pairs produced by ionization along the way. The electrons are drifted by an electric
field to a Gas Electron Multiplier (GEM), amplified and eventually collected by a pixellated
detector without changing the shape of the track. What is eventually returned by the instrument is
the image of the photoelectron path projected on the plane of the detector, and this image is used
to reconstruct the emission direction. Actually, only the azimuthal direction of emission is
measured, namely the angle $\varphi$ in Figure~\ref{fig:PhSketch}, whereas the instrument is not
sensitive to the polar distribution, i.e. $\theta$. Other instruments which exploit different
geometries for collecting the ionization charges exist, e.g. the Time Projection Chamber
\citep{Black2007}, but as a matter of fact they all perform an equivalent projection of the track on
a plane. This plane, which is orthogonal to the direction of incidence of the photons during the
normal use of the instrument in the focal plane of a telescope, will be named \emph{detection plane}
in the following.

Compton polarimeters usually can reconstruct the geometry of the event with the same limitations
as photoelectric instruments, that is, only the direction of scattering projected on the detection
plane is actually measured. To see it, we will take as an example the Compton polarimeter design
sketched in Figure~\ref{fig:ComptonSketch} because all the possible arrangements used to measure the
direction of the scattered photons are fundamentally equivalent at this regard. The instrument
comprises of an array of scintillator rods and it is sensitive to those events which are scattered
in a rod and interact with another one. The direction of scattering is derived as the line
connecting the center of the first (\emph{scatterer}) and of the second hit rods (\emph{absorber}).
The scintillator material can be the same for all rods or not, as in the design reported in
the figure. In the first case, each rod can work as the scatterer or as the absorber
\citep{Produit2005,Kamae2008,Bloser2009}, whereas in the second case each element is optimized for
one of the two tasks by choosing low or high atomic number materials, respectively
\citep{Sakurai1991,Costa1995}. An instrument like that in Figure~\ref{fig:ComptonSketch} is
sensitive only to the azimuthal distribution of the events, that is to the angle $\varphi$, because
the angle $\theta$ would be derived only by knowing the z-coordinate of the interaction in both the
scatterer and in the absorber. This is practically rather demanding, especially for the scatterer
for which the energy deposit is of the order of a few keV for photons below 100~keV. Therefore, even
if there are noteworthy exceptions such as Compton telescopes, polarimeters for hard X-rays are
usually designed to ignore the scattering angle of the event which is not necessary for their task
and in this sense the direction of the event is reconstructed only on the detection plane. Such
instruments can put only weak constrains on the angle of scattering which derive from the assumption
that the scattering direction of all accepted events must intercept both the scatterer and the
absorber.

The information on the angle and on the degree of polarization of the incident photons is retrieved
in a similar way for both photoelectric and Compton polarimeters. Firstly, the histogram is
constructed of the azimuthal angles that the direction of the photoelectron or of the scattered
photon forms with a reference direction. Such a histogram, which is named \emph{modulation curve},
is supposed to retain the same azimuthal dependency as the differential cross section. Therefore,
a cosine square modulation is expected to appear in case of partially polarized photons, whereas
the modulation curve should be flat for unpolarized radiation, except for the fact that the number
of entries in each angular bin of the histogram will never be exactly identical because of
statistical fluctuations. To quantify the amplitude of the possible cosine square contribution, the
modulation curve is fitted with a function, which we will call \emph{modulation function} $\M$, that
is 
\begin{equation}
\M(\varphi) = A + B\cos^2(\varphi-\varphi_0) \; . \label{eq:Cos2}
\end{equation}
The free parameters in the fit, $A$, $B$ and $\varphi_0$, allow to derive the polarization of the
incident photons. The phase of the cosine $\varphi_0$ in Equation~(\ref{eq:Cos2}) singles out the
direction where the emission is more probable and therefore it is the angle of polarization in case
of photoelectric polarimeters, and it is the angle of polarization plus or less $\pi/2$ in case of
Compton instruments. The degree of polarization $\P$ is linearly proportional to the amplitude of
the measured cosine square modulation $\A$, which is defined as
\begin{equation}
\A = \frac{\M_{\max}-\M_{\min}}{\M_{\max}+\M_{\min}}=\frac{B}{2A+B}
\; \notag , 
\end{equation}
where $\M_{\max}$ and $\M_{\min}$ are the maximum and minimum values of the modulation function,
respectively. To derive $\P$, the value of $\A$ has to be rescaled for the amplitude of the
modulation for completely polarized photons, which is named \emph{modulation factor} $\mu$. Then 
\begin{equation}
\P = \A/\mu \notag
\end{equation}
with
\begin{equation}
\mu =\frac{B_{1}}{2A_{1}+B_{1}} \; . \label{eq:Mu}
\end{equation}
The detection of the polarization is statistically significant only if it exceeds the \emph{Minimum
Detectable Polarization} (MPD), which is the signal expected from statistical fluctuations only
\citep{Weisskopf2010,Strohmayer2013}.

\begin{figure*}[htbp]
\begin{center}
\includegraphics[angle=0,width=12cm]{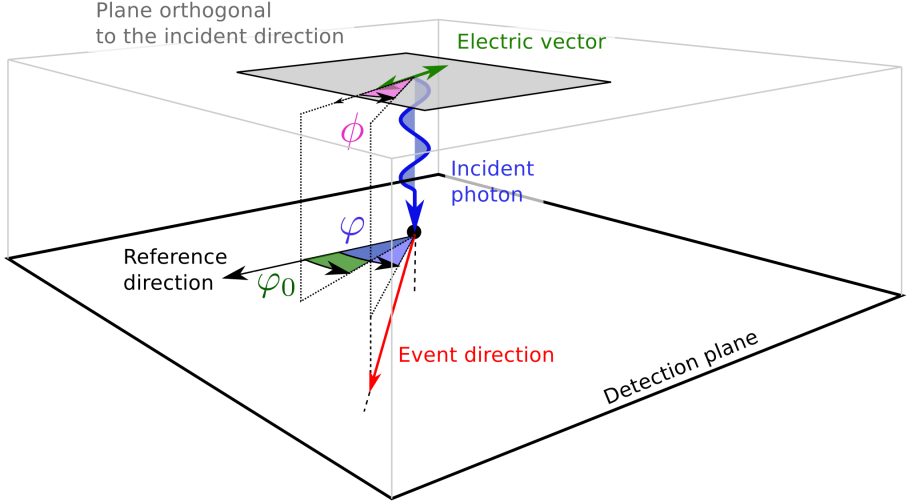}
\end{center}
\caption{Relation between $\phi$, $\varphi$ and $\varphi_0$ when photons are incident on-axis.
For the sake of simplicity, we show in the figure only the case of a photoelectric polarimeter, so
that the phase $\varphi_0$ coincides with the direction of polarization.}
\label{fig:PhiVarPhi}
\end{figure*}

It is worth noting that in Equation~(\ref{eq:Cos2}) and in Figure~\ref{fig:Sketch} we intentionally 
used the angle $\varphi$ instead of $\phi$. Although both characterize the azimuth of the event
direction after the interaction, the latter is the angle to the polarization vector measured on the
plane orthogonal to the incident direction, the former is measured from some axis of reference of
the instrument on the detection plane. The relation between the two is obvious on-axis because in
this case the plane orthogonal to the photon direction is parallel to the detection plane (see
Figure~\ref{fig:PhiVarPhi}) and then:
\begin{subequations}
\begin{eqnarray}
\phi =& \varphi-\varphi_0 & \mbox{for photoelectric polarimeters} \label{eq:PhiPh} \\
\phi =& \varphi-\varphi_0+\frac{\pi}{2} &\mbox{for Compton polarimeters}
\end{eqnarray}
\end{subequations}
where $\varphi_0$ is the phase of the modulation function, cf. Equation~(\ref{eq:Cos2}). Although
it may appear superfluous to stress the difference between $\varphi$ and $\phi$ at this stage, we
will see that in case of inclined radiation they have to be clearly distinguished.

\section{The method} \label{sec:Method}

We have seen that the differential cross section for photoabsorption and Compton scattering shows a
cosine square modulation in case of polarized photons and therefore the measured modulation is
usually fitted with a cosine square function. We are going to prove that such a conclusion can be
drawn only as long as photons are incident on-axis. To do this, we have to define a formal
analytical method to calculate the modulation function. We present our procedure in this section,
applying it firstly to on-axis radiation to obtain the well-known cosine square behavior and then
extending the result to an off-axis beam.

\subsection{Application to the on-axis case} \label{sec:OnAxis}

The modulation function is practically the modulation curve expected to be measured by an instrument
for a certain polarization degree and angle. We are going to calculate it by counting how many
events per azimuthal bin are emitted and eventually detected by the instrument according to the
distribution $\D$ given by the differential cross section for a certain polarization state. In our
treatment, we will neglect all of the causes which may deviate the shape of the modulation curve
from its intrinsic azimuthal dependency, that is, that given by the underlying physics of the
interaction. Therefore, we will neglect any possible systematic effect due to instrumental
nonuniformities in the azimuthal response.

The number $\d\N$ of events whose emission direction is in the elementary solid angle $\d
\Omega$ in the $(\phi\,;\theta$) direction is proportional to their angular distribution $\D$:
\begin{equation}
\d \N(\phi,\theta) = \kappa \; \D(\phi,\theta) \sin\theta \d\theta \d\phi\; , \notag
\end{equation}
where $\kappa$ is a constant of proportionality, $\d\Omega=\sin\theta \d\theta \d\phi$ and $\D$ is
the generic angular distribution of the events in case of photoelectric absorption or Compton
scattering, for polarized or unpolarized incident radiation. We already pointed out in
Section~\ref{sec:Polarimeters} that state-of-the-art photoelectric and Compton polarimeters are not
designed to be sensitive to the polar direction of the event. This limitation implies that, when we
calculate the number $\N(\phi)$ of events \emph{emitted} in the azimuthal direction $\phi$, all
events with the same $\phi$ are summed regardless of the polar angle and then
\begin{equation}
\N(\phi) = \int_{\theta_{\min}}^{\theta_{\max}} {\d \N(\theta,\phi)} = \kappa
\int_{\theta_{\min}}^{\theta_{\max}} \D(\theta,\phi) \sin\theta \d\theta \; .
\label{eq:NTheta}
\end{equation}

The two integration limits on $\theta$ in Equation~(\ref{eq:NTheta}) are a peculiar characteristic
of the instrument because, basically, they define the polar interval of the events which are
accessible to the device. For photoelectric polarimeters, a reasonable approximation is to assume
that $\theta_{\min}=0$ and $\theta_{\max}=\pi$. Instead, in case of Compton polarimeters, the values
of $\theta_{\min}$ and $\theta_{\max}$ are usually constrained by the geometry of the detector, see
for example \citet{Muleri2012}. It is out of the scope of this paper to discuss a specific design
and therefore we will make use of well-tuned values for $\theta_{\min}$ and $\theta_{\max}$ only
when we will compare our results with those obtained for particular instruments by other authors.
For the time being, we will assume that $\theta_{\min}=0$ and $\theta_{\max}=\pi$ also for Compton
polarimeters. Such an assumption is more than adequate to qualitatively discuss the effect of the
inclined incidence of the photons on the modulation function and it has the advantage to simplify
the discussion that follows. Therefore
\begin{equation}
\N(\phi) = \kappa\int_{0}^{\pi} \D(\theta,\phi) \sin\theta \d\theta \; . \notag
\end{equation}

Let us first consider the case of photoelectric polarimeter. We have for completely polarized
radiation that
\begin{align}
\N^{\Pol}(\phi) &= \kappa\int_{0}^{\pi} \DPhPol \sin\theta \d\theta =
\nonumber\notag\\
&= \kappa\int_{0}^{\pi} \frac{\sin^2\theta\,\cos^2\phi}{\left(1-\beta\cos\theta\right)^4}
\sin\theta \d\theta =\notag\\
&= \kappa\left[\int_{0}^{\pi}
\frac{\sin^3\theta}{\left(1-\beta\cos\theta\right)^4}\d\theta\right]\;\cos^2\phi =\notag\\
& = \kappa\,F(\beta) \cos^2\phi \; ,
\label{eq:NPhPol}
\end{align}
where $F(\beta)$ is a function of the incident photon energy through $\beta$,
\begin{equation}
F(\beta)= \int_{0}^{\pi} \frac{\sin^3\theta}{\left(1-\beta\cos\theta\right)^4}\d\theta \; . \notag
\end{equation}
Although it is possible to calculate explicitly the function $F(\beta)$, it is not much relevant
for the discussion that follows. In fact, we will find more convenient to rewrite $F(\beta)$ by
introducing the total number of events collected, $\N^\Pol_\tot$, where the $\Pol$ superscript is
because we are dealing with polarized photons. Then, it holds that 
\begin{equation}
\int_{0}^{2\pi} \N^{\Pol}(\phi) \d\phi= \N^\Pol_\tot \label{eq:NPhNorm} \; ,
\end{equation}
which simply imposes that the sum of the events detected in all azimuthal bins is
$\N^\Pol_\tot$. Substituting Equation~(\ref{eq:NPhPol}) in Equation~(\ref{eq:NPhNorm}), we derive
that $\kappa F(\beta)=\N^\Pol_\tot/\pi$. Then,
\begin{equation}
\N^{\Pol}(\phi) = \frac{\N^\Pol_\tot}{\pi}\cos^2\phi \; .
\label{eq:NPhPol2}
\end{equation}

Analogously, in case of completely unpolarized radiation we have that
\begin{align}
\N^{\UnP}(\phi) &= \kappa\int_{0}^{\pi}
\frac{1}{2}\frac{\sin^2\theta}{\left(1-\beta\cos\theta\right)^4} \sin\theta\d\theta
=\notag\\
& = \kappa\,\frac{F(\beta)}{2} = \frac{\N^\UnP_\tot}{2\pi} \; ,
\label{eq:NPhUnp}
\end{align}
where we have used a condition equivalent to Equation~(\ref{eq:NPhNorm}) but naming $\N^\UnP_\tot$
the
number of collected events.

The number of events emitted for completely polarized and completely unpolarized radiation,
expressed in Equation~(\ref{eq:NPhPol}) and Equation~(\ref{eq:NPhUnp}) respectively, can be linearly
combined to deal with the case of partially polarized radiation. If $\P$ is the degree of
polarization and $\N_\tot$ the total number of events, we have that by definition $\P =
{\N^\Pol_\tot}/{\N_\tot}$ and $\N^\UnP_\tot = \N_\tot-\N^\Pol_\tot$. Therefore, the number of
events $\N(\phi)$ emitted per azimuthal angle $\phi$ in case of partially polarized radiation is
\begin{align}
\N(\phi) & = \N^{\Pol}(\phi) + \N^{\UnP}(\phi) = \notag \\
& = \frac{\N^\Pol_\tot}{\pi}\cos^2\phi + \frac{\N^\UnP_\tot}{2\pi} = \notag \\
& = \N_\tot\left[\P\frac{\cos^2\phi}{\pi} + \frac{(1-\P)}{2\pi}\right] \; .
\label{eq:NPh}
\end{align}
It is worth stressing that Equation~(\ref{eq:NPh}) formalizes a result which, although obvious, is
fundamental for the discussion below. The azimuthal distribution of the emission directions
comprises two contributions, one due to the polarized component and one due to that unpolarized, and
their magnitude depends on the degree of polarization.

We obtained from Equation~(\ref{eq:NPh}) that the emitted azimuthal distribution of the
photoelectrons is cosine square modulated and that the modulation is complete when $\P=1$.
Notwithstanding, it is well-known that the modulation curve of any real photoelectric (or
Compton) polarimeter is not completely modulated for 100\% polarized photons, even in the case it
should be because, e.g., the contribution from spherical symmetric shells is absolutely predominant.
The ultimate reason for this is that the measurement of the event direction is naturally
affected by some uncertainty. As a matter of fact, a photoelectron emitted in a certain direction
may instead be reconstructed in an another one because of a number of causes which are often
intrinsic in the measurement process. Just as a few examples, the diffusion of the charges during
the drift blurs the photoelectron track thus making more difficult the reconstruction of the initial
direction or the elastic scatterings with the gas mixture atomic nuclei may change the direction of
the photoelectron significantly. A similar conclusion holds in case of Compton polarimeters, that
is, the initial direction of scattering may not be measured correctly because, for example, the
photon had a second scattering in the same element of the scatter. As a consequence, the modulation
curve, which by definition is the number of directions \emph{reconstructed} in a certain azimuthal
bin, does no coincide with the azimuthal distribution $\N(\phi)$ of the \emph{emitted} events which
we calculated in Equation~(\ref{eq:NPh}). The most favorable condition, which we will assume
verified hereafter, is when the ``probability of error'' does not depend on the azimuthal angle and
then the instrument does not introduce any systematic effect. In this case we can assume that the
azimuthal dependence of the modulation curve remains identical to that of $\N(\phi)$ and the only
contribution of the events incorrectly reconstructed is an additional constant term. Therefore, the
modulation function of a real instrument can be rewritten as 
\begin{align}
\M(\phi) &= \N(\phi)+\K = \notag \\
&=\N\left[\P\frac{\cos^2\phi}{\pi} + \frac{(1-\P)}{2\pi}\right] + \K \; ,
\label{eq:ffactor}
\end{align}
where we used $\N$ instead of $\N_\tot$ because now this quantity represents only a
\emph{fraction} of the collected photons and $\K$ is the constant which takes into account of the
events not correctly reconstructed.

The non-ideal response of actual polarimeters, either photoelectric or Compton, is usually taken
into account by introducing the modulation factor $\mu$. We defined it with Equation~(\ref{eq:Mu}),
but conceptually it can be seen as the ratio between the number $\N^{\mathrm{csq}}$ of the
events which are cosine square modulated over the total $\N_{\tot}$ for completely polarized
incident radiation. Such a definition is consistent with that already given because, assuming that
the modulation function is represented by the function $A_{1}+B_{1}\cos^2(\phi)$ as in
Section~\ref{sec:Polarimeters}, we have that 
\begin{align}
\mu &=\frac{\N^{\mathrm{csq}}}{\N_\tot}=\frac{\int_0^{2\pi} B_{1}\cos^2\phi\,\d\phi}{\int_0^{2\pi}
\left(A_{1}+B_{1}\cos^2\phi\right)\d\phi} = \notag \\
&= \frac{\pi B_{1}}{2\pi A_{1}+\pi B_{1}} = \frac{B}{2A+B} \; . \notag
\end{align}
Unfortunately, this definition becomes ambiguous if photons are incident off-axis because the
response to polarized radiation is no more a cosine square. Therefore, we find convenient to cope
with the incomplete response of real polarimeters by introducing a new quantity $f$ or f-factor.
To introduce it, let us assume that completely polarized photons are incident on a instrument
and that $\N_\tot$ photons are detected. As a working hypothesis, we can think that among them the
direction of $\N_\mathrm{good}$ events is ``perfectly'' reconstructed and therefore they show a
modulation identical to the emitted one; instead, the remaining $\N_\tot-\N_\mathrm{good}$ events
are reconstructed with a direction which is randomly distributed over the azimuthal angle. Such a
description is an effective representation of the fact that, in reality, the measurement of all
event directions is affected by some uncertainty which can deviate the measured direction to that of
emission of a smaller or larger amount. We define the f-factor as the fraction
$\N_\mathrm{good}/\N_{\tot}$, so that the quantity $\N_\mathrm{good}$ plays for $f$ a role similar
to $\N^{\mathrm{csq}}$ for $\mu$.

In our working hypothesis, \emph{all} and \emph{only} the events correctly reconstructed will
contribute to the component of the modulation function proportional to the emitted distribution and
therefore the factor $\N$ in Equation~(\ref{eq:ffactor}) is exactly $\N_\mathrm{good}=f \N_{\tot}$.
Then
\begin{equation}
\M(\phi) =f \N_{\tot} \left[\P\frac{\cos^2\phi}{\pi} + \frac{(1-\P)}{2\pi}\right] + \K \; .
\notag
\end{equation}
The sum of the events in all azimuthal bins must be equal to $\N_{\tot}$,
\begin{align}
\int_0^{2\pi} \left\{f \N_{\tot}\left[\P\frac{\cos^2\phi}{\pi} + \frac{(1-\P)}{2\pi}\right] +
\mathcal{K}\right\}\d\phi = \N_{\tot} \; ,
\label{eq:Closure}
\end{align}
and then $\mathcal{K}=\N_\tot\frac{1-f}{2\pi}$. The relation between $f$ and the modulation factor
will be better clarified at the end of this section.

The line of reasoning we followed eventually ends up with the conclusion that, in case of a
photoelectric polarimeter, the modulation curve expected to be measured for a certain polarization
degree $\P$ and polarization angle $\varphi_0$ can be written as
\begin{align}
\M(\varphi) &= f \N_{\tot} \left[\P\frac{\cos^2(\varphi-\varphi_0)}{\pi} + 
\frac{(1-\P)}{2\pi}\right] + \notag \\
&+ \N_\tot\frac{1-f}{2\pi}\; ,
\label{eq:MPh2}
\end{align}
where we eventually made use of Equation~(\ref{eq:PhiPh}) to substitute the variable $\phi$, which
is the angle to the direction of polarization, with $\varphi$ which instead is the angle
to some axis of reference of the instrument. The expression in Equation~(\ref{eq:MPh2}) is the
analogous in our treatment of Equation~(\ref{eq:Cos2}). Although the former may appear more
complicated than the latter, they are mathematically equivalent in the sense that both comprises a
cosine square contribution plus a constant term. Rather, the expression obtained in our treatment
has two advantages. The first is that the angle and the degree of polarization, which are explicit
in Equation~(\ref{eq:MPh2}), are the only parameters to be estimated by the fitting procedure. In
fact, $f$ is a ``quality parameter'' of the instrument and it has to be measured by means of
calibration measurements as the modulation factor, and the number of events collected $\N_\tot$ is
known independently by the fit. Instead, the expression given in Equation~(\ref{eq:Cos2}) depends on
three parameters, $A$, $B$ and $\varphi_0$, basically because it does not exploit the ``closure''
condition in Equation~(\ref{eq:Closure}). The second advantage is that Equation~(\ref{eq:MPh2})
allows to distinguish at least formally the contribution to the modulation curve of the events
``correctly'' reconstructed, that is, it discriminates between the constant contribution due to
the unpolarized component and that which is still constant but produced by events whose direction
was not correctly reconstructed. This is not important for the on-axis response but it becomes
essential when inclined incident radiation is considered. To put into evidence the two contributions
due to the events correctly reconstructed coming from polarized and unpolarized radiation, we
rewrite Equation~(\ref{eq:MPh2}) as
\begin{align}
\M(\varphi) & = f \N_{\tot} \left[\P \Phi^\Pol(\varphi) + (1-\P)\Phi^\UnP(\varphi) \right] +
\notag \\
& + \N_\tot\frac{1-f}{2\pi}\; ,
\label{eq:M}
\end{align}
where $\Phi^\Pol(\varphi)$ and $\Phi^\UnP(\varphi)$ are the \emph{normalized} azimuthal
distributions of the emitted events in case of completely polarized and unpolarized radiation:
\begin{subequations}
\label{eq:M_Phi}
\begin{align}
\Phi^\Pol(\varphi) &=
\frac{\int_0^{\pi}\D^\Pol(\varphi,\theta)\sin\theta\d\theta}
{\int_0^{2\pi}\left[\int_0^{\pi}\D^\Pol(\varphi,\theta)\sin\theta\d\theta\right]
\d\varphi} \; ; \\
\Phi^\UnP(\varphi) &= 
\frac{\int_0^{\pi}\D^\UnP(\varphi,\theta)\sin\theta\d\theta}
{\int_0^{2\pi}\left[\int_0^{\pi}\D^\UnP(\varphi,\theta)\sin\theta\d\theta\right]
\d\varphi} \; .
\end{align}
\end{subequations}
Such a definition sums up the procedure that we followed to arrive at Equation~(\ref{eq:NPhPol2})
and Equation~(\ref{eq:NPhUnp}).

Equations~(\ref{eq:M}) and (\ref{eq:M_Phi}) can be trivially extended to the case of Compton
polarimeters. The only practical difference when calculating explicitly the modulation function is
that, on the contrary to photoelectric absorption, in the case of Compton scattering there is an
azimuthal constant contribution in the number of emitted events even in the case of completely
polarized photons. This is due to the well-known fact that the amplitude of the cosine square
modulation with polarization depends on the energy but it is never complete except in the Thomson
limit and for $\theta=\pi/2$. Then, a further azimuthal constant contribution is present in the
modulation function which sums to that due to unpolarized radiation and to that due to the events
not reconstructed correctly. The resulting explicit expression of the modulation function is
qualitatively equivalent to that of photoelectric polarimeters, i.e. Equation~(\ref{eq:MPh2}),
although it is algebraically more complicated than the latter. Therefore, we will not treat the
Compton case explicitly, but in the next section we will start from Equation~(\ref{eq:M}) to derive
the modulation function for off-axis incident radiation in case of both photoelectric and Compton
polarimeters.

It is helpful to conclude this section clarifying the difference between the modulation factor and
the f-factor. As discussed above, the azimuthal response of real instruments does not show a
complete cosine square modulation even in case of completely polarized radiation. In principle,
this is due to a combination of two distinct effects. On the one hand, the modulation of the
\emph{emission} directions may be not complete even for 100\% polarized photons, on the other, the
measurement of the event direction introduces some uncertainty which inherently reduces the
amplitude of the modulation eventually detected. The former is an ``intrinsic'' effect of the
interaction process, whereas the latter is a pure instrumental effect which would be absent
for an ideal device able to reconstruct ``perfectly'' all of the event directions. The modulation
factor mixes up both of these contributions because it is defined starting from the cosine square
modulation eventually measured by the instrument. Therefore, it does not distinguish if a certain
modulation amplitude is obtained with a instrument performing an effective event reconstruction in a
condition of low intrinsic modulation or vice versa, that is, with an instrument performing a poor
event reconstruction in a condition of, e.g., favorable event selection on the polar angle. Let us
assume for example that the modulation factor for two different Compton polarimeter designs is 0.40
at 100~keV, but in one case the instrument is sensitive to all events, whereas the geometry of the
other is such that only events scattered in the interval $\theta=\left[\pi/4,3\pi/4\right]$ are
accepted. Since the intrinsic modulation is higher for the second design but the measured modulation
is identical, it is clear that the second instrument must have a higher probability to not correctly
reconstruct the event direction.

We defined the f-factor to take into account only for the non-ideal response of the instrument to
polarization. In fact, the method we put forward to calculate the modulation function already
includes the possibility that the intrinsic modulation may be not complete; this is implicit
in the definition of the normalized azimuthal distribution of the events which is basically
calculated from the differential cross section of the interaction. Therefore, an ideal instrument
is always characterized by having $f=1$, regardless the amplitude of the ``measured'' modulation,
whereas an ideal instrument will have $\mu=1$ only if the intrinsic modulation is complete.

The f-factor for a real instrument will in general depend on the energy because, usually,
the capability to correctly reconstruct the event direction usually do. This is particularly true
for photoelectric polarimeters because higher is the energy of the photoelectron, longer and easier
to reconstruct is the track. The value of $f$ at a certain energy can be derived by the
corresponding value of the modulation factor with a simple relation, which can be obtained by means
of Equation~(\ref{eq:M}). In fact, if we pose $\P=1$ and name $B_{1}$ the coefficient of the cosine
square contribution and $A_{1}$ the constant term, the modulation factor is as usual
$\mu=\frac{B_{1}}{2A_{1}+B_{1}}$. For example, the modulation function for photoelectric
polarimeters is given in Equation~(\ref{eq:MPh2}) and then we have that
\begin{align}
\mu_{\Ph}&=\frac{B_{1}}{2A_{1}+B_{1}}= \notag \\
&= \frac{\frac{f \N_{\tot}}{{\pi}}}{2\left[
\N_\tot\frac{1-f}{2\pi} \right] + \frac{f \N_{\tot}}{{\pi}}} =f \; . \notag
\end{align} 
The fact that the f-factor coincides with the modulation factor for photoelectric polarimeters is
not surprising. We have restricted ourselves to the case of photoelectric absorption in the K-shell
and then the distribution of the emission directions is intrinsically completely cosine square
modulated for polarized photons. As a consequence, the fraction of events whose emission direction
is correctly reconstructed, which is the f-factor, coincides with that of the events which are
modulated as a cosine square, that is the modulation factor.

\begin{figure}[bp]
\begin{center}
\includegraphics[angle=0,width=8.5cm]{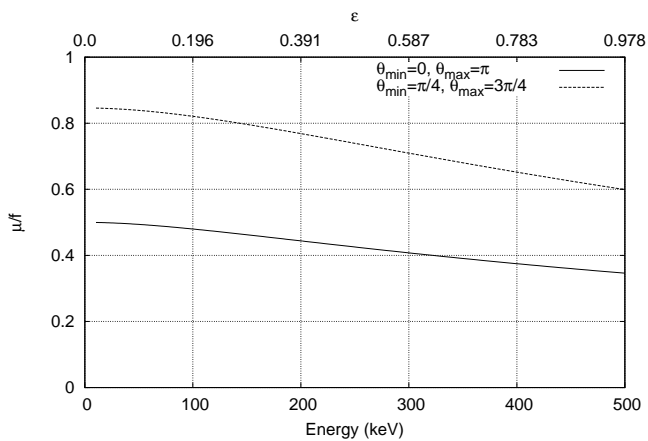}
\end{center}
\caption{Dependency on the energy of the ratio between the modulation factor $\mu$ and the f-factor
$f$ for a Compton polarimeter in case $\theta_{\min}=0$ and $\theta_{\max}=\pi$ (solid line) or
$\theta_{\min}=\pi/4$ and $\theta_{\max}=3\pi/4$ (dashed line). In the latter case the constant of
proportionality between $\mu$ and $f$ is larger because the intrinsic modulation with the
polarization is higher.}
\label{fig:MufRel_Cm}
\end{figure}

The same procedure can be applied also to Compton polarimeters. After some algebra which we
carried out with the help of the Computer Algebra System
\textsc{Maxima}\footnote{\url{http://maxima.sourceforge.net/}}, the result in case there is no
selection on the event polar angle, that is $\theta_{\min}=0$ and $\theta_{\max}=\pi$, is that
\begin{widetext}
\begin{equation}
\mu=\frac{\left(
8{\varepsilon}^{3}+16{\varepsilon}^{2}+10\varepsilon+2\right) \log\left|
2\varepsilon+1\right| -16{\varepsilon}^{3}-16{\varepsilon}^{2}-4\varepsilon}{\left(
4{\varepsilon}^{4}-4{\varepsilon}^{3}-15{\varepsilon}^{2}-10\varepsilon-2\right)
\mathrm{log}\left| 2\varepsilon+1\right|
+2{\varepsilon}^{4}+18{\varepsilon}^{3}+16{\varepsilon}^{2}+4\varepsilon} f \; .
\label{eq:muCompton}
\end{equation}
\end{widetext}
Therefore, the modulation factor and the f-factor are related by a simple linear dependency.
The constant of proportionality takes into account of the fact that the intrinsic modulation of the
Compton scattering with polarization decreases with the energy. Its value, which basically is
the modulation factor for an ideal device, is reported in Figure~\ref{fig:MufRel_Cm}. In the
Thomson limit, Equation~(\ref{eq:muCompton}) becomes $\mu=f/2$, which is the well-known result that
for an ideal Compton polarimeter, i.e., if $f=1$, the modulation factor is 0.5 when there is not any
selection on the scattering angle \citep[see, for example, Figure~5 in][]{Krawczynski2011}.
 
Let us use Equation~(\ref{eq:muCompton}) to derive the f-factor for the two Compton
polarimeters with $\mu=0.40$ taken as an example above. In the first design, all of the scattered
events are accepted and therefore $f\approx0.40/0.48\approx0.83$, where 0.48 is approximately the
value of the constant of proportionality between $\mu$ and $f$ at 100~keV (see the solid line in
Figure~\ref{fig:MufRel_Cm}). The second instrument is sensitive only to the events scattered in the
interval $\theta=\left[\pi/4,3\pi/4\right]$ (see the dashed line in Figure~\ref{fig:MufRel_Cm}) and
therefore $f\approx0.40/0.83\approx 0.49$.

The role of the f-factor becomes relevant as soon as one applies our results to a real instrument.
Although this will be the aim of a future work, in this paper we are mainly interested in
highlighting the effects of the inclined incidence of the photons on the intrinsic polarimeter
response. To avoid mixing them up with those derived from the not-ideal behavior of real
polarimeters, we will assume hereafter that $f=1$ unless otherwise specified.

\subsection{Off-axis generalization} \label{sec:OffAxisProcedure}

The procedure we described in Section~\ref{sec:OnAxis} can be straightforwardly applied also in the
case of photons which are inclined with respect to the instrument, but the crucial point is to
use in Equation~(\ref{eq:M}) the correct angular distribution of the emitted event directions. We
have already seen that it descends by definition from the differential cross section of the
interaction and this obviously still holds in case of inclined photons. Nonetheless, we have to
remember that the expressions we used above, i.e. Equations~(\ref{eq:DPh}) and
Equations~(\ref{eq:DCm}), are valid only when the angles $\theta$ and $\phi$ are defined in a frame
of reference that has the $z$-axis along the direction of incidence and the $x$-axis along the
direction of polarization (see Figure~\ref{fig:Angles}). In this frame of reference, that we will
call hereafter \emph{photon} frame of reference, the azimuthal distribution of the events always
shows a cosine square modulation in case of polarized photons. However, the modulation curve is
constructed by the azimuthal distribution in the \emph{instrument} frame of reference whose
$xy$-plane coincides with the detection plane and the $z$-axis is perpendicular to it. These two
frames of reference are equivalent only if the photons are incident orthogonal to the detection
plane and therefore only in this assumption the instrument ``sees'' the azimuthal distribution as it
is in the photon frame of reference. Instead in the general case of inclined photons, we have to
calculate how the angular distribution of the emitted events transforms in the instrument frame of
reference before being able to derive the modulation function.

\begin{figure*}[htbp]
\begin{center}
\subfloat[\label{fig:Plot_3d_onaxis}]{\includegraphics[width=8.2cm]{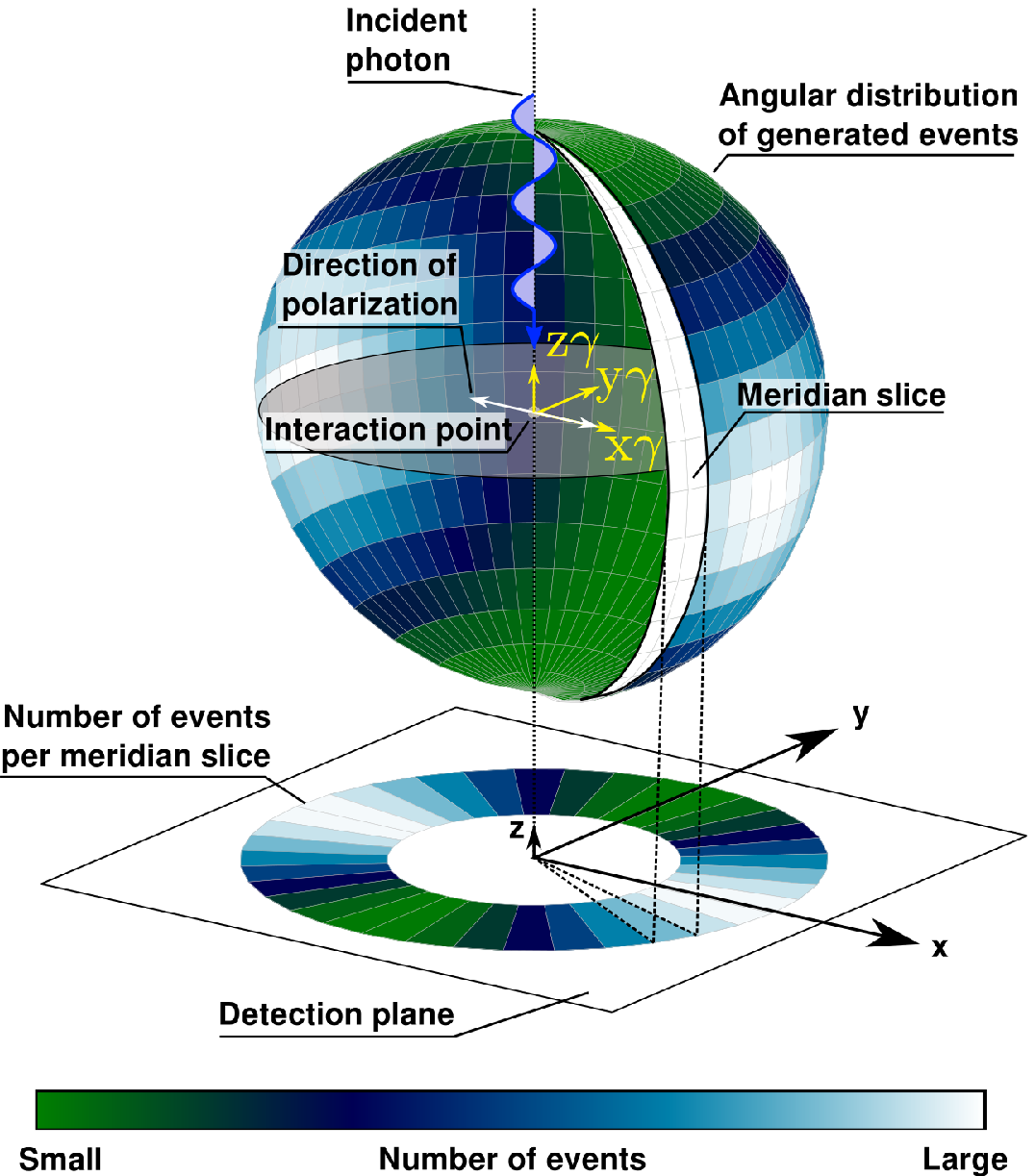}}
\hspace{5mm}
\subfloat[\label{fig:Plot_3d_offaxis}]{\includegraphics[width=8.2cm]{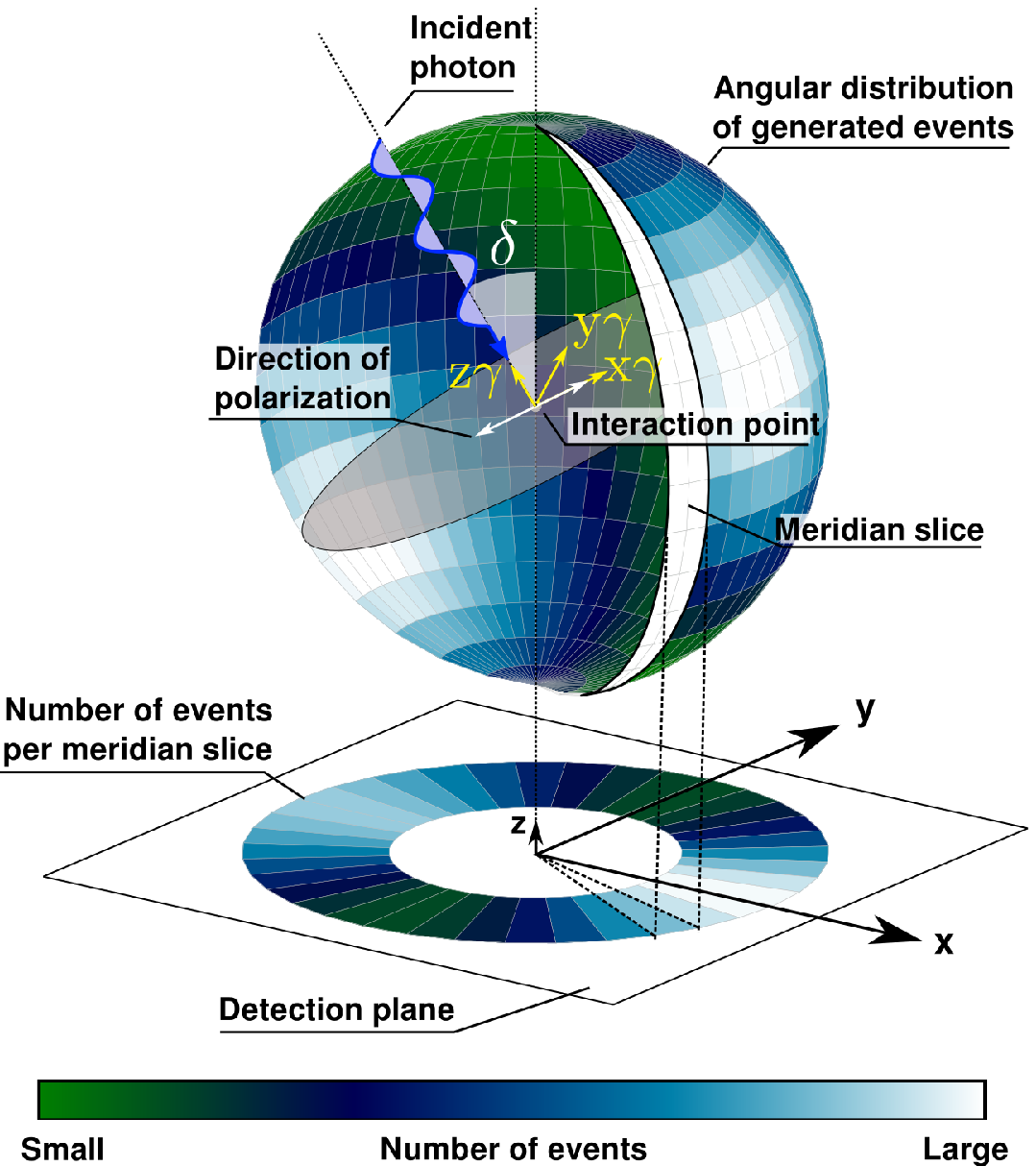}} 
\end{center}
\caption{Qualitative effect of the inclination of the incident photons on the modulation curve
measured by the instrument. The angular distribution of the events is plotted in color code on the 
surface of a sphere with at the center the interaction point. The modulation curve, obtained by
summing the events emitted in each meridian slice, is reported on the detection plane as an
annulus. This picture is referred to the particular case of polarized photons absorbed by
photoelectric effect, but the result for unpolarized radiation and Compton scattering are
qualitatively equivalent.}
\label{fig:Plot_3d}
\end{figure*}

The consequence of the off-axis incidence of the photons on the azimuthal distribution as it is seen
by the instrument is qualitatively illustrated in Figure~\ref{fig:Plot_3d} considering as an
example the case of photoelectric polarimeters and polarized radiation. The angular distribution
of the event directions is plotted in color code on the surface of a sphere with at the center the
interaction point. The modulation curve is constructed by counting how many events are emitted in
each azimuthal angular bin on the detection plane and this is equivalent to total the number of
events in each ``meridian slice''. In case of photons which impinge orthogonally to the detection
plane (see Figure~\ref{fig:Plot_3d_onaxis}), each meridian slice contains simply the events emitted
in the corresponding azimuthal interval in the photon frame of reference. Therefore, the content of
the slice is obtained by simply integrating Equations~(\ref{eq:DPh}) and (\ref{eq:DCm}) over the
appropriate polar interval as we did in Section~\ref{sec:OnAxis}, with the result that the
modulation curve, reported in color code as an annulus on the detection plane, is a cosine square.
Instead, when photons are incident off-axis (see Figure~\ref{fig:Plot_3d_offaxis}), the angular
distribution of the events is rotated with respect to the meridians and the polar interval of the
events which are contained in each slice depends in a complex way on the azimuth. The effect of the
forward bending is particularly relevant because it makes the northern and southern hemisphere of
the angular distribution not symmetric. This breaks the periodicity of the modulation function
modulo $180^\circ$ because the content of opposite meridian slides is different.

\begin{figure}[tbp]
\begin{center}
\includegraphics[angle=0,width=8.5cm]{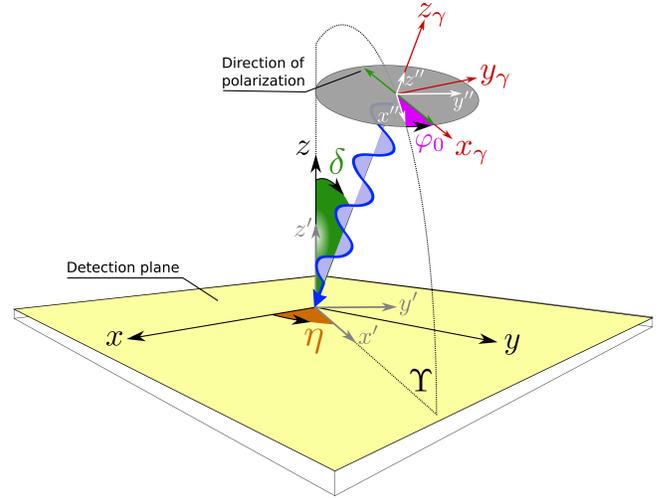}
\end{center}
\caption{Definition of the angles which identify the photon frame of reference
$x_\gamma y_\gamma z_\gamma$ with respect to the instrument one $xyz$. The inclination $\delta$ and
the azimuth $\eta$ characterize the incident direction, while $\varphi_0$ defines the angle of
polarization. $\Upsilon$ is the plane containing the incident direction and the normal to the
detection plane. As discussed in Appendix~\ref{app:CoordTransf}, we indicate as $x'y'z'$ the frame
of reference which is rotated with respect to $xyz$ of $\eta$ around $z$ and as $x''y''z''$ a frame
of reference which is also rotated of $\delta$ around $y'$.}
\label{fig:Inclination}
\end{figure}

The off-axis incidence of the photons can be characterized by means of three angles. Two of
them, the inclination $\delta$ and the azimuth $\eta$, are necessary to describe the incident
direction, while the third one is the angle of polarization (see Figure~\ref{fig:Inclination}). We
will indicate it by $\varphi_0$ as in the previous section, but in this case the angle of
polarization is not measured with respect to some axis of reference of the instrument but with
respect to the meridian plane on which the incident direction lies. These three angles allow to
completely identify the photon frame of reference with respect to the instrument one and, therefore,
the angular distribution of the events $\D$ which is known in the former can be calculated in the
latter by an opportune change of coordinates involving $\varphi_0$, $\delta$ and $\eta$. Basically,
we have to express the spherical coordinates $(\phi;\theta)$ defined in the photon frame of
reference as a function of those $(\varphi;\vartheta)$ defined in the instrument one. The procedure
that we followed requires only standard algebra and we describe it in the
Appendix~\ref{app:CoordTransf}.

Whenever the angular distribution of the event directions in the instrument frame of reference is
known, it is possible to derive the modulation function by applying Equations~(\ref{eq:M}) and
(\ref{eq:M_Phi}). The only caution to take is that, when calculating the normalized azimuthal
distribution of the emitted events, we have to integrate over the polar angle in the
\emph{instrument} frame of reference, that is $\vartheta$ instead of $\theta$ as in
Equation~(\ref{eq:M_Phi}). In the previous section we implicitly assumed the equivalence of
$\vartheta$ and $\theta$ to avoid unnecessary complications in case of on-axis photons. Therefore,
the correct expressions for the normalized distribution of the emitted events are
\begin{subequations}
\label{eq:M_Phi2}
\begin{align}
\Phi^\Pol(\varphi) &=
\frac{\int_0^{\pi}\D^\Pol(\varphi,\vartheta)\sin\vartheta\d\vartheta}
{\int_0^{2\pi}\left[\int_0^{\pi}\D^\Pol(\varphi,\vartheta)\sin\vartheta\d\vartheta\right
]
\d\varphi} \; ; \\
\Phi^\UnP(\varphi) &= 
\frac{\int_0^{\pi}\D^\UnP(\varphi,\vartheta)\sin\vartheta\d\vartheta}
{\int_0^{2\pi}\left[\int_0^{\pi}\D^\UnP(\varphi,\vartheta)\sin\vartheta\d\theta\right]
\d\varphi} \; .
\end{align}
\end{subequations}
Equations~(\ref{eq:M}) and (\ref{eq:M_Phi2}) will be used in the next section to derive the
modulation function for photoelectric and Compton polarimeters in the general case of off-axis
photons.

\section{Modulation function for inclined sources} \label{sec:ModInclined}

\subsection{A ``simple'' scenario} \label{sec:SimpleScenario}

\begin{figure}[tbp]
\begin{center}
\includegraphics[angle=0,width=8.5cm]{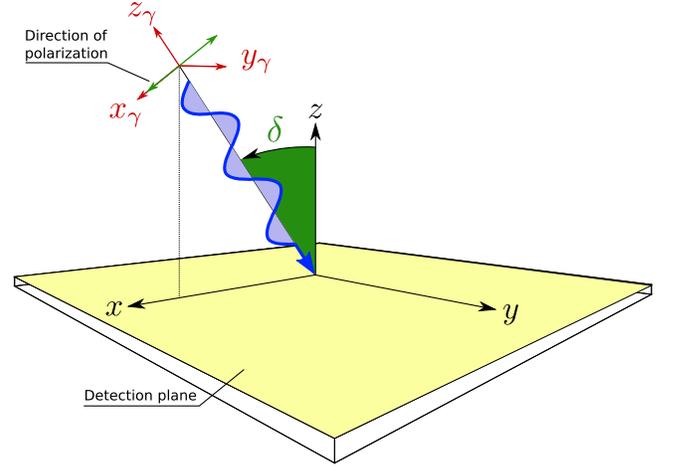}
\end{center}
\caption{Geometry assumed throughout Section~\ref{sec:SimpleScenario} and
Appendix~\ref{app:SimpleScenario}. In this configuration, the values of the angle
of polarization and of the azimuth of the incident beam are zero, $\varphi_0=0$ and $\eta=0$.}
\label{fig:InclinationSimple}
\end{figure}

The procedure described in Section~\ref{sec:OffAxisProcedure} and its results can be better
appreciated if we consider firstly a ``simple'' case whose algebra can be handled explicitly.
Therefore, in this section we will discuss the modulation function of photoelectric and Compton
polarimeters in case both the incident direction and the direction of polarization lay on the $xz$
plane. According to the notation introduced by Figure~\ref{fig:Inclination}, we will change
the inclination $\delta$ keeping the angle of polarization and the azimuth of the incident beam
constant, $\varphi_0=0$ and $\eta=0$ (see Figure~\ref{fig:InclinationSimple}). The modulation
function in such a configuration was already studied by \citet{Muleri2008c} but it is still useful
to face it with the formalism developed in Section~\ref{sec:Method}.

The calculus of the modulation function in case of photoelectric polarimeters is developed in
details in the Appendix~\ref{app:SimpleScenario}. Here, we report only the result:
\begin{align}
\M_\Ph(\beta,\varphi,\delta) &= f \N_{\tot} \left\{\P\left[
-\frac{9\beta\cos^{2}\delta\sin\delta}{8} {\cos^{3}\varphi} +
\right. \right. \notag \\ & \left.\left.
+ \frac{{\cos^{2}\delta}}{\pi}{\cos^{2}\varphi} +
\right. \right. \notag \\ & \left.\left.
+\frac{3\beta\left(3{\cos^{2}\delta}-1\right)\sin\delta }{8}\cos\varphi + \frac{\sin^{2}\delta}
{2\pi}
\right]+\right. \notag \\ &\left.+(1-\P)\left[ 
\frac{9\beta{\sin^3\delta}}{16}{\cos^3\varphi}-\frac{{\sin^2\delta}}{2\pi}{\cos^2\varphi}+ 
\right. \right. \notag \\ & \left. \left.
+\frac{3\beta(3{\cos^2\delta}-4)\sin\delta}{16}\cos\varphi +
\right. \right. \notag \\ & \left.\left.
+\frac{3-{\cos^2\delta}}{4\pi} 
\right]\right\} + \N_\tot\frac{1-f}{2\pi} \; .
\label{eq:PhMSimple}
\end{align}
As we discuss in the appendix, this solution is not \emph{exact} because for the sake of simplicity
we developed the energy dependence of the angular distribution at the first order in the
photoelectron velocity $\beta$, that is, we assumed that
\begin{align}
\DPhPol &\approx \sin^2\theta\,\cos^2\phi\left(1+4\beta\cos\theta\right) \; ; \notag \\
\DPhUnP &\approx \frac{1}{2}\sin^2\theta\left(1+4\beta\cos\theta\right) \; . \notag
\end{align}
Such a linear approximation is adequate to qualitatively illustrate the wealth of effects caused by
the inclined incidence of the photons, which is the primary aim of this paper. Nevertheless, in case
of a real instrument it is necessary to carefully evaluate how many terms of the Maclaurin series
are required to model with sufficient accuracy the modulation function in the whole energy range of
the instrument. As we discuss at the end of Appendix~\ref{app:SimpleScenario}, the first
order approximation may be inadequate for most applications because it provides a good
precision only at very low energy.

Equation~(\ref{eq:PhMSimple}), which collapses as expected to Equation~(\ref{eq:MPh2}) for
$\delta=0$, gives an insight of how more complex is the response of the instrument in case of
inclined photons, even in the simple geometry assumed in this section. We can in principle
distinguish two classes of effects, one due to the inclination of the incident direction which
depends on $\delta$ and the other caused by the forward bending which is therefore energy dependent.
In the low energy limit, that is if $\beta=0$, the odd powers of $\cos\varphi$ vanish and the
modulation function becomes
\begin{align}
\M_\Ph(\beta=0,\varphi,\delta) &= f \N_{\tot} 
\left\{\P
\left[\frac{{\cos^{2}\delta }}{\pi}{\cos^{2}\varphi} +\frac{{\sin^{2}\delta }}{2\pi }\right]+\right.
\notag \\
&\left.+ (1-\P)
\left[\frac{-{\sin^2\delta}}{2\pi}{\cos^2\varphi}+ 
\right.\right. \notag \\ & + \left. \left.
\frac {3-{\cos^2\delta}}{4\pi}
\right]\right\} + \N_\tot\frac{1-f}{2\pi} \; .
\label{eq:PhMSimpleLE}
\end{align}
Such a function shows a cosine square modulation in case of polarized radiation exactly as the
modulation function on-axis, in fact if $\P=1$ we obtain that
\begin{align}
\M_\Ph(\beta=0,\varphi,\delta,\P=1) &= f \N_{\tot} 
 \left[\frac{{\cos^{2}\delta }}{\pi}{\cos^{2}\varphi} + \notag \right. \\ &
\left. +\frac{{\sin^{2}\delta }}{2\pi}\right] +
\N_\tot\frac{1-f}{2\pi}.
\notag
\end{align}
The only differences with Equation~(\ref{eq:MPh2}) are the presence of a $\cos^2\delta/\pi$
factor in front of the cosine square modulation and the $\sin^{2}\delta/(2\pi)$ constant term. Such
additional contributions have the net effect of decreasing the amplitude of the modulation of a
factor which, in case of an ideal device with $f=1$ or $\mu=1$ on-axis, is $\cos^2\delta$. Using the
standard on-axis analysis, this would be interpreted as a reduction of the modulation factor due to
the fact that the instrument ``works worse'' when photons are incident off-axis. Instead, in our
view this effect is not instrumental but it is intrinsic and unavoidable as long as the polar
direction of the event is unknown. Nonetheless, the most interesting result of
Equation~(\ref{eq:PhMSimpleLE}) is the presence of a $\cos^2\varphi$ term in the modulation function
even if $\P=0$ and then a modulated cosine square signal has to be expected also for completely
unpolarized radiation if it is incident off-axis. The amplitude of such a contribution increases
with $\delta$ and, interesting enough, the negative sign makes its phase opposite with respect to
the signal obtained for polarized radiation. The fact that, at least in the low energy limit, the
modulation function for unpolarized and inclined photons has exactly the same azimuthal dependency
as that obtained for polarized radiation poses a serious issue. It suggests that there is a certain
degeneracy among the different parameters on which the modulation function depends and this puts
into question the capability to derive unambiguously the degree and the angle of polarization. We
will discuss more on such a degeneracy in Section~\ref{sec:Discussion}.

If we now drop the assumption of $\beta=0$, the modulation function reported in
Equation~(\ref{eq:PhMSimple}) loses the usual cosine square dependency for both polarized and
unpolarized radiation because $\cos\varphi$ terms with other powers are present. Such
contributions are linear in $\beta$ only because we have stopped the Maclaurin series at the first
order. In case of better approximations, terms with higher powers in $\beta$ and $\cos\varphi$
contributions with powers higher than the third are to be expected, with the same effect of breaking
the periodicity of the modulation function modulo $180^\circ$. This is caused ultimately by the fact
that when the forward bending is combined with the projection on the detection plane, opposite
azimuthal bins in the instrument frame of reference \emph{appear} not equivalent to the instrument,
although being obviously still physically equivalent in the photon frame of reference.

\begin{figure*}[tbp]
\begin{center}
\subfloat[\label{fig:PhPolVsDelta}]{\includegraphics[angle=0,totalheight=5.6cm]{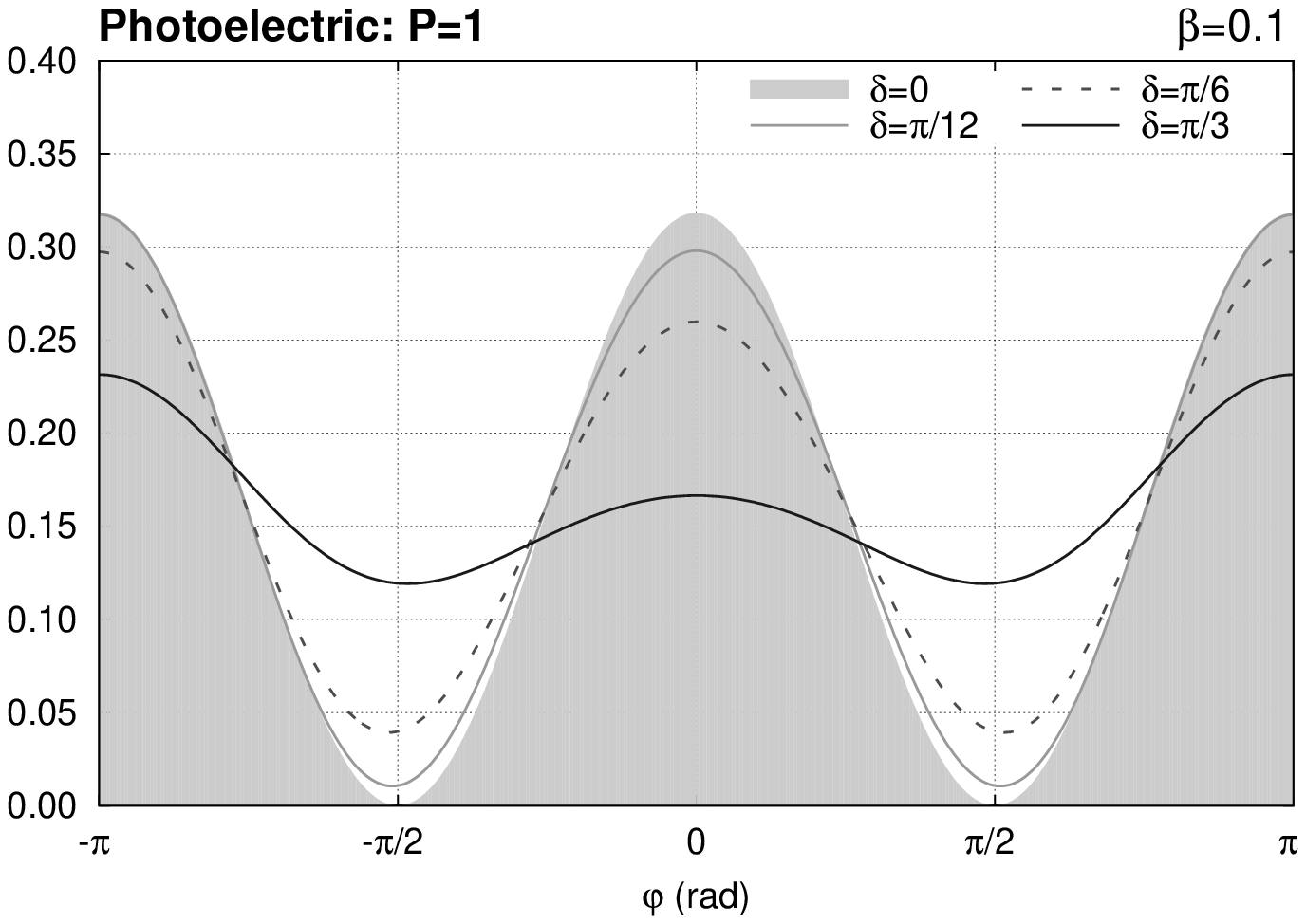}}
\hspace{1mm}
\subfloat[\label{fig:PhUnPVsDelta}]{\includegraphics[angle=0,totalheight=5.6cm]{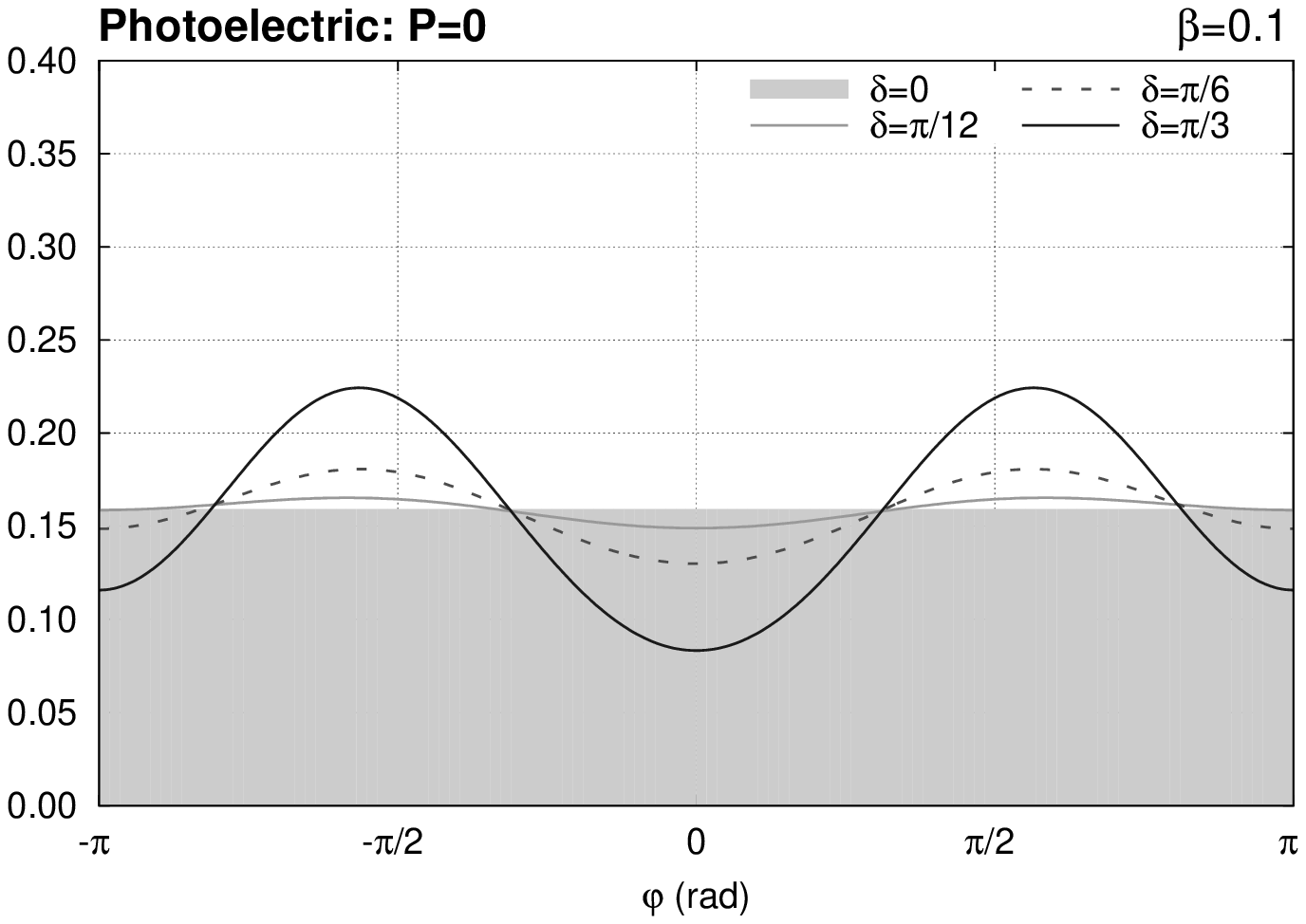}}
\end{center}
\caption{Modulation function for increasing values of the inclination $\delta$ in case of
photoelectric polarimeters. The case of completely polarized and unpolarized radiation is in (a) and
(b), respectively. It is assumed the geometry in Figure~\ref{fig:InclinationSimple}, that is
$\varphi_0=0$ and $\eta=0$, and that $\beta=0.1$, $f=1$ and $\N_\tot=1$.}
\label{fig:PhVsDelta}
\end{figure*}

The off-axis modulation function for $\beta=0.1$, that corresponds to photoelectrons of about
$2.6$~keV, is compared with that on-axis in Figure~\ref{fig:PhVsDelta} assuming an ideal instrument
with $f=1$. The on-axis response is the light-gray filled function, while the response
for increasing values of the inclination $\delta$ is reported as solid or dashed lines. The cases of
completely polarized and unpolarized radiation are reported in Figure~\ref{fig:PhPolVsDelta} and
Figure~\ref{fig:PhUnPVsDelta}, respectively. As discussed above, the amplitude of the modulation
for polarized photons decreases with the inclination and the two peaks are no more identical,
because of the departure from the cosine square behavior, also at relatively low energies and
inclinations. The emergence of a modulation for unpolarized radiation off-axis is equally
evident in Figure~\ref{fig:PhUnPVsDelta}, although in this case its asymmetry is less manifest.

\begin{figure*}[htbp]
\begin{center}
\subfloat{\includegraphics[angle=0,totalheight=5.6cm]{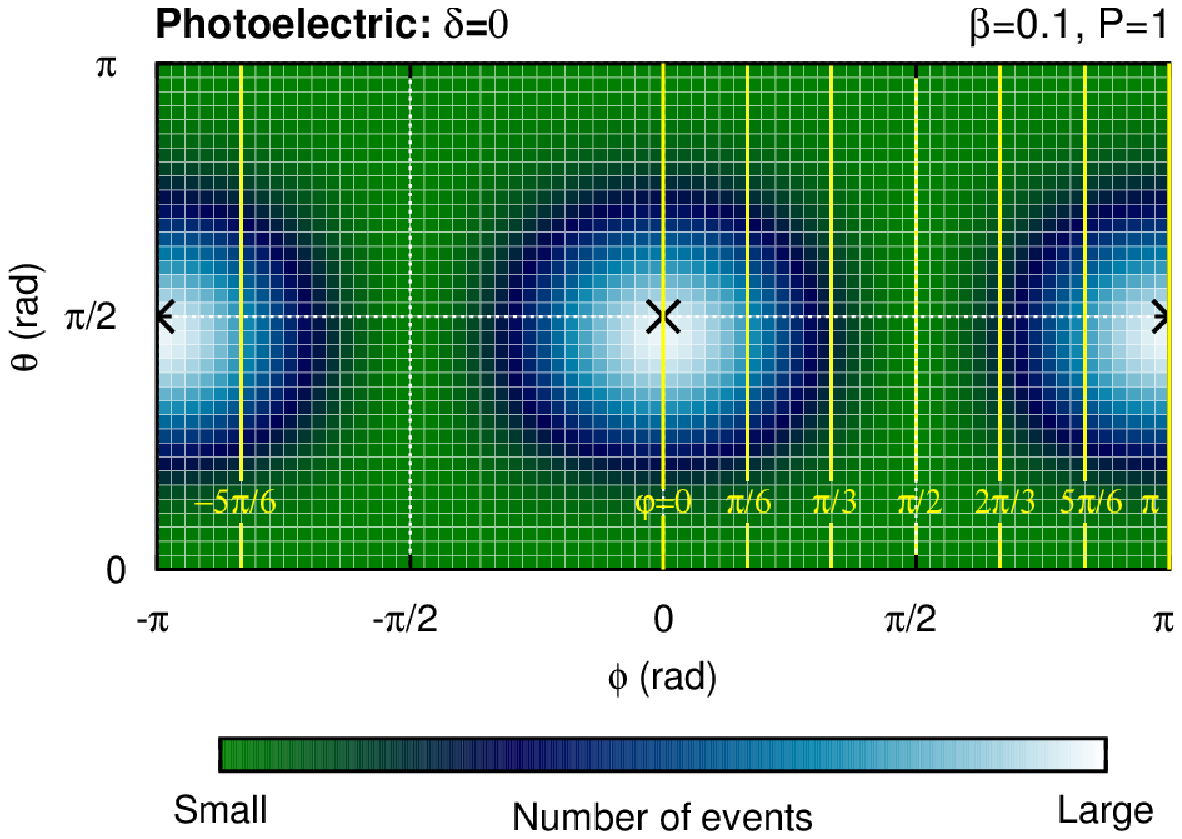}}\hspace{1mm}
\subfloat{\includegraphics[angle=0,totalheight=5.6cm]{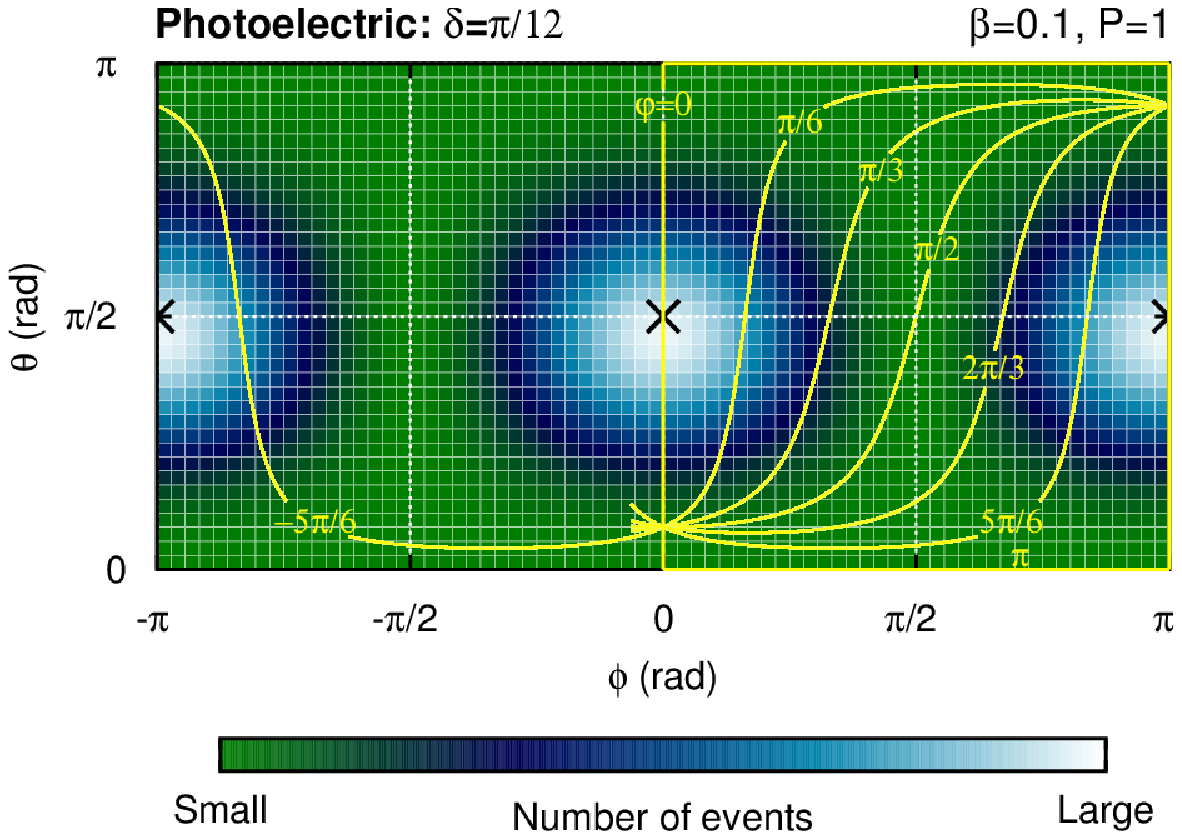}}
\\
\subfloat{\includegraphics[angle=0,totalheight=5.6cm]{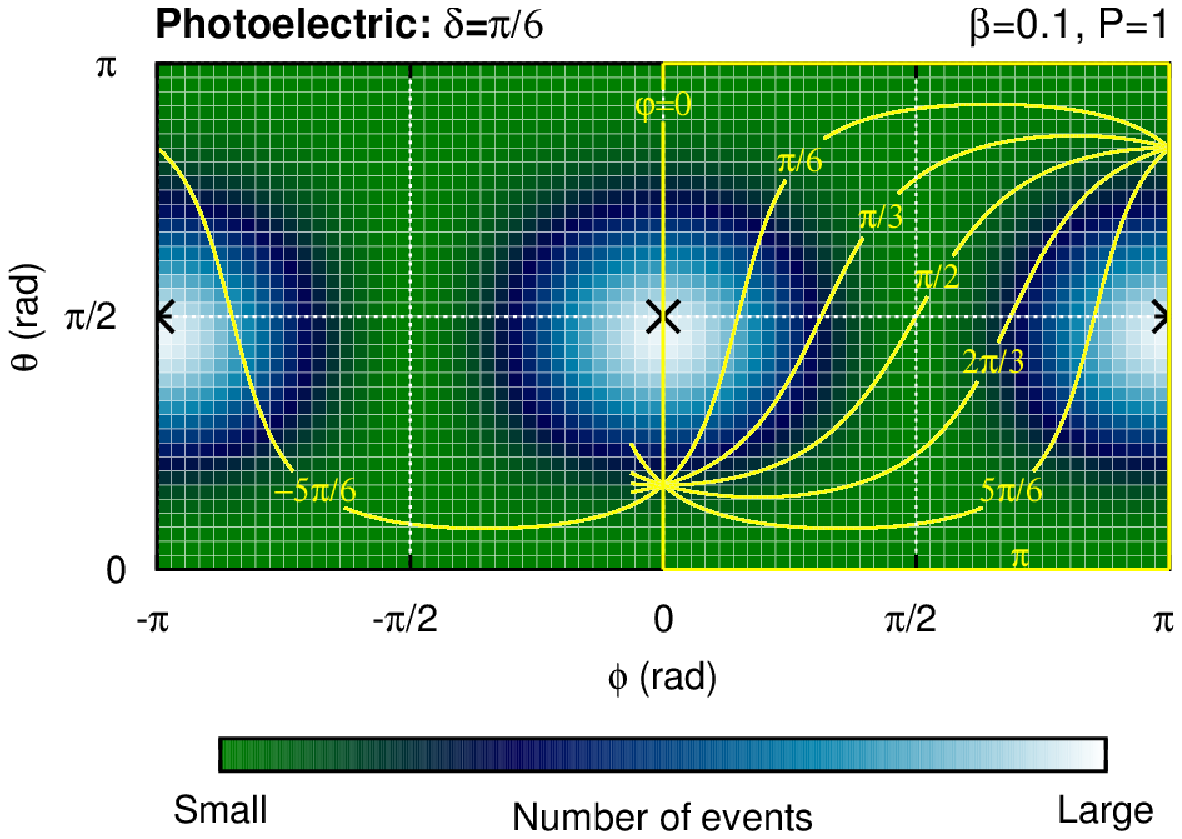}}\hspace{1mm}
\subfloat{\includegraphics[angle=0,totalheight=5.6cm]{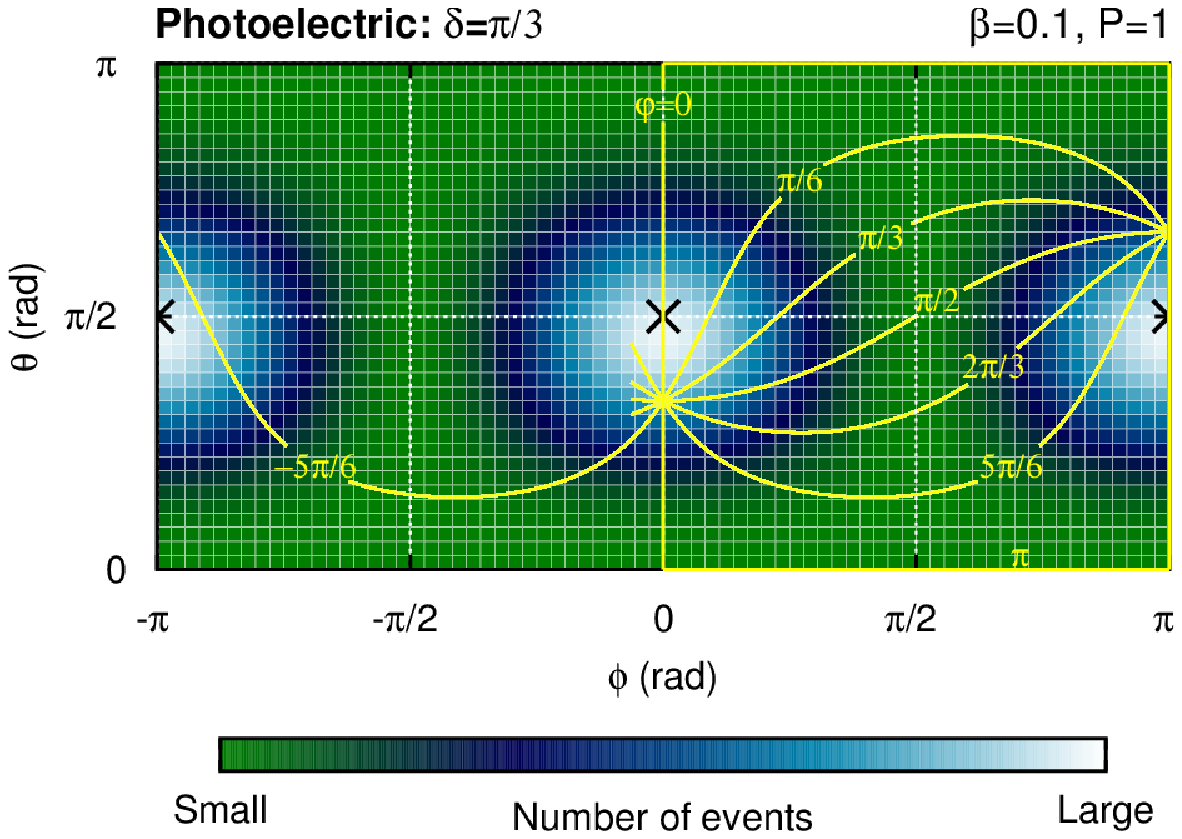}}
\end{center}
\caption{Angular distribution of the events plotted in color code in the photon frame of reference
in case of photoelectric polarimeters and polarized radiation. The emission is concentrated along
the direction of the electric field, distinguished by the black thick crosses, and on the plane
$\theta=\pi/2$ except for the effect of the forward bending. We plotted over it some of the
meridians in which we can think to divide the emitted events to derive the modulation curve. The
fact that the meridians pass from vertical to complex curves allows to qualitatively explain the
behavior of the modulation function off-axis. See the text for further details.}
\label{fig:PhEvents}
\end{figure*}

The behavior of the modulation function can be qualitatively understood by looking at
Figure~\ref{fig:PhEvents}, where we take as an example the case of polarized photons and
photoelectric polarimeters. The angular distribution of the events in the photon frame of reference,
that is $\D_\Ph\sin\theta$, is plotted using the color code on the $\phi\theta$ plane, where $\phi$
and $\theta$ are as usual the spherical coordinate in the photon frame of reference. As discussed in
Section~\ref{sec:IntProc}, the emission is concentrated along the direction of polarization,
characterized by the black thick crosses, and on the plane $\theta=\pi/2$, that is that orthogonal
to the incident direction of the photons, except for the forward bending effect. The latter is
evident in the figure as the peaks of the distribution are not coincident with the black thick
crosses. We have already seen at the beginning of Section~\ref{sec:OffAxisProcedure} that an
effective way to visualize how the modulation curve is constructed is to imagine that the events are
emitted on a sphere centered in the absorption point (see Figure~\ref{fig:Plot_3d}). Then, the value
of the modulation function in the bin $\varphi_\mathrm{i}<\varphi<\varphi_\mathrm{i+1}$ is the total
number of events emitted in the meridian slice limited by the meridians $\varphi=\varphi_\mathrm{i}$
and $\varphi=\varphi_\mathrm{i+1}$. We plotted in Figure~\ref{fig:PhEvents} some of such meridians
as they appear in the photon frame of reference. If we want to know the value of the modulation
function for $\varphi=\bar{\varphi}$, we have to just follow the meridian $\varphi=\bar{\varphi}$
and sum the number of events emitted in the strip around of it. When photons are not inclined (top
left panel in the figure), the photon and the instrument frame of reference are equivalent and
so are the spherical coordinates, that is $\varphi\equiv\phi$ and $\vartheta\equiv\theta$. The
result is that the meridians $\varphi=\mathrm{constant}$ are parallel and vertical lines in the
$\phi\theta$ plane and the content of each meridian slices, and then the modulation function, is
simply a mirror of the azimuthal cosine square modulation in the photon frame of reference. The
meridians obtained if $\delta\neq0$ are no more simple vertical lines and this allows to explain
both the reduction of the modulation for completely polarized photons and the loss of periodicity
modulo 180$^\circ$ that we discussed above. The first effect is a indication of the fact that, even
orthogonal to the polarization, there is always a certain number of emitted events and to understand
why we have to simply follow the meridian $\varphi=\pi/2$ on the $\phi\theta$ plane. On-axis, this
meridian is equivalent to the line $\phi=\pi/2$ and, since there is not emission at $\phi=\pi/2$ for
any value of $\theta$, the total number of events emitted at $\varphi=\pi/2$ is zero and the
modulation is complete. If $\delta\neq0$, the meridian $\varphi=\pi/2$, as any other, crosses
regions in which the emission is not zero and then when we sum the number of events emitted in the
around of it we obtain a non-zero value of the modulation function. As $\delta$ increases, the
meridian passes closer and closer to the peak of the distribution and then the number of events
emitted at $\varphi=\pi/2$ increases. Analogously, the loss of periodicity over 180$^\circ$ can be
understood by following two opposite meridians, e.g. that corresponding to $\varphi=\pi/6$ and
$\varphi=-5\pi/6$ depicted in Figure~\ref{fig:PhEvents}. On-axis, they cross regions in the
$\phi\theta$ plane which are different, but the number of the emitted events on the two paths is
exactly the same for symmetry reasons, essentially the angular distribution of the events is
identical for $\phi=\pi/6$ and $\phi=-5\pi/6$. Off axis the path of the two meridians retains a
certain degree of symmetry, basically they can be regarded as being specular with respect to the
horizontal axis $\theta=\pi/2$, but in this case the symmetry around such an axis of the angular
distribution is lost because of the forward bending effect. As a consequence, the number of events
emitted in the around of two opposite meridians is different and so is the value of the modulation
function in two corresponding bins which are $180^\circ$ out of phase.

\begin{figure}[htbp]
\begin{center}
\includegraphics[angle=0,totalheight=5.6cm]{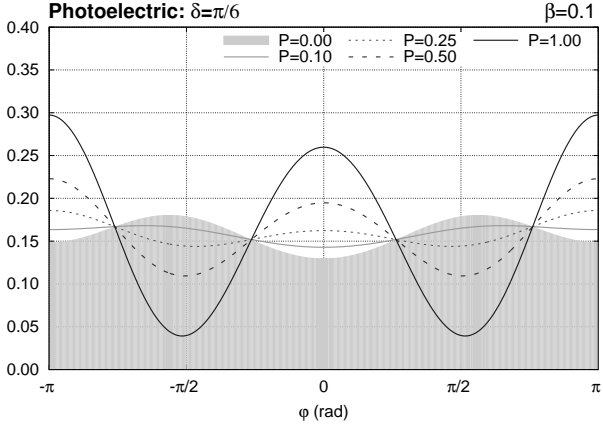}
\end{center}
\caption{Evolution of the modulation function for an increasing degree of polarization in case
of photoelectric polarimeters. The inclination and the value of $\beta$ are fixed to $\pi/6$ and
0.1, respectively. As above, $f=1$ and $\N_\tot=1$.}
\label{fig:PhModVsPol}
\end{figure}

We report in Figure~\ref{fig:PhModVsPol} how the modulation function changes by increasing the
degree of polarization but maintaining constant the inclination and the energy, $\delta=\pi/6$ and
$\beta=0.1$. In this figure, the light-gray filled curve is the modulation function for $\P=0$ and
the solid and dashed lines refers to increasing values of the polarization. We have already seen
that the modulation function in case of off-axis unpolarized radiation shows a nearly cosine square
contribution, whose amplitude increases with the inclination and whose phase is opposite to that due
to the polarized component. As a consequence, we obtain a (nearly) flat modulation not for
unpolarized radiation, but when the modulation due to the polarized component nearly compensates for
the unpolarized one. The value of $\P$ at which the former takes over of the latter depends on the
inclination, because the higher the inclination, the larger the modulation for unpolarized radiation
and the lower that for polarized photons, but in general a small modulated signal for off-axis
incident radiation does not imply a small polarization.

\begin{figure*}[htbp]
\begin{center}
\subfloat[\label{fig:CmPolVsDelta}]{\includegraphics[angle=0,totalheight=5.6cm]{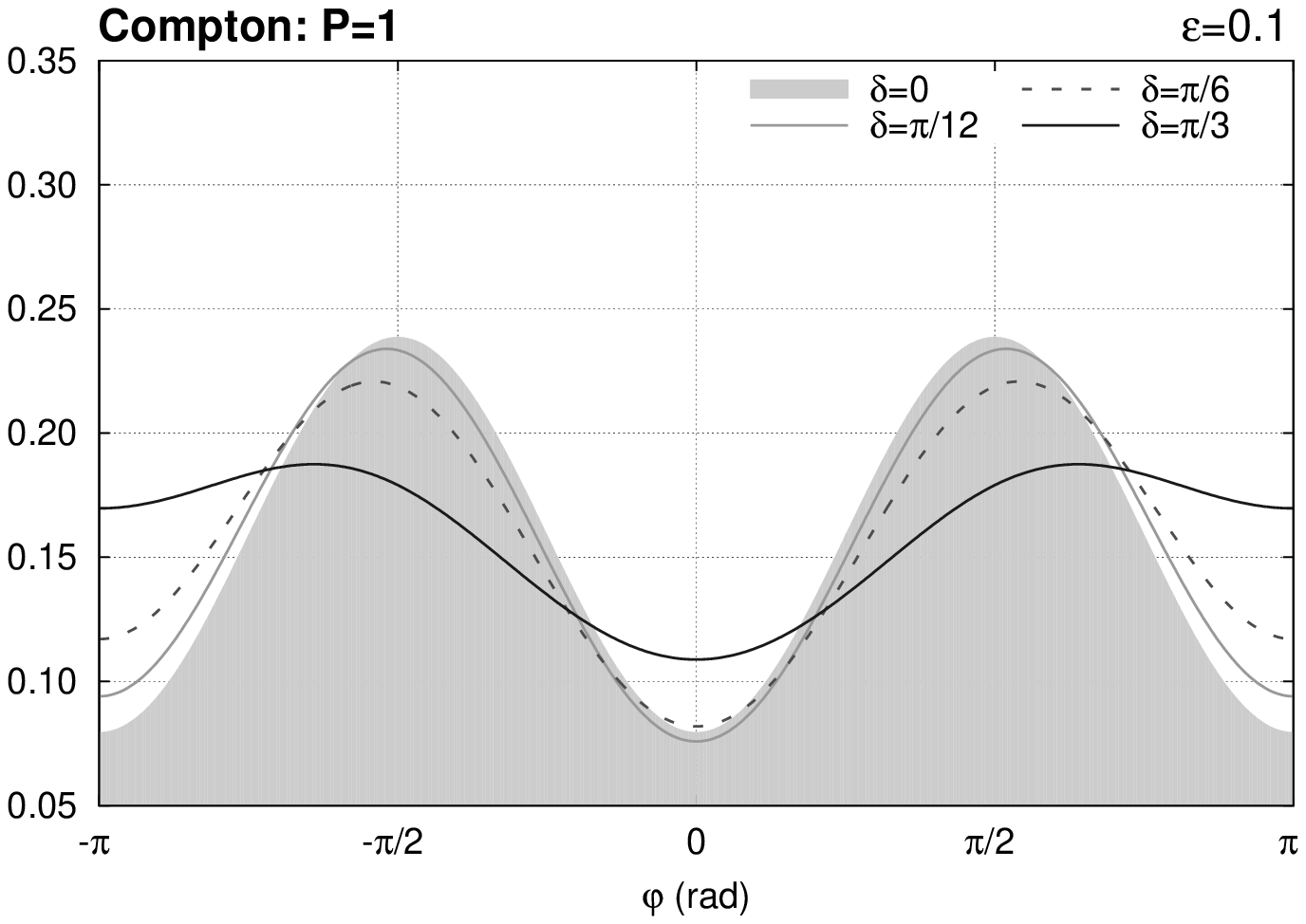}}
\hspace{1mm}
\subfloat[\label{fig:CmUnPVsDelta}]{\includegraphics[angle=0,totalheight=5.6cm]{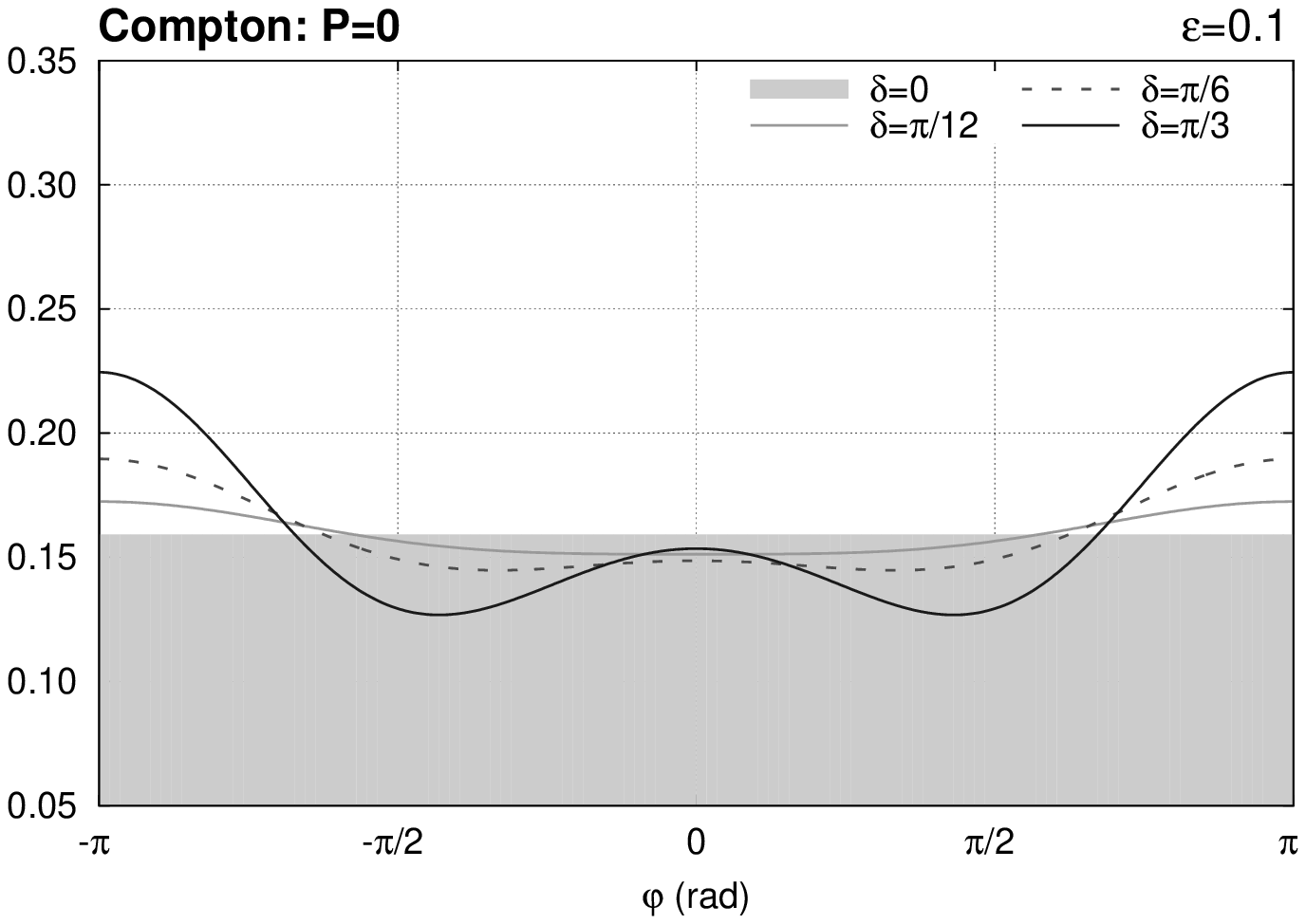}}
\end{center}
\caption{The same as Figure~\ref{fig:PhVsDelta} but in the case of Compton polarimeters.
The modulation functions for completely polarized (a) and unpolarized (b) radiation impinging on
the detector as in Figure~\ref{fig:InclinationSimple} is calculated assuming $\varepsilon=0.1$,
$f=1$ and $\N_\tot=1$.}
\label{fig:CmVsDelta}
\end{figure*}

The response of a Compton polarimeter in the simple geometry reported in
Figure~\ref{fig:InclinationSimple} can be derived by repeating the same procedure that we followed
for photoelectric instruments, with the only difference of using the event distribution for
scattering. Also in this case, it is convenient  for the sake of simplicity to develop $\D_\Cm$ at
the first order with respect to the energy $\varepsilon$ so that
\begin{align}
\DCmPol &\approx 2(1-2\varepsilon\cos\theta) +
\notag \\
& -2\sin^2\theta\cos^2\phi[1-2\varepsilon(1-cos\theta)]
\; \notag \\
\DCmUnP &\approx 2(1-2\varepsilon\cos\theta)-\sin^2\theta[1-2\varepsilon(1-cos\theta)] \; . \notag
\end{align}
In this assumption, the resulting modulation function is
\begin{align}
\M_\Cm(\varepsilon,\varphi,\delta) &= f \N_{\tot} \left\{\P\left[
-\frac{9\varepsilon\cos^2\delta\sin\delta}{32(2\varepsilon-1)}\cos^3\varphi
+ \right. \right. \notag \\ & \left. \left. 
-\frac{\cos^2\delta}{2\pi}\cos^2\varphi+
\right. \right. \notag \\ & \left.\left. 
+\frac{9\varepsilon(\cos^2\delta+1)\sin\delta}{32(2\varepsilon-1)}\cos\varphi
+\frac{2+\cos^2\delta}{4\pi}
\right]+\right. \notag \\
&\left.+(1-\P)\left[ 
\frac{9\varepsilon\sin^3\delta}{64(2\varepsilon-1)}\cos^3\varphi+
\right. \right. \notag \\ & \left.\left.
+\frac{\sin^2\delta}{4\pi} cos^2\varphi+
\frac{3\varepsilon(3\cos^2\delta+4)\sin\delta}{64(2\varepsilon-1)}\cos\varphi+
\right. \right. \notag \\ & \left. \left. +\frac{\cos^2\delta+3}{
8\pi}
\right]\right\} + \N_\tot\frac{1-f}{2\pi} \; . \notag
\end{align}
We report in Figure~\ref{fig:CmVsDelta} the modulation function for increasing inclinations $\delta$
and $\varepsilon=0.1$, that is when the incident photon energy is about 50~keV. In principle, the
effect of the inclination on the modulation function is very similar to that already discussed for
photoelectric polarimeters, that is, there is a reduction of the modulation for polarized
radiation and the emergence of a modulation even for unpolarized photons. Also in the case of
Compton polarimeters, the modulation function is a cosine square only in the low energy limit and
the periodicity over 180$^\circ$ is broken when energy-dependent terms are introduced. At this
regard, it is worth noting an important result that we will discuss in more detail in the next
section, that is, the peak of the modulation function for off-axis polarized photons does not occur
for $\varphi=\pi/2$ (see Figure~\ref{fig:CmPolVsDelta}). Nonetheless, it is evident that the
amplitude of the effect is somehow reduced with respect to photoelectric polarimeters, at least if
we restrict ourselves to inclinations lower than $\pi/6$. Unfortunately, this does not mean that the
modulation function of Compton polarimeters is less affected by the off-axis incidence of the
photons and, as we will see in the next section, Compton polarimeters behave much like photoelectric
instruments.

\begin{figure}[htbp]
\begin{center}
\includegraphics[angle=0,totalheight=5.6cm]{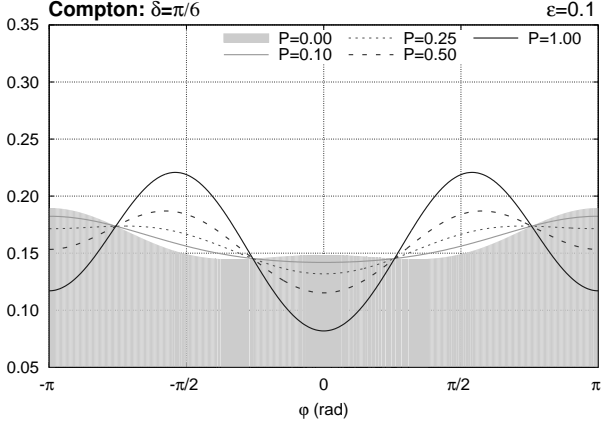}
\end{center}
\caption{The same as Figure~\ref{fig:PhModVsPol} but for Compton polarimeters. The inclination and
the value of $\varepsilon$ are fixed to $\pi/6$ and 0.1, respectively. As usual, $f=1$ and
$\N_{\tot}=1$.}
\label{fig:CmModVsPol}
\end{figure}

Eventually, we report in Figure~\ref{fig:CmModVsPol} the modulation function for a fixed inclination
and energy but increasing polarization. As in the case of photoelectric polarimeters, the response
is never flat and the polarized signal becomes dominant with respect to the unpolarized one when the
polarization degree is a few tens of percent.

\subsection{Unrestrained off-axis incident direction} \label{sec:Generic}

The modulation function becomes much more complex to handle when all of the three angles which
describe the incident geometry, that is $\delta$, $\eta$ and $\varphi_0$, are not constrained in a
particular configuration. Therefore, we will not try to write down $\M$ explicitly but we will
characterize its behavior with the help of the \textsc{Maxima} software.

\begin{figure*}[htbp]
\begin{center}
\subfloat[]{\includegraphics[angle=0,totalheight=5.6cm]{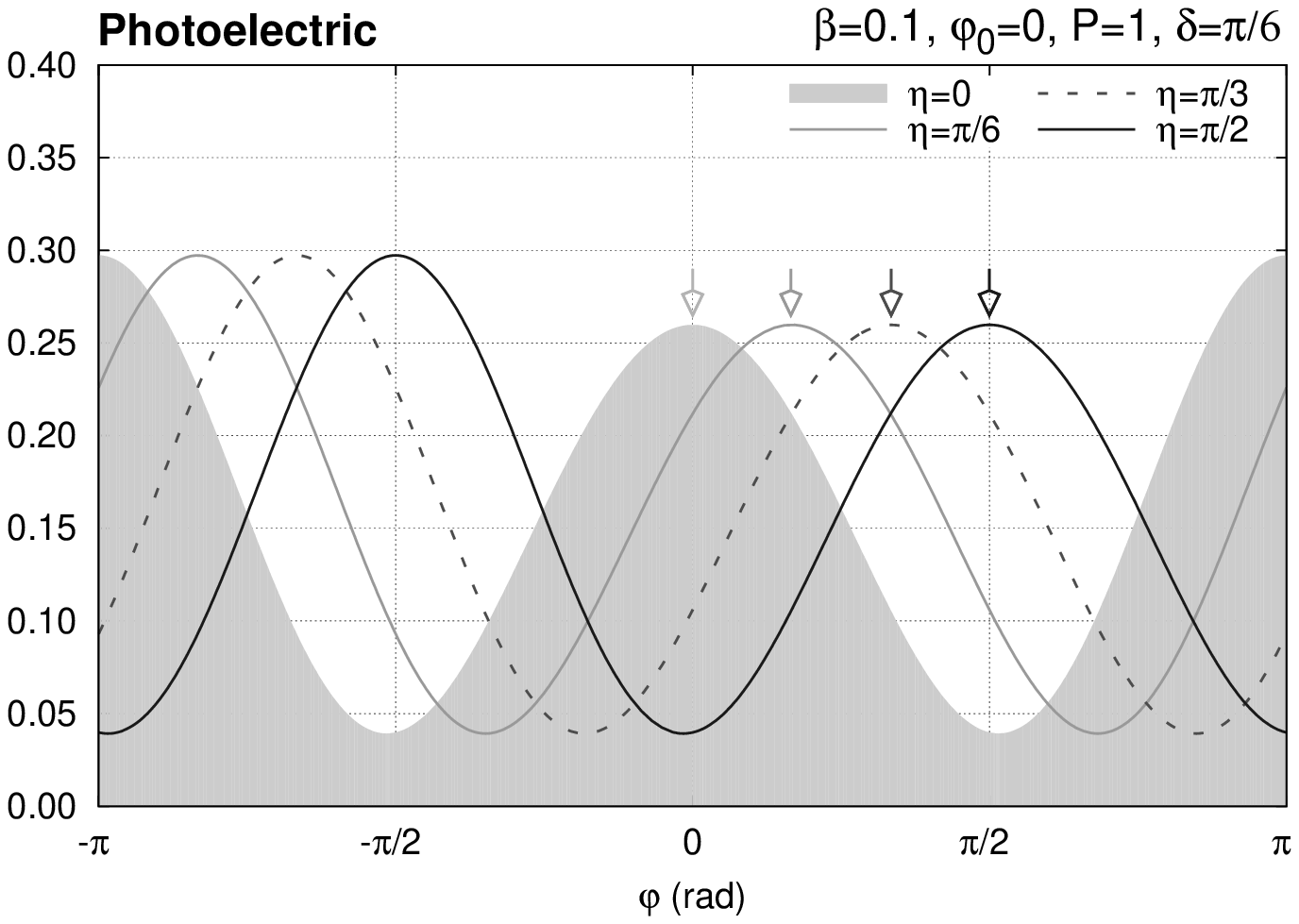}}\hspace{1mm}
\subfloat[]{\includegraphics[angle=0,totalheight=5.6cm]{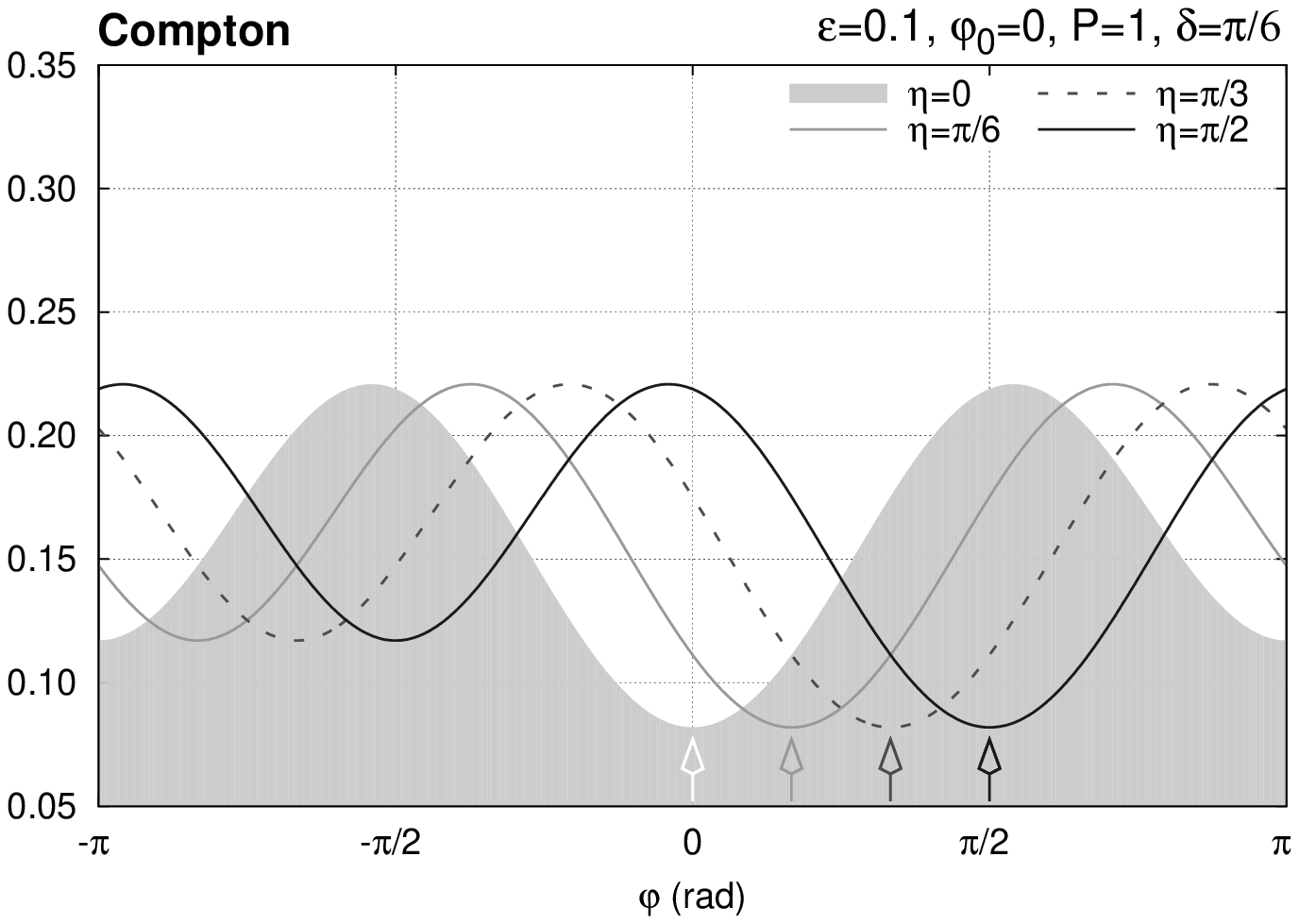}}
\end{center}
\caption{Dependency of the modulation function for photoelectric (a) and Compton (b) polarimeters
on the azimuthal direction of the impinging photons $\eta$. The net effect is just a shift of
the modulation function. For illustration purposes only, it is assumed that the radiation is
polarized, the inclination is $\pi/6$ and the angle of polarization is $0$. The energy is
$\beta=0.1$ and $\varepsilon=0.1$ for photoelectric and Compton polarimeters, respectively, $f=1$
and $\N_\tot=1$. The empty arrows represent the values of $\eta$ of each curve.}
\label{fig:PolVsEta}
\end{figure*}

A first general property of the modulation function is that the curve obtained when the photons
are impinging from an azimuthal direction $\eta=\bar{\eta}$ is just horizontally shifted of a phase
$\bar{\eta}$ with respect to the behavior for $\eta=0$ (see Figure~\ref{fig:PolVsEta}). Therefore,
the dependency on $\varphi$ and $\eta$ is only through their difference, that is 
\begin{equation}
\M(\varphi,\eta)\equiv\M(\varphi-\eta,0) \; . 
\label{eq:MEta}
\end{equation}
Such a result, which holds for any value of the other parameters, is not surprising because when
$\eta\neq0$ the modulation function is fundamentally obtained by rotating it of an angle $\eta$
around the $z$ axis which is orthogonal to the detection plane, see Appendix~\ref{app:CoordTransf}.
Nonetheless, it is interesting to note that Equation~(\ref{eq:MEta}) replaces a similar property of
the on-axis modulation function which, as a matter of fact, shifts for a change of the angle of
polarization $\varphi_0$.

\begin{figure*}[htbp]
\begin{center}
\subfloat{\includegraphics[angle=0,totalheight=5.6cm]{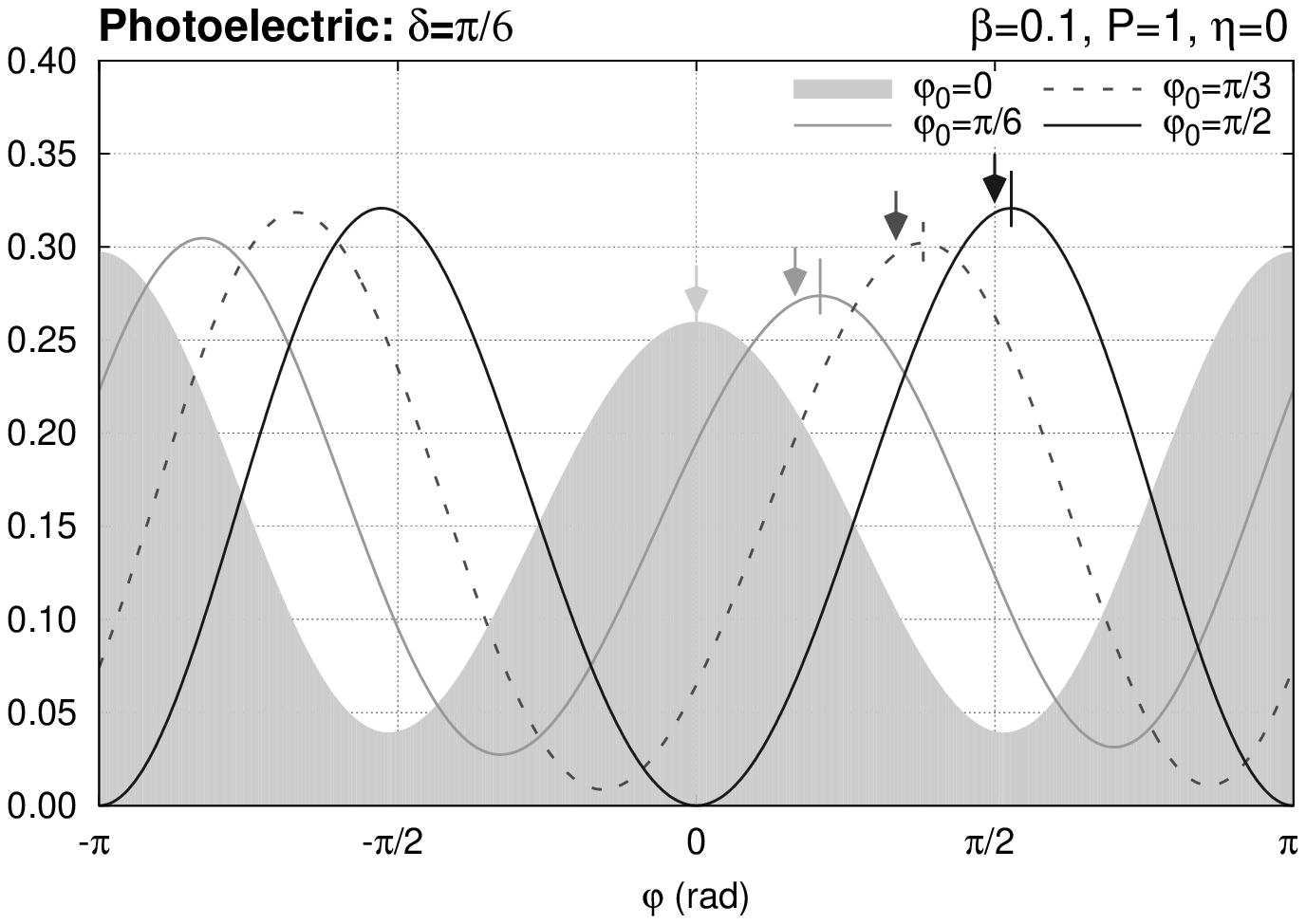}}\hspace{1mm}
\subfloat{\includegraphics[angle=0,totalheight=5.6cm]{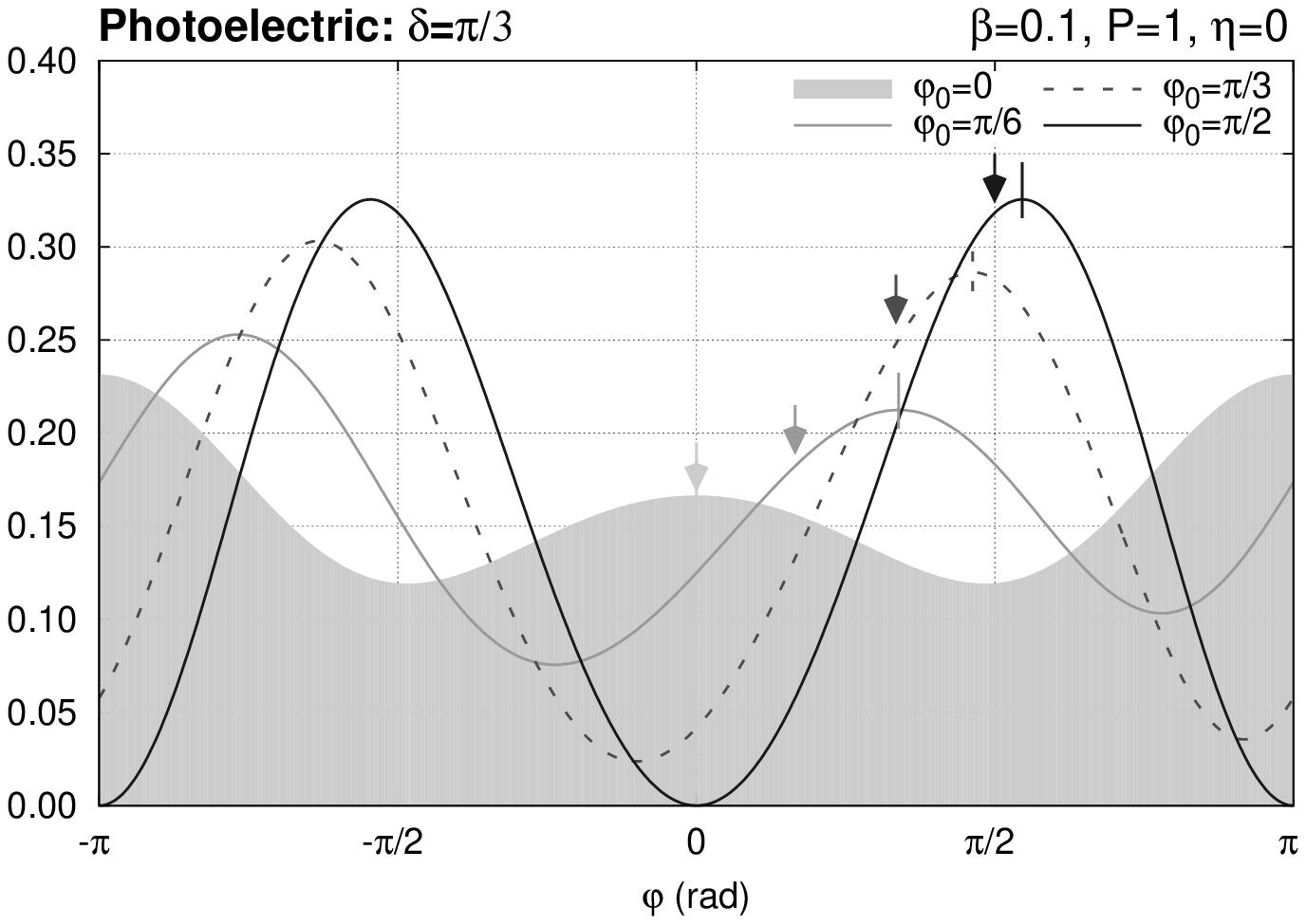}}
\\
\subfloat{\includegraphics[angle=0,totalheight=5.6cm]{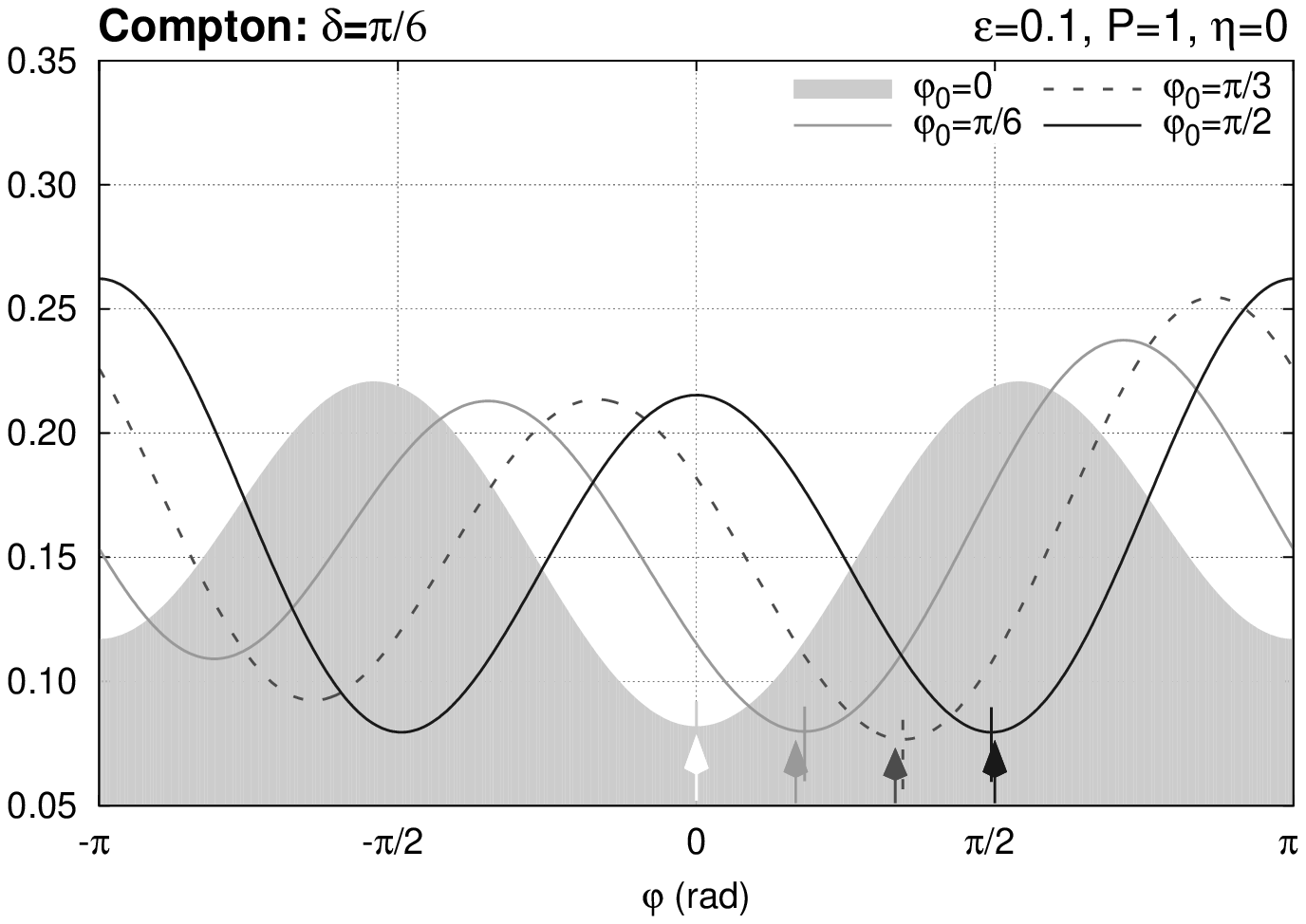}}\hspace{1mm}
\subfloat{\includegraphics[angle=0,totalheight=5.6cm]{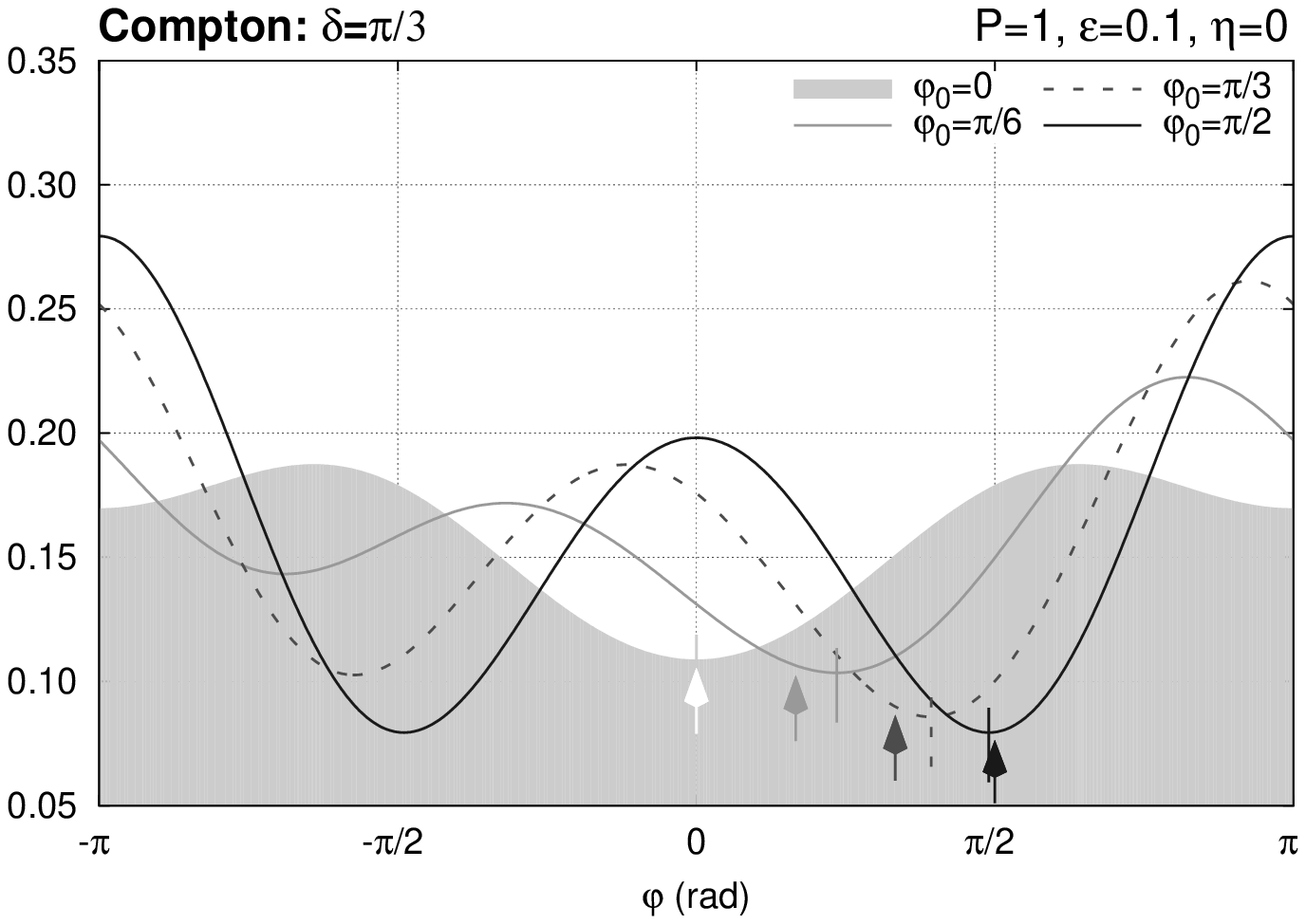}}
\end{center}
\caption{Dependency of the off-axis modulation function on the direction of polarization
$\varphi_0$ for photoelectric (top) and Compton polarimeters (bottom). The inclination $\delta$ is
$\pi/6$ and $\pi/3$ for the left and right panels, respectively. The filled arrows distinguish the
value of the angle of polarization of each modulation function, whereas the vertical lines
highlight the peak of the modulation function or its minimum for photoelectric or Compton
instruments, respectively. The energy is $\beta=0.1$ and $\varepsilon=0.1$ for top and lower
panels. We assumed that $\eta=0$, $f=1$ and $\N_\tot=1$.}
\label{fig:PolVsPhi0}
\end{figure*}

The dependency of the modulation function on $\varphi_0$ when photons are incident off-axis is much
more complex than that on $\eta$. This is ultimately due to the fact that the polarization angle is
measured in the photon frame of reference, while the azimuthal angle $\varphi$ which $\M$ depends on
is defined in the instrument one. The evolution of the modulation function for an increasing angle
of polarization is reported in Figure~\ref{fig:PolVsPhi0} for photoelectric (top) and Compton
(bottom) polarimeters. We assumed an inclination of $\pi/6$ and $\pi/3$ in the left and right
panels, respectively. As a matter of fact, the direction of polarization $\varphi_0$ affects the
shape of the modulation function and not only its phase. Conceptually, the modulation function for
both kinds of instruments evolves for intermediate values of the angle of polarization between two
limit curves, those corresponding to $\varphi_0=0$ and $\varphi_0=\pi/2$. The former case, already
discussed in Section~\ref{sec:SimpleScenario}, is shown in the figure as the light-gray filled
function, the latter is the black solid line. Interestingly enough, one of the two limit curves
shows more prominently the effects of the inclined incidence of the photons, that is the presence of
a constant contribution and a different height of the two peaks, whereas the other does not show any
of these two signatures and it is much more like to a pure cosine square modulation. This holds true
for photoelectric polarimeters and, to some extent, also for Compton instrument, although there is a
sort of ``opposite'' correspondence between the conditions in which each of the two cases occurs. In
case of photoelectric polarimeters the modulation function is more or less affected by inclined
incidence of the photons if $\varphi_0=0$ or $\varphi_0=\pi/2$, respectively, whereas the opposite
happens for Compton instrument. This difference arises from the fact that the most probable
direction of event emission is along the direction of the electric field for photoelectric effect,
whereas is it is perpendicular to it in case of Compton scattering. Taking this into account, it
becomes evident that the two limit modulation functions correspond to when the large part of the
events are emitted parallel or orthogonal to the plane $\Upsilon$, defined by the direction of
incidence and the orthogonal to the detection plane (see Figure~\ref{fig:Inclination}). In
particular, if most of the events are emitted parallel to $\Upsilon$, which happens if $\varphi_0=0$
and $\varphi_0=\pi/2$ for photoelectric and Compton polarimeters respectively, the modulation
function is more affected, while it is cosine-square like if they are produced orthogonal to this
plane ($\varphi_0=\pi/2$ and $\varphi_0=0$ for photoelectric and Compton polarimeters). 

\begin{figure*}[htb]
\begin{center}
\subfloat{\includegraphics[angle=0,totalheight=6.2cm]{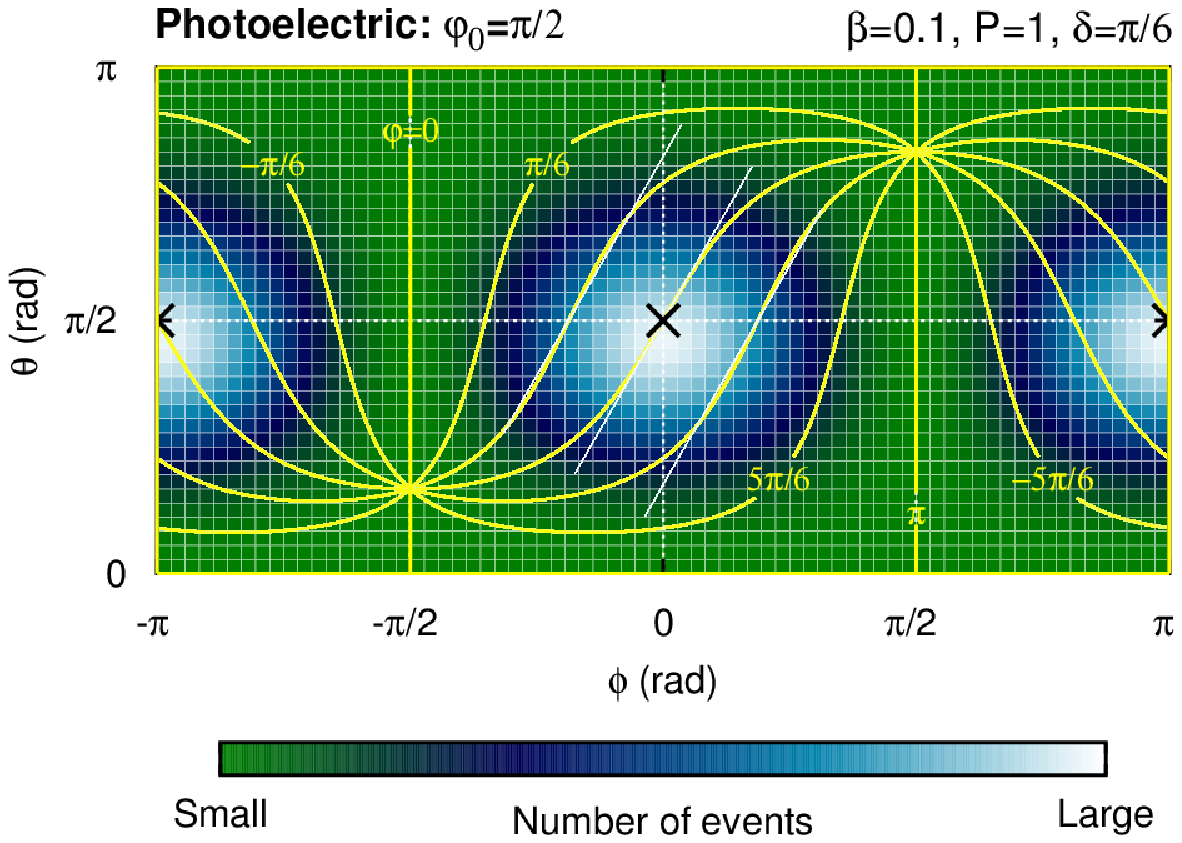}}\hspace{1mm}
\subfloat{\includegraphics[angle=0,totalheight=6.2cm]{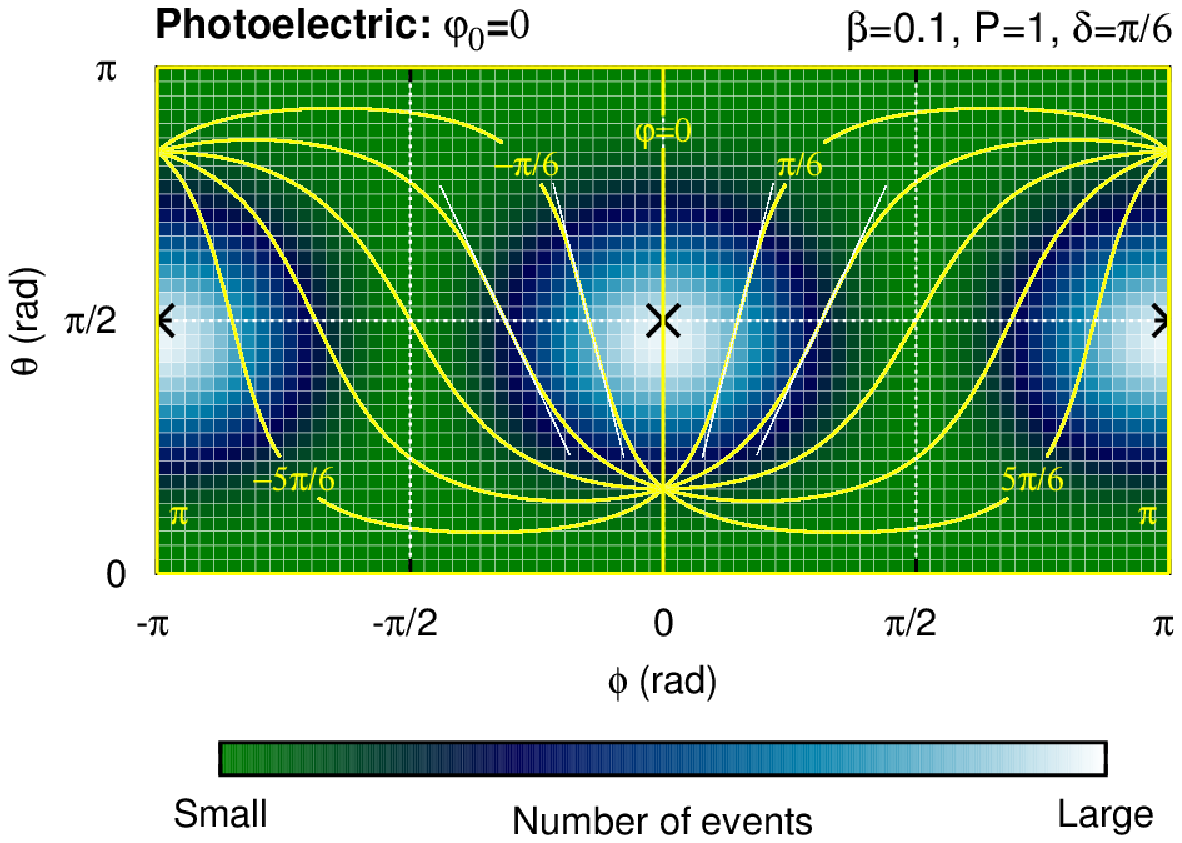}}
\\
\subfloat{\includegraphics[angle=0,totalheight=6.2cm]{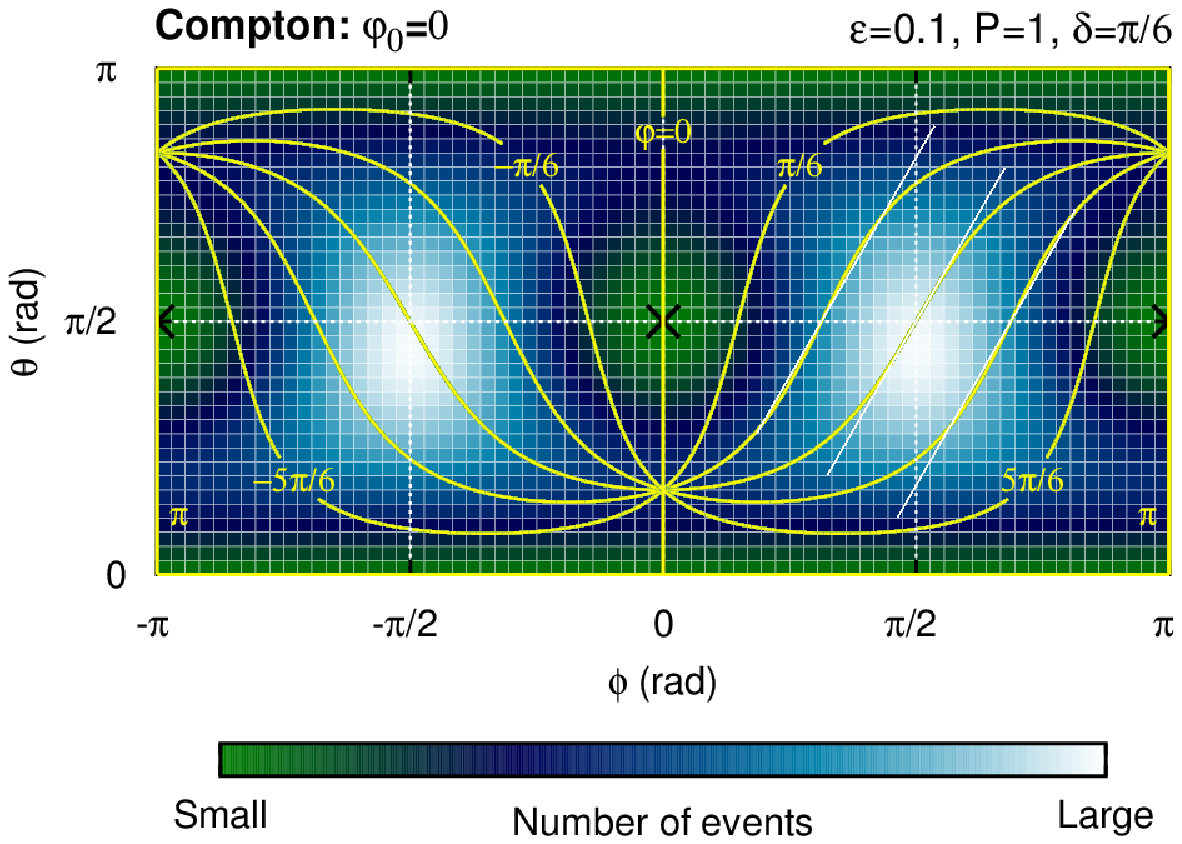}}\hspace{1mm}
\subfloat{\includegraphics[angle=0,totalheight=6.2cm]{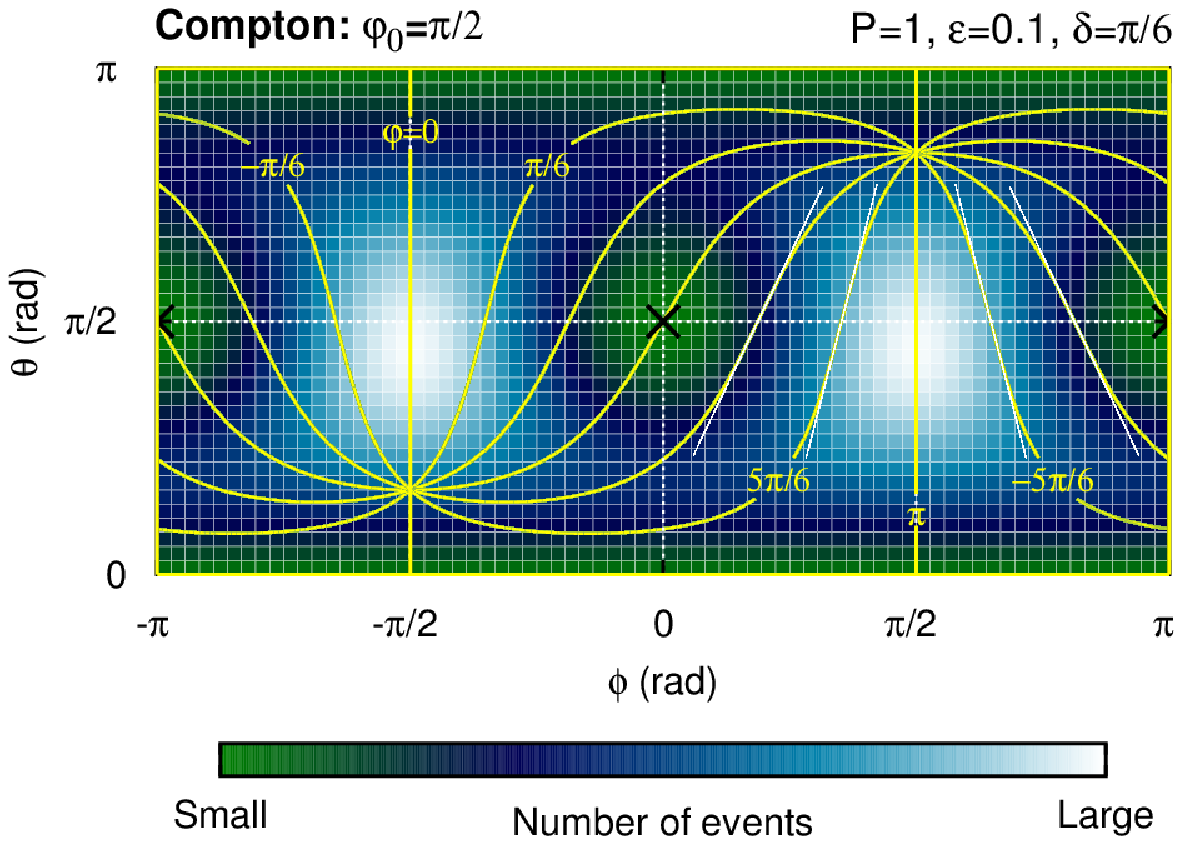}}
\end{center}
\caption{The same as Figure~\ref{fig:PhEvents}, but for photoelectric (top panels) and Compton
polarimeters (bottom panels). Left panels refer to the condition for which the
large part of the events are produced orthogonal to the plane $\Upsilon$ (see
Figure~\ref{fig:Inclination}), that is $\varphi_0=\pi/2$ for photoelectric polarimeters and
$\varphi_0=0$ for Compton instruments, whereas the panels on the right represent the case in which
the events are more probably emitted parallel to it, that is $\varphi_0=0$ and $\varphi_0=\pi/2$ for
photoelectric and Compton polarimeters respectively. The white lines guide the eye to distinguish
the direction of the meridians in one of the two regions where the emission is concentrated. The
black thick crosses characterize the direction of the electric field.}
\label{fig:VsPhi0}
\end{figure*}

The evolution of the modulation function with the angle of polarization can be qualitatively
understood by looking at Figure~\ref{fig:VsPhi0}, where we report as in Figure~\ref{fig:PhEvents}
the distribution of the events in the photon frame of reference and the meridians in $\phi\theta$
coordinates. The photoelectric absorption case and the Compton scattering one are reported in top
and bottom panels, respectively, and left and right panels refer to the condition for which the
large part of the events are produced  orthogonal or parallel to $\Upsilon$. We have already seen in
Section~\ref{sec:SimpleScenario} that the value of the modulation function in a certain point
$\varphi=\bar{\varphi}$ is the sum of the events emitted in the $\phi\theta$ plane along the
corresponding meridian $\bar{\varphi}$. At this regard, the most important contribution to the
eventual value comes from the behavior of the meridian in correspondence of the regions where the
emission is more probable because these areas ``weigh more'' in the sum. Therefore, the fact that
the modulation function is cosine square modulated on-axis can be viewed as a consequence of the
fact that in this case the meridians cross such regions as parallel vertical lines, see the
top left panel in Figure~\ref{fig:PhEvents}. The meridians are never vertical lines when
$\delta\neq0$, but accidentally when the events are more probably produced orthogonal to the
$\Upsilon$ plane they are nearly parallel in the regions where the emission is concentrated (see the
left panels in Figure~\ref{fig:VsPhi0}). This is sufficient to produce a nearly cosinusoidal square
modulation because the event distribution has a certain degree of cylindric symmetry around the peak
of the emission. Such a symmetry is evident in particular for photoelectric absorption, for which in
fact the limit modulation function for $\varphi_0=\pi/2$ is more similar to a cosine square. On the
contrary, when the large part of the events are emitted parallel to the $\Upsilon$ plane (see the
right panels in Figure~\ref{fig:VsPhi0}), the meridians have different slopes and this causes the
effects of the inclination to be evident.

\begin{figure}[htbp]
\begin{center}
\subfloat[]{\includegraphics[angle=0,totalheight=6cm]{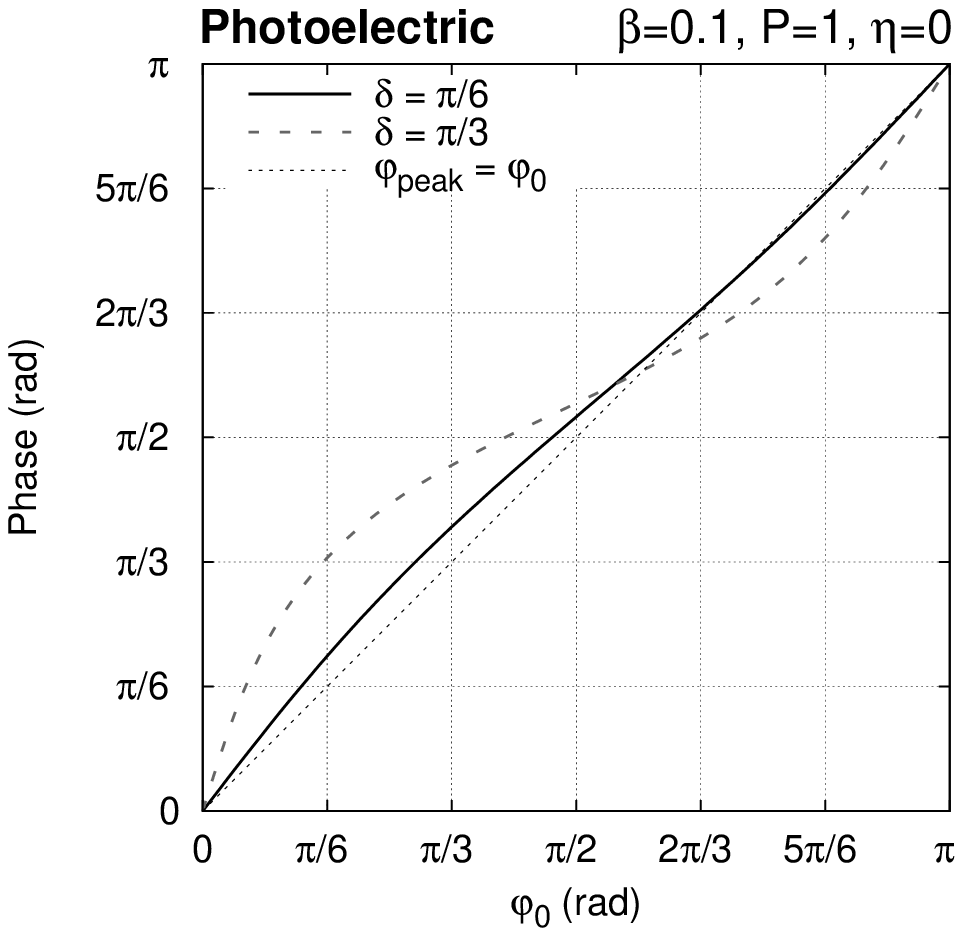}}\hspace{1mm}
\subfloat[]{\includegraphics[angle=0,totalheight=6cm]{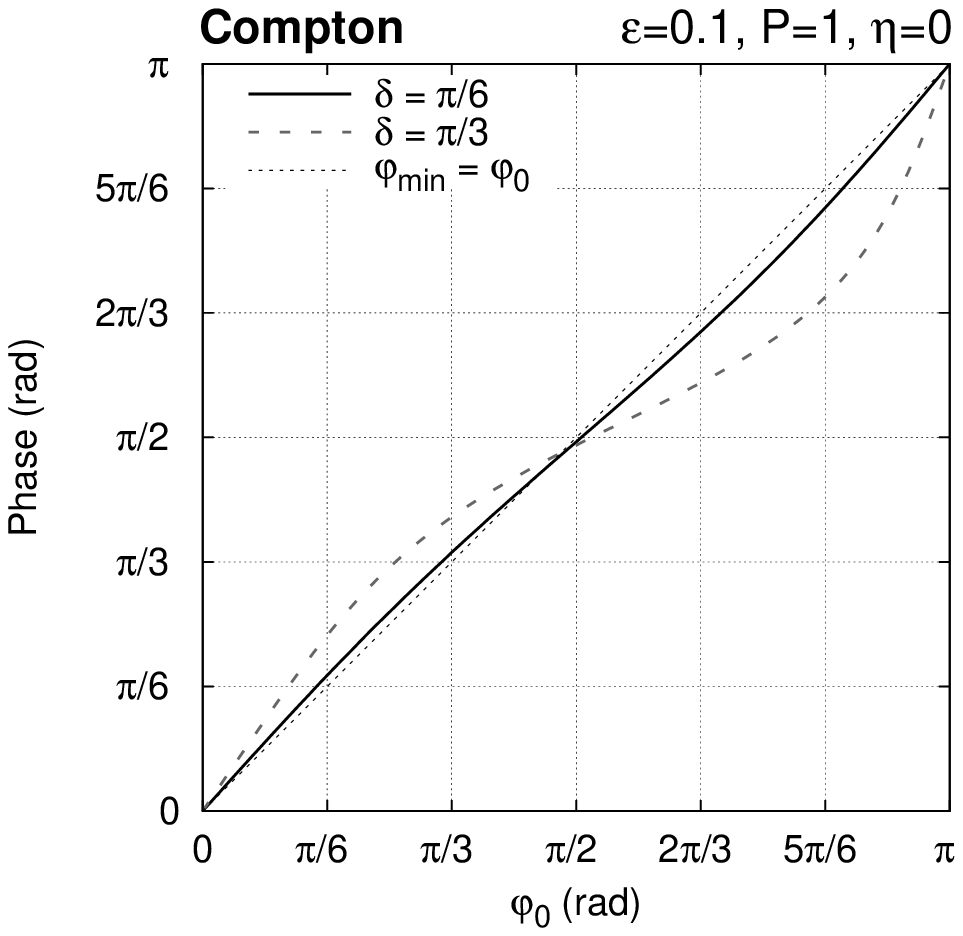}}
\end{center}
\caption{(a) Relation between the peak of the modulation function for photoelectric polarimeters
and the angle of polarization $\varphi_0$ when photons are incident off-axis. The inclination
$\delta$ is $\pi/6$ and $\pi/3$ for the solid and dashed curves, respectively. The relation
$\mathrm{Phase}=\varphi_0$ that holds when photons are incident on-axis is also reported for
comparison. (b) The same as (a) but for Compton polarimeters. In this case $\varphi_0$ has to be
put in relation with the minimum of the modulation function because the probability of scattering is
minimum along the direction of the electric field.}
\label{fig:ModPeakVsPhi0}
\end{figure}

It is worth noting that, although the modulation function can appear quite similar to a cosine
square under certain circumstances, the usual on-axis behavior is actually never recovered. For
example, the phase of the modulation function, intending the value corresponding to its peak for
photoelectric polarimeters or to its minimum for Compton ones, is the angle of polarization in case
of on-axis radiation, whereas when $\delta\neq0$ it is never equal to $\varphi_0$ except for a few
very special cases. The difference between the two is evident in Figure~\ref{fig:PolVsPhi0}, where
for each modulation function we indicated the corresponding value of the polarization angle with a
filled arrow and that of the phase with a vertical line. In general, the relation between the phase
and $\varphi_0$ is not linear when $\delta\neq0$ and we reported it for $\delta=\pi/6$ and
$\delta=\pi/3$ in Figure~\ref{fig:ModPeakVsPhi0}. We take as a reference the peak which is in
$\varphi=0$ for $\varphi_0=0$, but this is purely conventional and in fact the position of the other
peak has a specular behavior. The non-linearity of such dependency is a direct consequence of the
inclination of the impinging photons, whereas the effect of the forward bending is to make the
curves not symmetric with respect to $\varphi_0=\pi/2$.

\begin{figure*}[htbp]
\begin{center}
\subfloat[\label{fig:PhModVsPolDeg}]{\includegraphics[angle=0,totalheight=5.6cm]{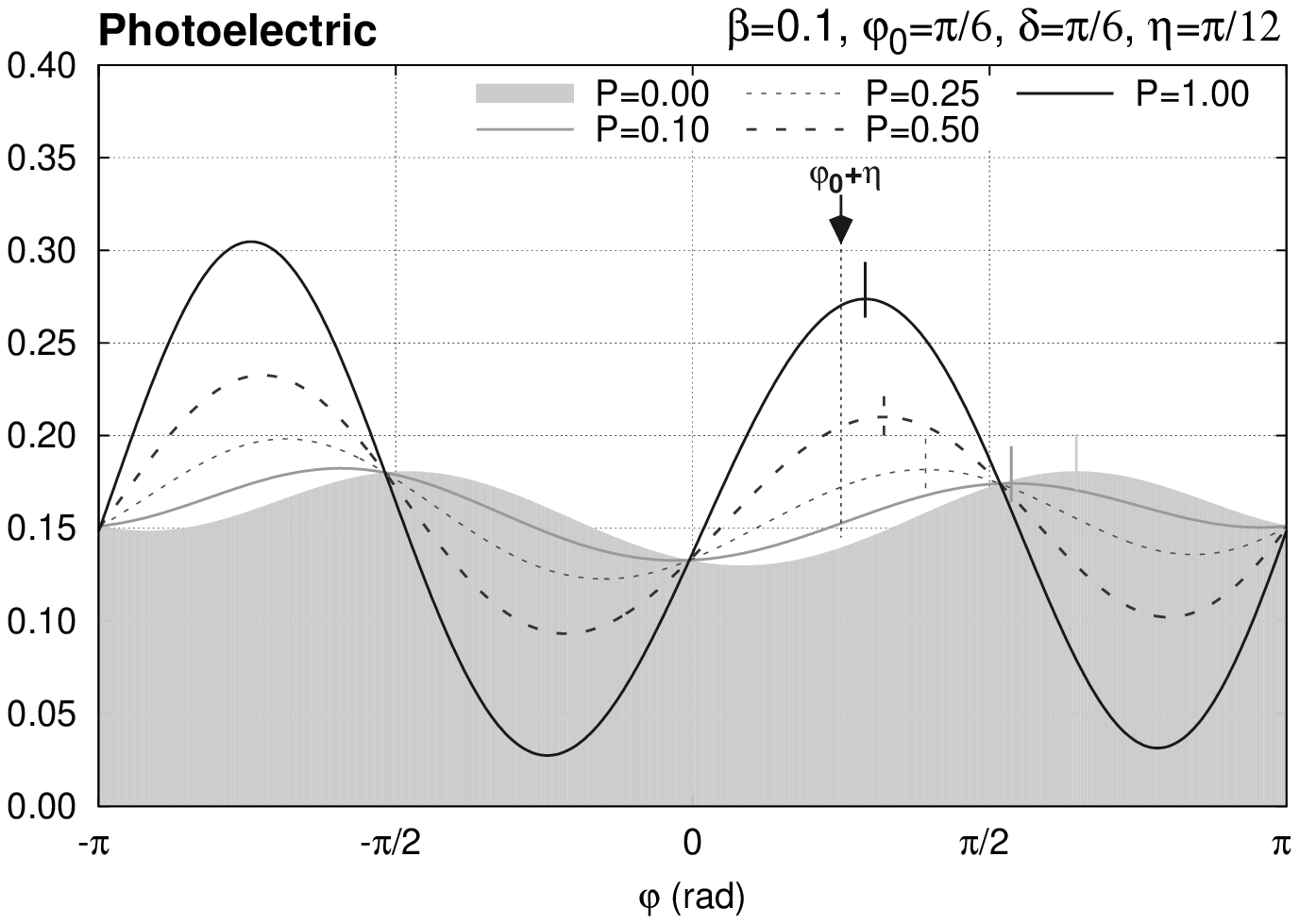}}
\hspace{1mm}
\subfloat[\label{fig:CmModVsPolDeg}]{\includegraphics[angle=0,totalheight=5.6cm]{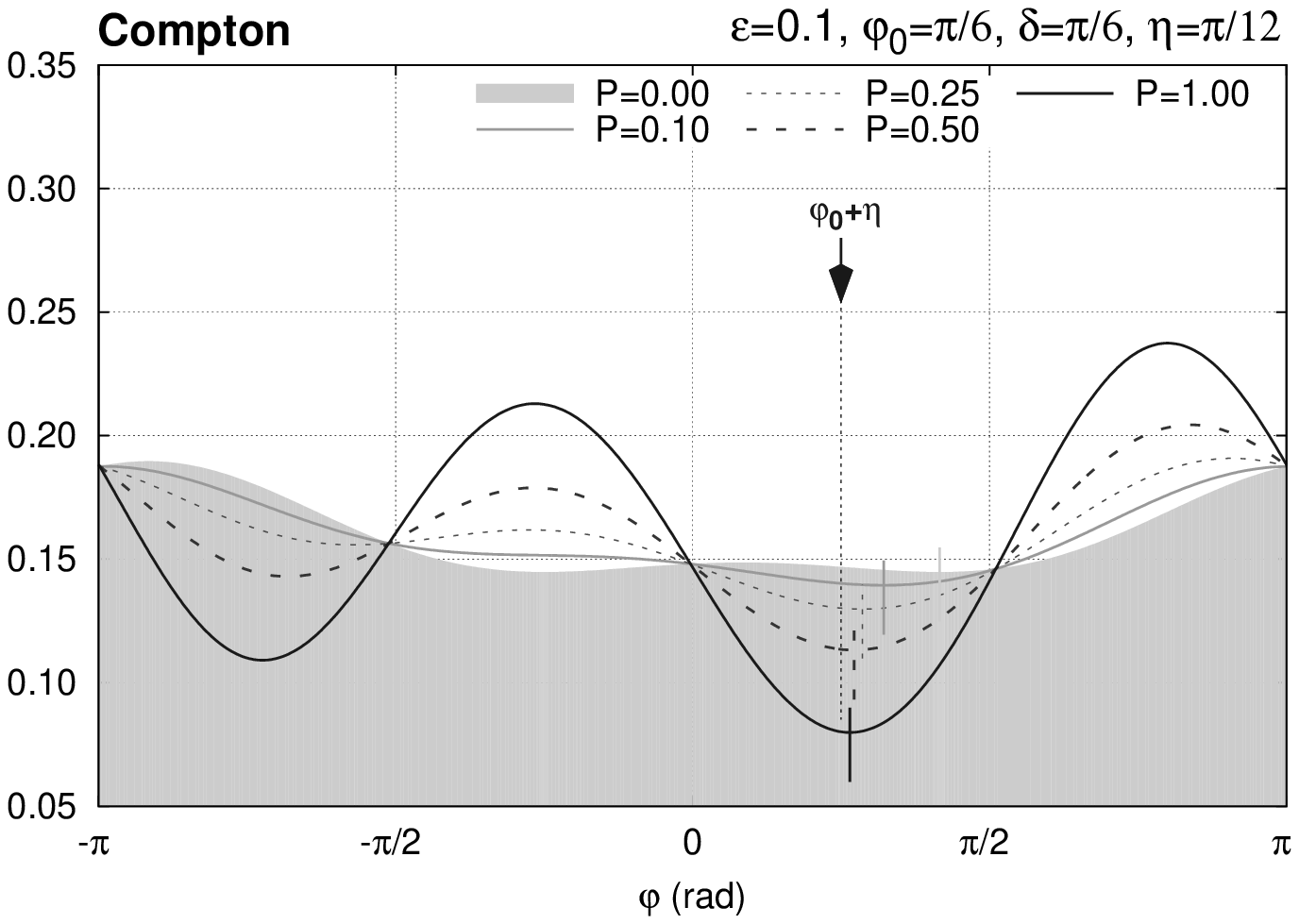}}
\end{center}
\caption{Dependence of the modulation function on the polarization degree for off-axis impinging
photons and photoelectric (a) and Compton (b) polarimeters. The inclination and the azimuth
of the incident beam is $\delta=\pi/6$ and $\eta=\pi/12$, respectively, whereas the angle of
polarization is $\pi/6$. The energy is $\beta=0.1$ in (a) and $\varepsilon=0.1$ in (b). The vertical
lines highlights the phase of each curve and the black arrow distinguishes the angle
$\varphi_0+\eta$. As usual, $\N_{\tot}=1$ and $f=1$.}
\label{fig:ModVsPolDeg}
\end{figure*}

The last relevant dependency of the modulation function is that on the degree of polarization,
which is shown in Figure~\ref{fig:ModVsPolDeg} for a fixed incident direction of the beam,
$\delta=\pi/6$, $\eta=\pi/12$, and an angle of polarization $\varphi_0=\pi/6$. As we discussed in
Section~\ref{sec:SimpleScenario}, the modulation function evolves smoothly from that for
unpolarized photons, reported in the figure as the light-gray filled curve, to that of completely
polarized photons, which instead is the black solid line. The features of the latter when the
inclination is not zero were already discussed above and we have only to remember that according to
Equation~(\ref{eq:MEta}) the curve for a beam which is incident from an azimuth $\eta=\pi/12$
is just shifted of $\pi/12$ with respect to the modulation functions for $\eta=0$ reported in
Figure~\ref{fig:PolVsPhi0}. The modulation function for unpolarized photons obviously depends only
on the direction of the incident photons and, as a matter of fact, the dependency on $\delta$ is
decoupled from that on $\eta$. The latter is also in this case that expressed by
Equation~(\ref{eq:MEta}), whereas the dependency on $\delta$ is the same one that we discussed in
the simple scenario presented in the previous section, that is, the presence of a modulation with
amplitude increasing with the inclination. 

The important result suggested by Figure~\ref{fig:ModVsPolDeg} is that it is fundamentally not
correct to relate the phase of the modulation function to the angle of polarization $\varphi_0$ when
the photons are incident off-axis. We have already found out that a change in $\varphi_0$ does not
correspond to an equal shift in phase for \emph{completely polarized photons}, see
Figure~\ref{fig:ModPeakVsPhi0}. Here we are pointing out that the angle at which the modulation
functions have a maximum or a minimum, which is highlighted in Figure~\ref{fig:ModVsPolDeg} by a
vertical line for each curve, changes also by increasing the polarization degree. The reason for
this is implicit in the fact that the modulation function for partially polarized photon is
intermediate between the two limit curves corresponding to completely polarized and unpolarized
radiation. As results from Figure~\ref{fig:ModPeakVsPhi0} and Equation~(\ref{eq:MEta}), the phase of
the former is, roughly speaking, $\approx\varphi_0+\eta$, whereas that of the latter is
$\approx\pi/2+\eta$, see Figure~\ref{fig:PhUnPVsDelta} and Figure~\ref{fig:CmUnPVsDelta}. Therefore,
the phase of the modulation function will naturally range between these two values for partially
polarized photons and its actual value will depend on the polarization degree.

For the sake of completeness, we clarify that the results above were discussed for values of the
angle of polarization between $0$ and $\pi/2$ only for graphical clarity. The modulation function
if $\varphi_0$ is in the $(-\pi/2,0)$ interval can be derived from the curves presented above by
considering that
\begin{equation}
\M(\varphi,-\varphi_0,\eta)\equiv\M(-\varphi,\varphi_0,-\eta) \; . \notag
\end{equation}
For example, the modulation function for $\varphi_0=-\pi/6$ and $\eta=\pi/12$ coincides with the
curve obtained by making the symmetric about the y axis of the modulation function for
$\varphi_0=\pi/6$ and $\eta=-\pi/12$. Such a property derives from the invariance of the
transformation from the photon to the instrument frame of reference for the change of variables
$(\varphi,-\varphi_0,\eta)\rightarrow(-\varphi,\varphi_0,-\eta)$, which can be easily verified by
making explicit all the factors (see  Equations~(\ref{eq:Transf})). Analogously, it is easy to
verify that for negative values of the inclination
\begin{equation}
\M(-\delta,\eta)\equiv\M(\delta,\eta+\pi) \; . \notag
\end{equation}

\section{Discussion} \label{sec:Discussion}

The picture which emerges from the analysis carried out in Section~\ref{sec:ModInclined} is that the
modulation function of photoelectric and Compton polarimeters obtained when photons are incident
off-axis depends on a number of parameters. Among them there is obviously the state of polarization,
nonetheless, we have also seen that the dependency on the direction of the incidence is remarkable
and that the modulation function depends always on the energy. An evident question which arises
from such a result is whether the modulation function alone always allows to unambiguously derive
the state of polarization of the incident photons, which is the relevant observable quantity of this
kind of instruments. We have already argued in Section~\ref{sec:SimpleScenario} that there may be
cases in which this is not true, since the modulation function for unpolarized and off-axis
radiation in the low energy limit is exactly identical to that of polarized photons. Here we want to
support such a claim, although a complete analysis is out of the scope of this paper and, rather, it
would be the subject of a future work.

\begin{figure}[htbp]
\begin{center}
\includegraphics[angle=0,totalheight=5.6cm]{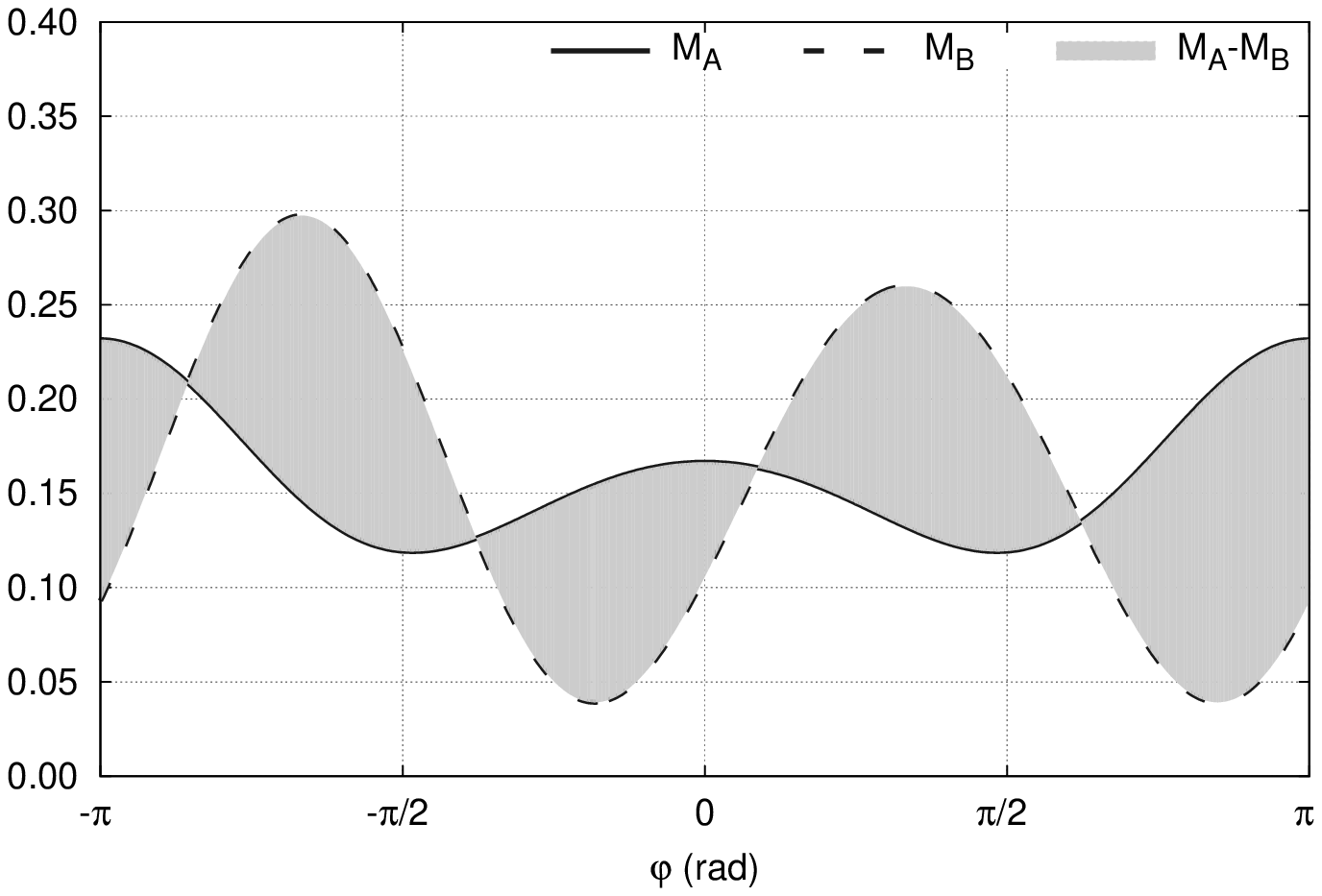}
\end{center}
\caption{We defined the normalized difference $\Delta$ between two modulation functions $\M_A$ and
$\M_B$ as the square root of the variance of the two curves, normalized to the area of the first.
Basically, it represents the area between the two modulation functions, which is the light-gray
region in the figure.}
\label{fig:Delta}
\end{figure}

The determination of the state of polarization of the incident photons is unique if there is a
unique correspondence between the modulation function and the photon parameters, that is the energy,
the polarization and the direction of the incident radiation. This is equivalent to say that any two
modulation functions have to be ``sufficiently different'' whenever corresponding to any groups
of ``sufficient different'' photon parameters $A$ and $B$, and that such a difference has to tend to
zero if and only if $A$ approaches to $B$. To quantitatively express such a condition, we introduce
the \emph{normalized difference} $\Delta$ between two modulation functions $\M_A$ and $\M_B$. Let us
assume that they are obtained with the same instrument in equivalent conditions, but $\M_i$ refers
to photons which (i) produce events of energy $\beta_i$ or $\varepsilon_i$ for photoelectric and
Compton polarimeters respectively, (ii) are characterized by an angle of polarization
$\varphi_{0,i}$ and by a degree of polarization $\P_i$, (iii) are incident with an inclination
$\delta_i$ and an azimuth $\eta_i$. Then, we define $\Delta$ as
\begin{equation}
\Delta = \dfrac{\sqrt{\int_{-\pi}^{\pi} \left[ \M_A(\varphi) - \M_B(\varphi) \right]^2 \d\varphi}}
{\int_{-\pi}^{\pi} {\M_A(\varphi) \d\varphi}} \; . \notag 
\end{equation}
The meaning of $\Delta$ is quite simple because it basically represents the area between the  two
modulation functions normalized to the area underlying the first (see Figure~\ref{fig:Delta}).
The latter is simply $\N_{\tot}$ (see Equation~(\ref{eq:Closure})) and then just a scale factor.

A complete analysis would require to study how $\Delta$ varies with the photon parameters of the
first and of the second modulation functions, and to verify that it vanishes only if the
corresponding parameters are identical. Nonetheless, our aim here is to argue if the difference
between two peculiar yet representative configurations would be appreciable with a real
instrumentation. We restrict ourselves in a quite specific case, fixing the large part of the
parameters and leaving free to vary only those which are the most relevant. In particular, we
assume that the first modulation function refers to unpolarized photons, $\P_A=0$, which are
incident with an inclination $\delta_A=\pi/6$ and an azimuth $\eta_A=0$. Then, we see if the same
modulation function can be reproduced with a different choice of the parameters, that is,
with partially polarized photons which are incident with a different inclination. We fix the angle
of polarization and the azimuth of the second configuration to $\varphi_{0,B}=\pi/2$ and $\eta_B=0$
because this choice minimizes the value of $\Delta$. Eventually, we assume for the moment that the
energy of the radiation is the same for both the configurations $A$ and $B$ and that it
produces events with $\beta_A=\beta_B=0.1$ and $\varepsilon_A=\varepsilon_B=0.1$ for photoelectric
and Compton polarimeters, respectively. We will let the energy to change further below.

\begin{figure*}[htbp]
\begin{center}
\subfloat{\includegraphics[angle=0,totalheight=7.5cm]{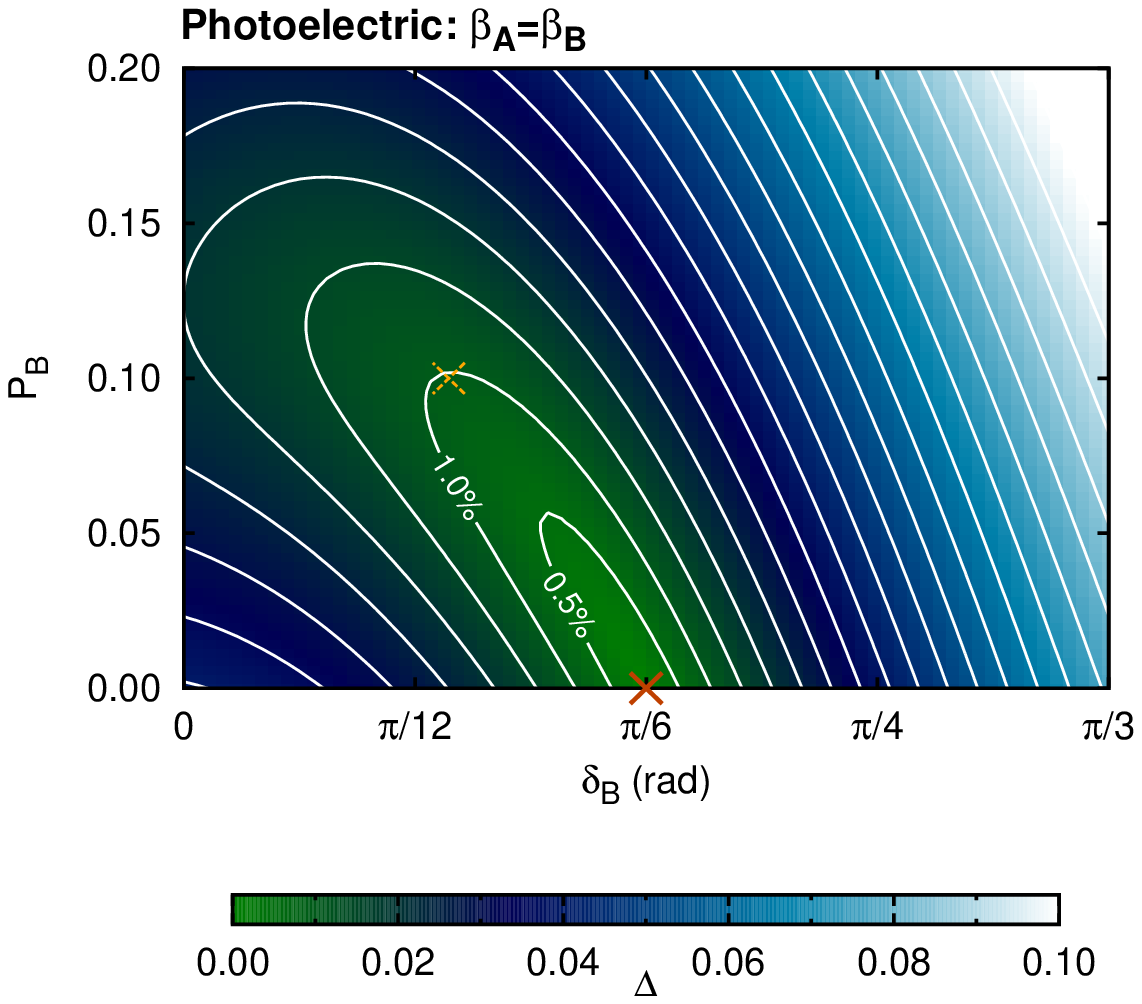}}
\hspace{1mm}
\subfloat{\includegraphics[angle=0,totalheight=7.5cm]{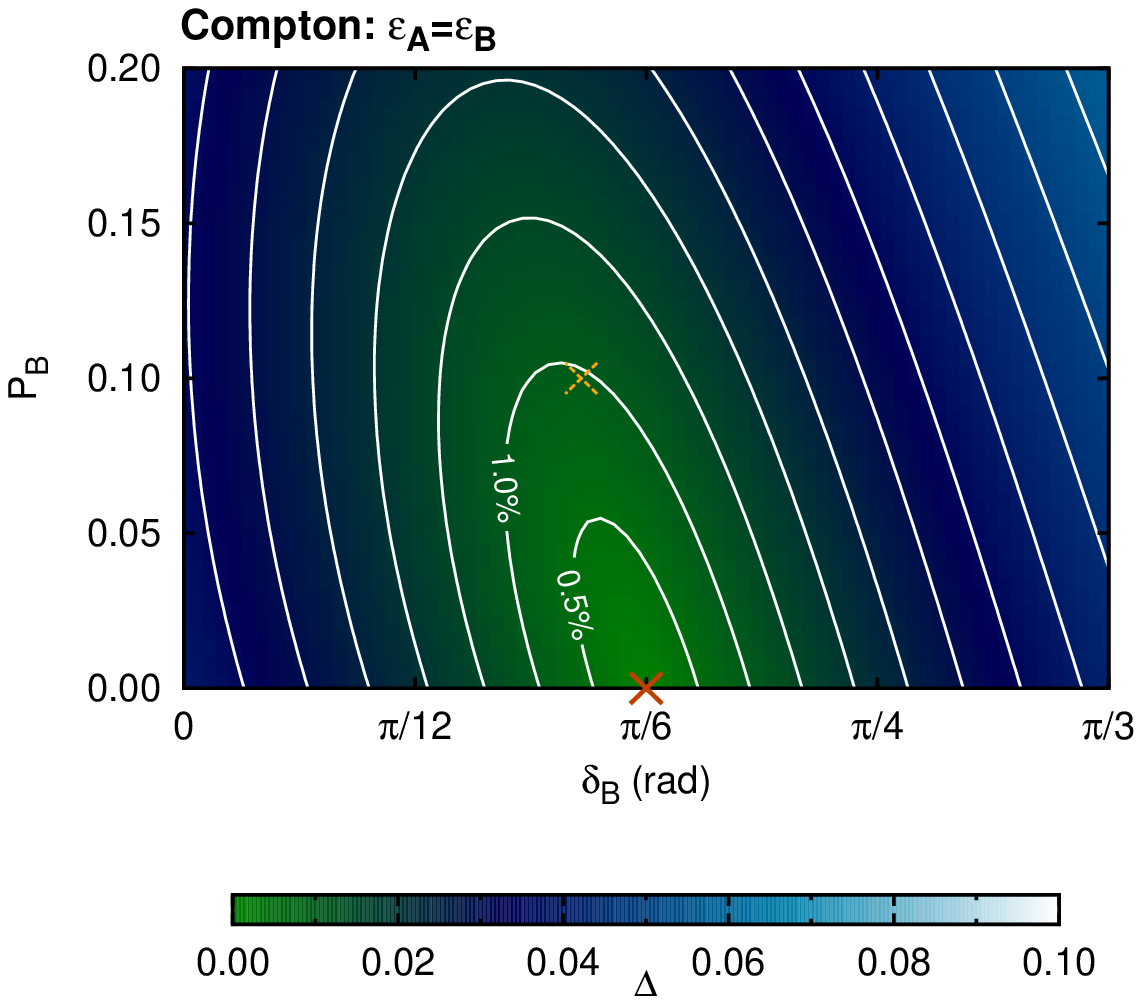}}
\end{center}
\caption{Normalized difference between two modulation functions $A$ and $B$ as a function of
the degree of polarization $\P_B$ and of the inclination $\delta_B$ of the latter. We assume that
$\beta_A,\varepsilon_A=\beta_B,\varepsilon_B=0.1$, $\varphi_{0,B}=\pi/2$, $\P_A=0$,
$\delta_A=\pi/6$, $\eta_A=\eta_B=0$. The thick solid cross spots the values of polarization and
inclination of the first modulation function, whereas the thin dashed one pinpoints the specific
configuration compared with $A$ in Figure~\ref{fig:UnicityMF_1}. Contour lines identify $\Delta$
increments of 0.5\%. As usual, $f=1$.}
\label{fig:UnicityPol_1}
\end{figure*}

\begin{figure*}[htbp]
\begin{center}
\subfloat{\includegraphics[angle=0,totalheight=5.6cm]{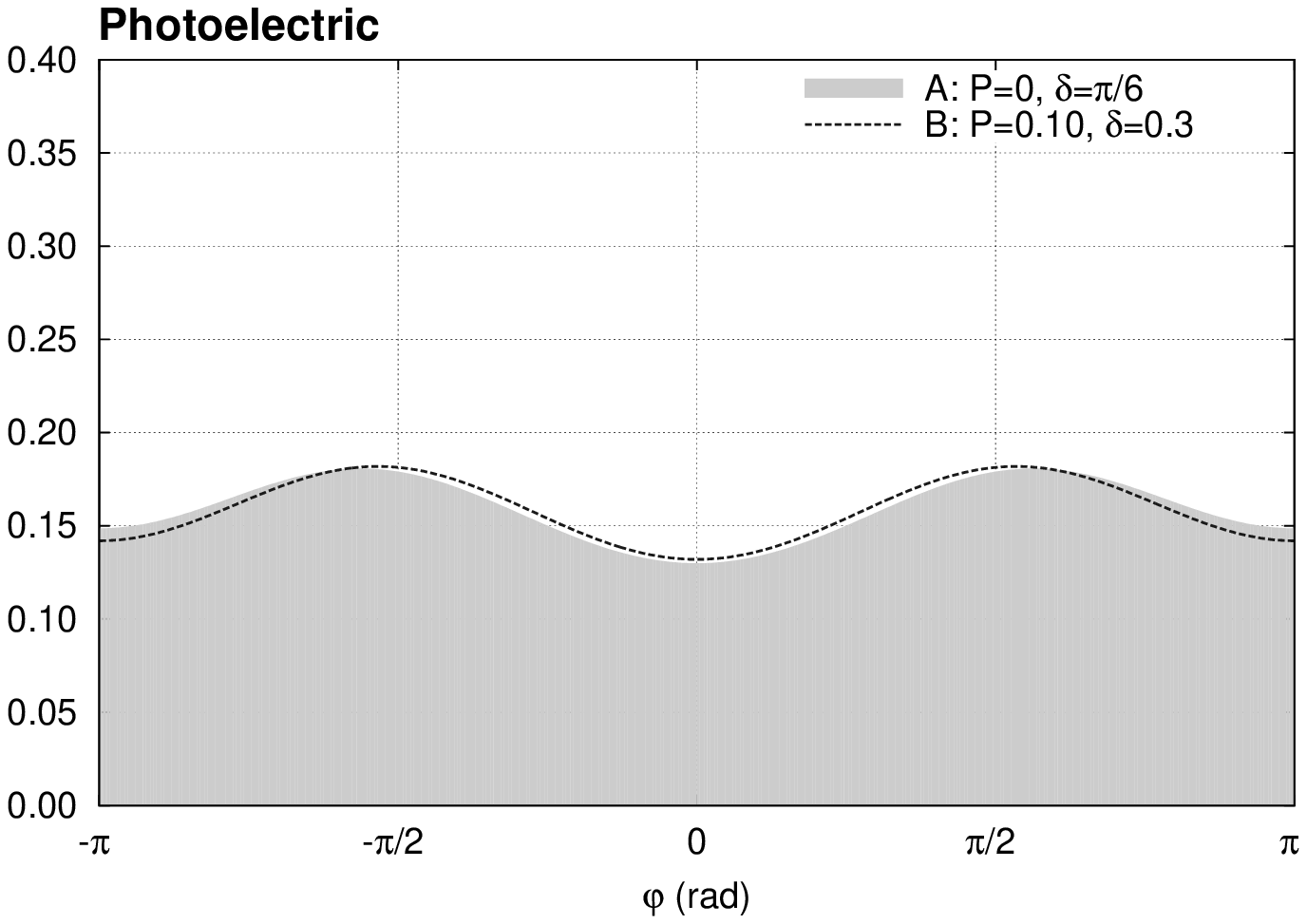}}
\hspace{1mm}
\subfloat{\includegraphics[angle=0,totalheight=5.6cm]{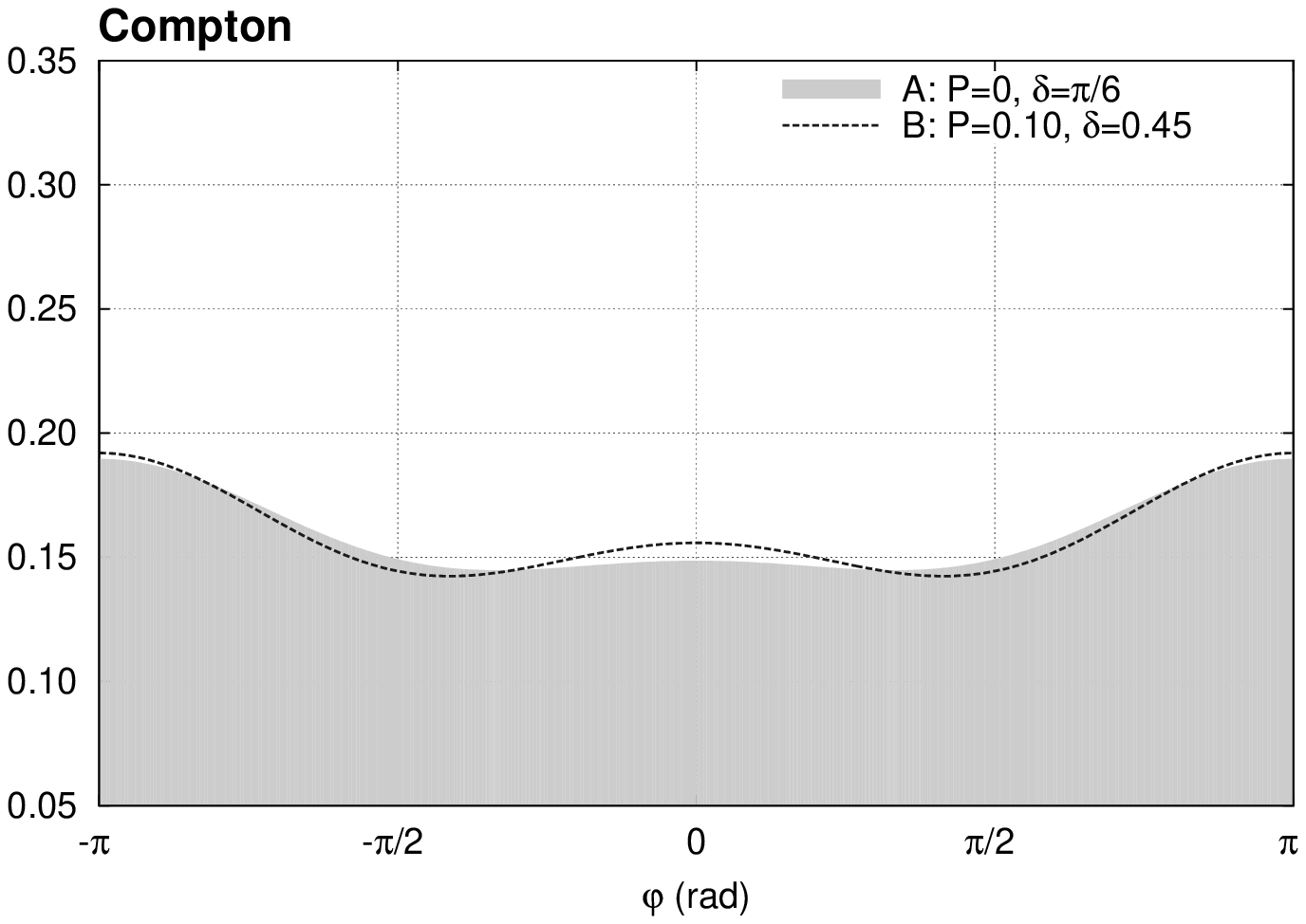}}
\end{center}
\caption{Comparison between two modulation functions $A$ and $B$ which differs of
$\Delta\approx1\%$. The first refers to unpolarized photons which are incident at $\delta=\pi/6$
(light-gray filled function), the second is obtained for 10\% polarized photons which are incident
at $\delta=0.30$ (left panel, photoelectric instruments) or $\delta=0.45$ (right panel, Compton
polarimeters). The other parameters of the two curves are
$\beta_A,\varepsilon_A=\beta_B,\varepsilon_B=0.1$, $\varphi_{0,B}=\pi/2$, $\eta_A=\eta_B=0$, $f=1$
and $\N_\tot=1$.}
\label{fig:UnicityMF_1}
\end{figure*}

The normalized difference between the modulation functions obtained in the two
configurations described above is reported in color code in Figure~\ref{fig:UnicityPol_1} as a
function of the polarization degree $\P_B$ and of the inclination $\delta_B$ of the second
configuration. The solid thick cross spots the values of the polarization and of the inclination of
the first modulation function, $\P_A=0$ and $\delta_A=\pi/6$, and the contour lines identify
$\Delta$ increments of 0.5\%. As expected, $\Delta$ vanishes only if the polarization and
the inclination of the two configurations is identical, that is in the region close to the solid
thick cross. Notwithstanding, the low absolute difference between configurations characterized by
even quite different parameters is striking. As a matter of fact, there is a large region in the
$\delta_B\P_B$ parameter space in which the modulation function $B$ would differ from $A$ less than
1\%. To give an insight of what this means, we report in Figure~\ref{fig:UnicityMF_1} the two
modulation functions in case we assume for $B$ the values of $\P_B$ and $\delta_B$ pinpointed by the
thin dashed cross in Figure~\ref{fig:UnicityPol_1}, so that the difference between $A$ and $B$ is
about 1\%. The curves are almost indistinguishable, although one refers to unpolarized photons and
the other is obtained from radiation which is 10\% polarized. For a real instrument, the small
difference between the two would be virtually undetectable because of statistical fluctuations in
the content of the modulation curve bins and the final result would be that of a severe
indetermination on the polarization degree. Obviously, the way of resolving such a degeneracy is
that of measuring the incident direction of the photons. Even a raw knowledge of the source position
would be sufficient to distinguish between the unpolarized and polarized curves because their
inclination differs of about 13$^\circ$ in case of photoelectric polarimeters ($\delta_B=0.30$) and
of about 4$^\circ$ in case of Compton instruments ($\delta_B=0.45$ in this case).

\begin{figure*}[htbp]
\begin{center}
\subfloat{\includegraphics[angle=0,totalheight=7.5cm]{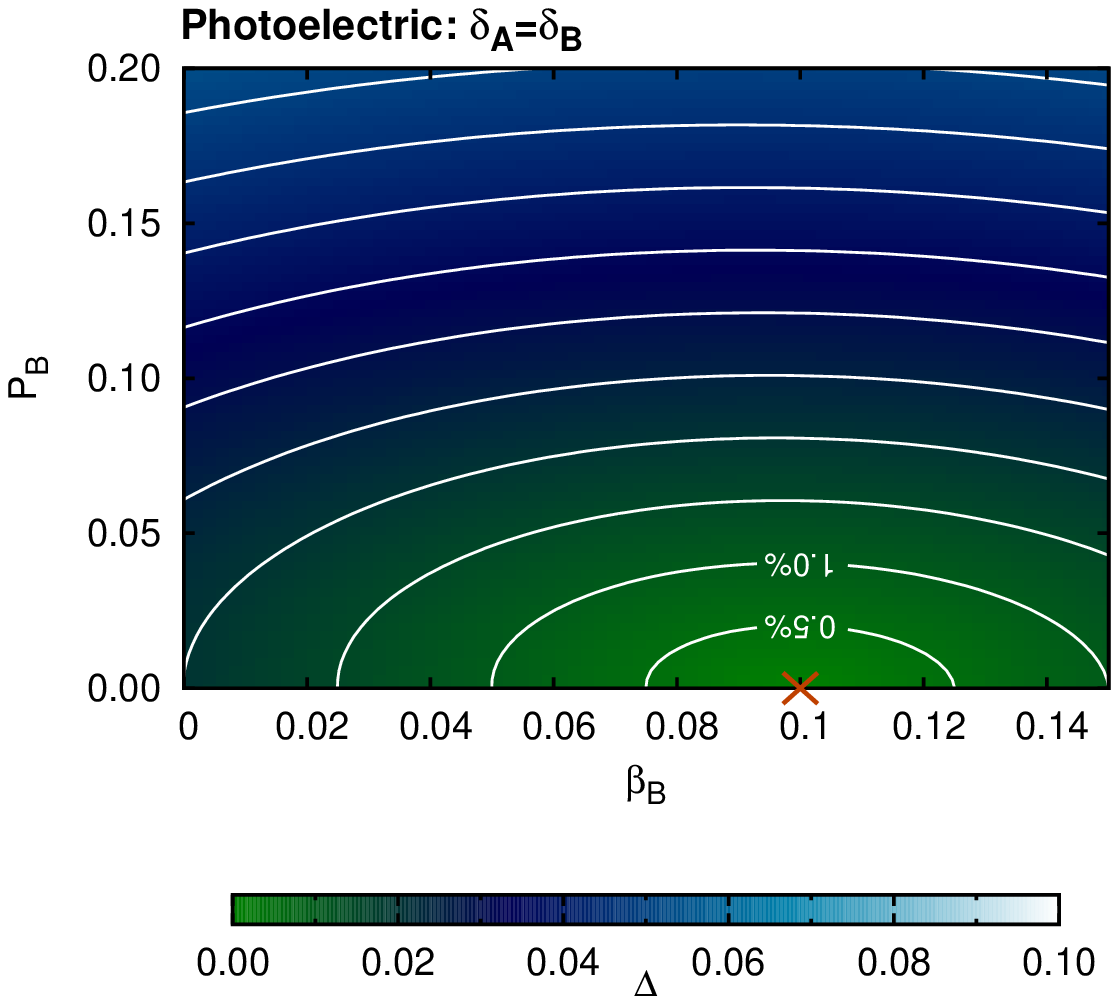}}
\hspace{1mm}
\subfloat{\includegraphics[angle=0,totalheight=7.5cm]{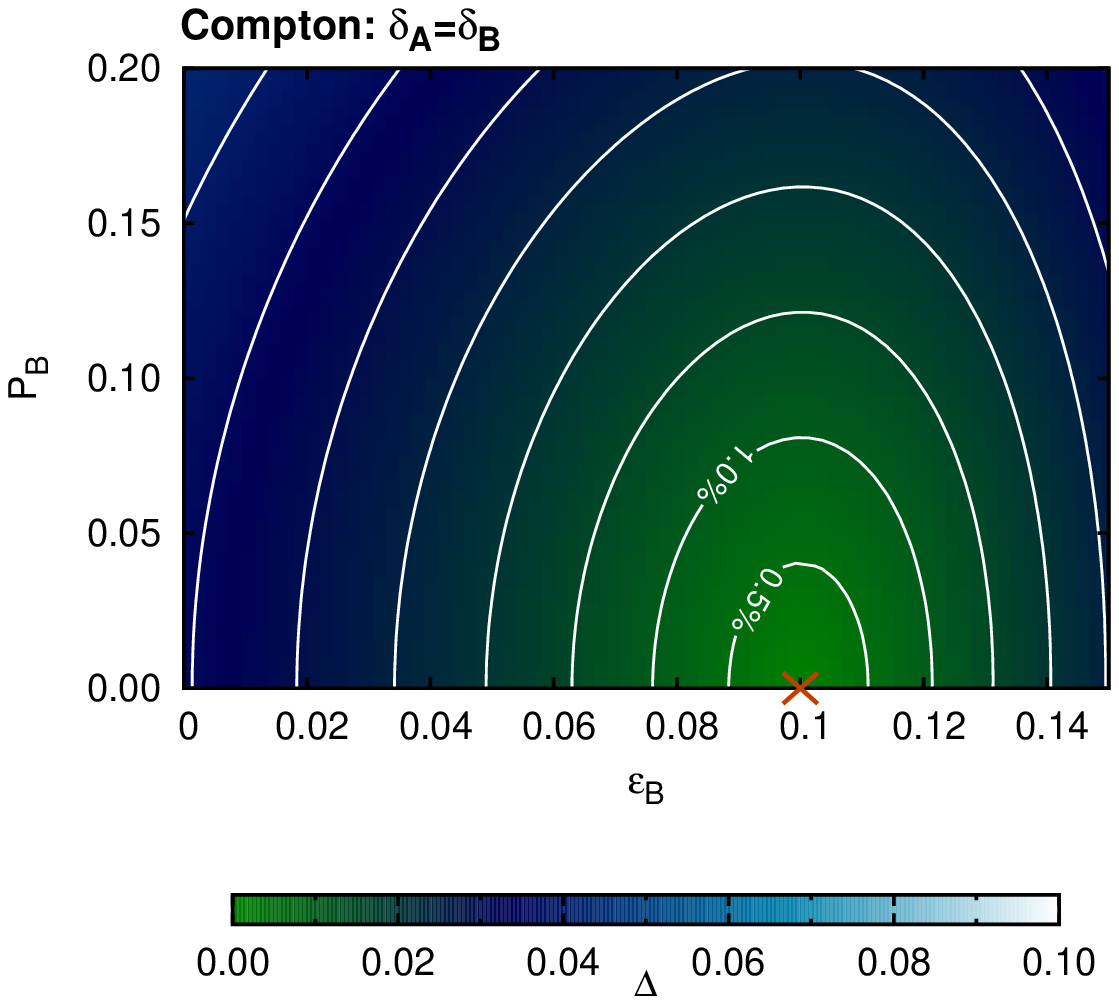}}
\end{center}
\caption{The same as Figure~\ref{fig:UnicityPol_1}, but in case the two configurations have
identical inclinations but different energy. In particular, we assumed that
$\beta_A,\varepsilon_A=0.1$, $\varphi_{0,B}=\pi/2$, $\P_A=0$, $\delta_A=\delta_B=\pi/6$,
$\eta_A=\eta_B=0$. The thick solid cross spots the values of polarization and energy of the first
modulation function. Contour lines identify $\Delta$ increments of 0.5\%. As usual, $f=1$.}
\label{fig:UnicityPol_2}
\end{figure*}

The measurement of the polarization is also degenerate with respect to the energy at some extent.
To show it, we take as an example two configurations $A$ and $B$ similar to those described above
but in this case we let the energy of the latter rather than its inclination free to vary. The value
of $\Delta$ as a function of $\P_B$ and $\beta_B,\varepsilon_E$ for $\delta_A=\delta_B=\pi/6$ and
$\beta_A,\varepsilon_A=0.1$ is reported in Figure~\ref{fig:UnicityPol_2} and, as anticipated, there
is a significant region in the $\P_B$-energy plane where the two configurations differ of a small
amount. Therefore, it would be difficult to appreciate the difference between two modulation
functions characterized by a different degree of polarization without knowing the energy of the
photons.

The discussion above makes evident that the knowledge of the source position and of its spectrum is
somehow necessary to unambiguously derive the state of polarization from the modulation function,
but at this stage it is difficult to evaluate the precision required. Obviously it is desirable that
the error related to the uncertainty in the photon incident direction and energy introduces a
negligible systematic effect on the polarization measurement. Although we will quantify such a
condition in a future work, it makes sense to us that the precision provided in the GRBs position
from Interplanetary Network (IPN), which is at level of $<1^\circ$, is adequate for small
instruments which are currently in-orbit \citep{Yonetoku2011} or in an advance stage of development
\citep{Orsi2011}. In case of larger detectors which allow for more precise measurements, it would
preferable to support the polarimeter with a small ancillary coded mask monitor, which can provide a
positioning at the level of arcminutes or less with a limited mass and volume \citep{Brandt2012}.
For what concerns the determination of the photon energy, an energy resolution of a few tens percent
may be sufficient, but we will quantify this claim in a future work as well.

\begin{figure*}[htbp]
\begin{center}
\subfloat[\label{fig:Yonetoku_1}]{\includegraphics[angle=0,totalheight=5.6cm]{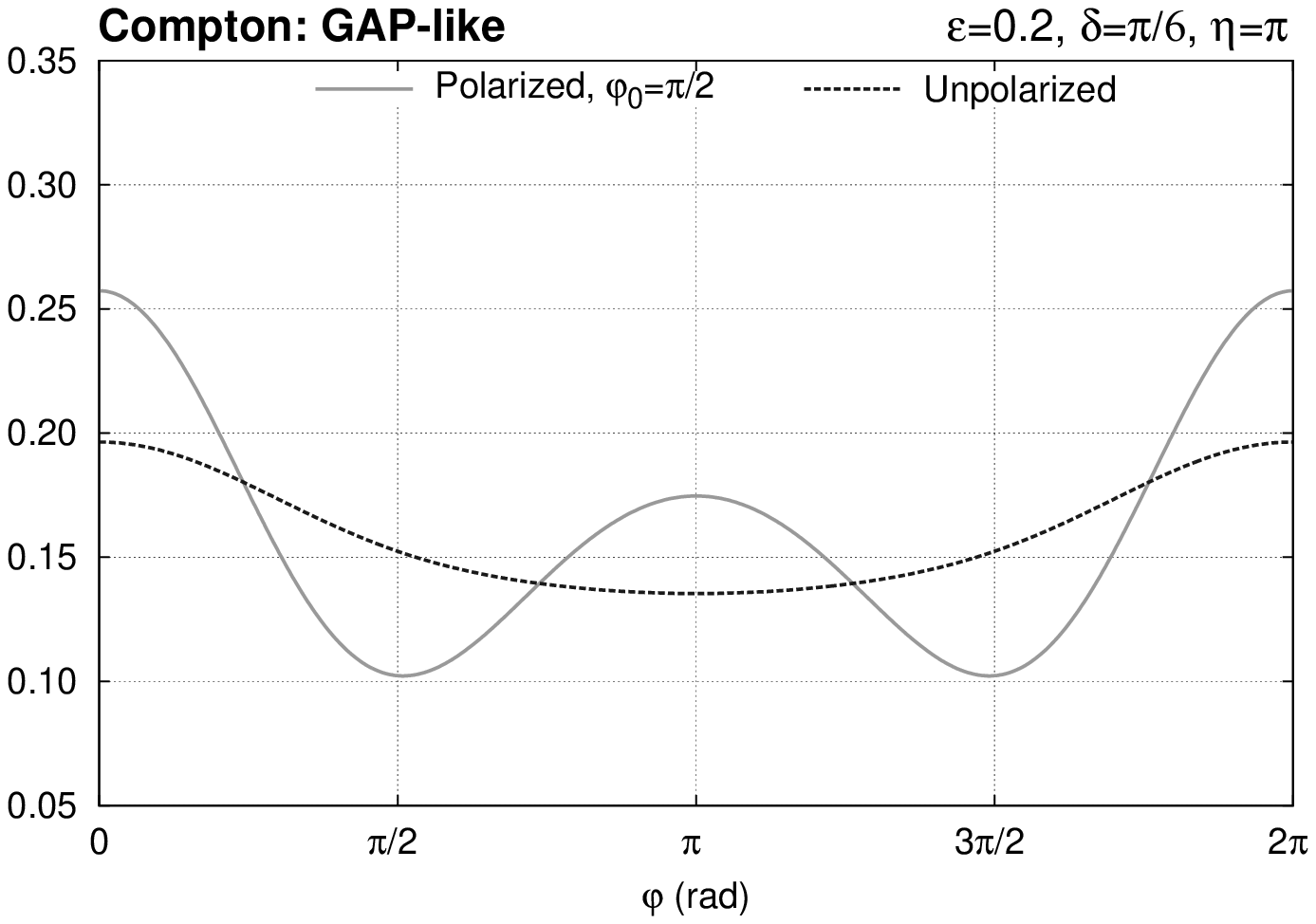}}
\hspace{1mm}
\subfloat[\label{fig:Xiong_1}]{\includegraphics[angle=0,totalheight=5.6cm]{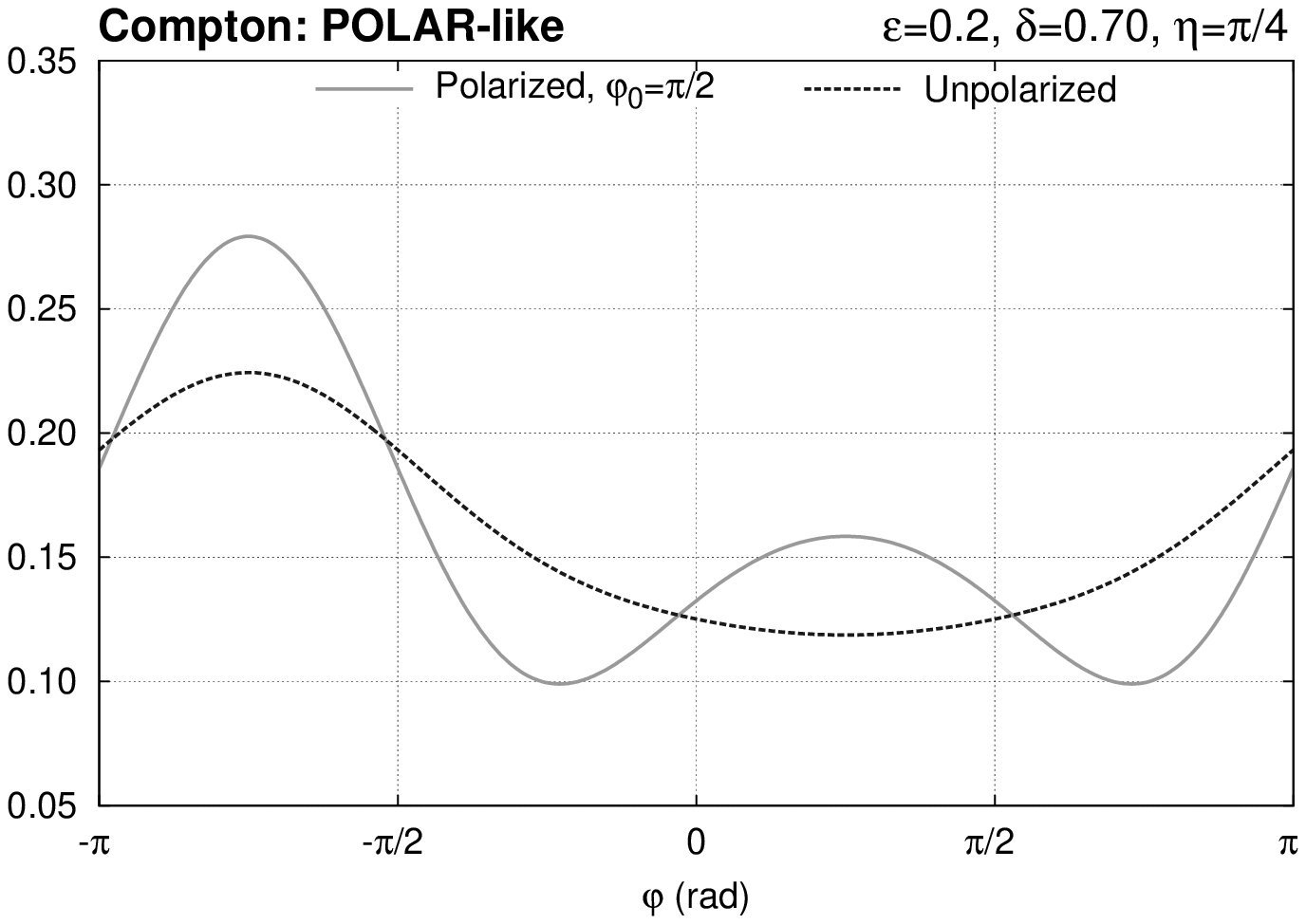}}
\end{center}
\caption{Modulation function of a Compton polarimeter for polarized and unpolarized photons
obtained with our treatment but assuming the same configuration discussed for actual instruments by
other authors. (a) Response in case of a GAP-like polarimeter to be compared with Figure~8 of
\citet{Yonetoku2006}. We estimated a $f$-factor of $0.50$ from the modulation factor reported by the
GAP team and from the assumption that valid events are scattered between $\theta_{\min}=\pi/3$ and
$\theta_{\max}=2\pi/3$. (b) The same as (a) but for a POLAR-like polarimeter. In this case we
assumed that $\theta_{\min}=0$ and $\theta_{\max}=\pi$, and evaluated $f=0.70$. This plot can not
be directly related to Figure~5 and Figure~6 of \citet{Xiong2009} because the actual response shows
a significant contribution from instrument systematic effects which are not included in our
treatment.}
\end{figure*}

We conclude this section by discussing how our results compare with those presented by other
authors. To our knowledge, only two groups have evaluated the off-axis response of X-ray
polarimeters and in both cases the results refer to Compton scattering instruments and are obtained
by Monte Carlo simulations. The modulation curve of the GAP instrument for polarized and unpolarized
photons which are incident at $\delta=\pi/6$ was reported in Figure~8 of \citet{Yonetoku2006}. Its
behavior can be compared with the modulation function derived with our treatment which is in
Figure~\ref{fig:Yonetoku_1}, where we assumed the same inclination as \citet{Yonetoku2006} and made
reasonable estimates on the other parameters which were not explicitly specified by the GAP team. In
particular, we assumed that photons are incident with energy $\varepsilon=0.2$,
$E\approx100~\mathrm{keV}$, and that valid events are those scattered between $\theta_{\min}=\pi/3$
and $\theta_{\max}=2\pi/3$, basically because of the geometry of the sensitive volume of the
instrument. We derived  an $f$-factor of 0.50 from the latter assumption and from the value of the
modulation factor at this energy. Notwithstanding these simplified assumptions, the modulation
function that we obtain for both polarized and unpolarized radiation is strikingly similar to the
modulation curves reported by \citet{Yonetoku2006}, suggesting a qualitative agreement between the
two results with an encouraging accuracy. An analogous comparison can be performed with the POLAR
instrument whose off-axis modulation curve was discussed by \citet{Xiong2009}, but unfortunately it
is less significant. In fact, the POLAR response is the result of the superimposition of the
dependency on the polarization and of a significant systematic effect due to the fact that the
sensitive volume is subdivided in square pixels \citep{Produit2005}. The latter effect
is not included in our treatment and it should be added before being able of comparing the POLAR
modulation curve with our results, but this is out the scope of this paper. For the sake of
completeness, we nevertheless report in Figure~\ref{fig:Xiong_1} the modulation function as derived
in our treatment in the same assumptions discussed in Figure~5 and Figure~6 of \citet{Xiong2009}. We
assumed for POLAR the same limits on $\theta$ used throughout the paper, that is $\theta_{\min}=0$
and $\theta_{\max}=\pi$, and evaluated $f=0.70$.

A technique proposed for deriving the polarization when the photons are incident off-axis
is to divide the measured modulation curve by that measured or derived from simulations for
unpolarized photons impinging from the same incident direction \citep{Xiong2009}. Such an ``off-axis
normalized'' modulation curve is supposed to recover the cosine square dependency so that the
subsequent analysis to derive the angle and the degree of polarization can proceed as that for
on-axis radiation. This approach is usually successful in removing from the modulation curve the
instrumental systematic effects which may deviate the modulation curve from the expected cosine
square behavior even on-axis \citep{Lei1997}. However, our findings indicate that this method is
effective only in the very first approximation when applied to off-axis radiation because,
actually, the cosine square dependency is not exactly recovered.

\begin{figure*}[htbp]
\begin{center}
\subfloat[\label{fig:VsNorm_1_1}]{\includegraphics[angle=0,totalheight=5.6cm]{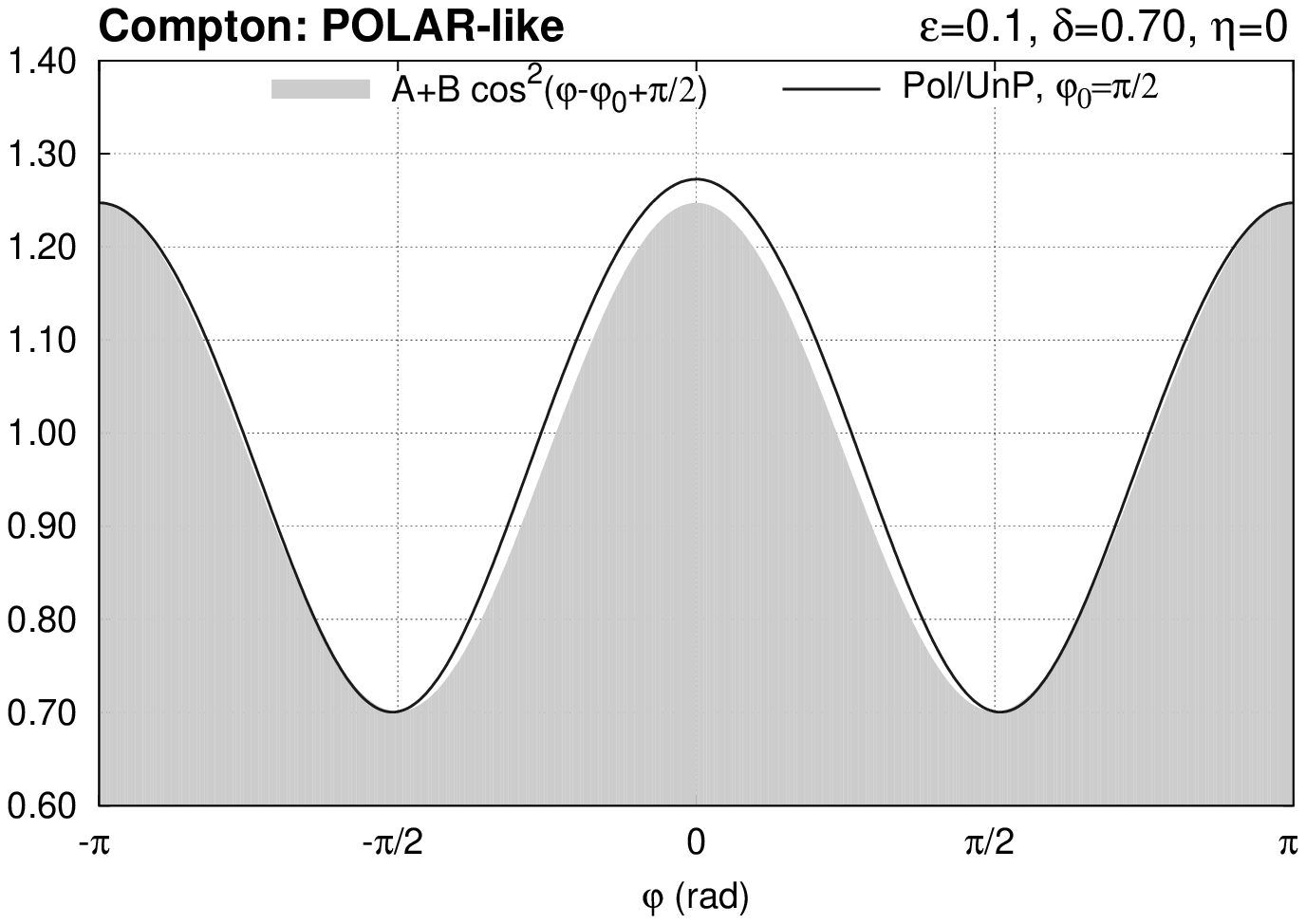}}
\hspace{1mm}
\subfloat[\label{fig:VsNorm_2_1}]{\includegraphics[angle=0,totalheight=5.6cm]{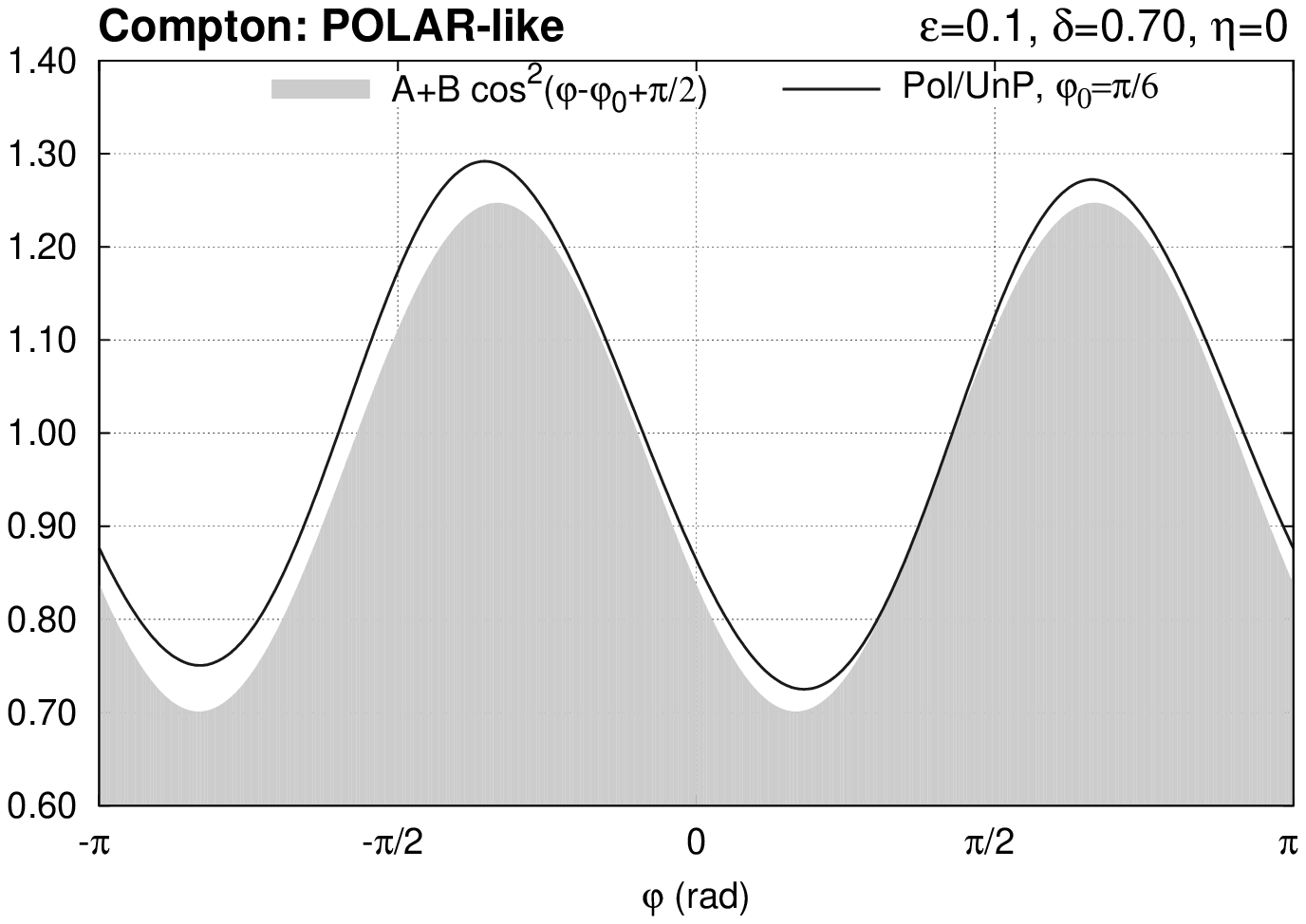}}
\end{center}
\caption{Comparison of the off-axis normalized modulation function for a Compton
polarimeter and completely polarized photons which are incident at $\delta=0.7$ with a cosine
square function for $\varphi_0=\pi/2$ (a) and $\varphi_0=\pi/6$ (b). The two cosine square function
are just shifted accordingly to the polarization angle. We assume a POLAR-like instrument with
$f$=0.70 and an energy of $\varepsilon=0.1$, $E\approx50$~keV.}
\label{fig:VsNorm__1}
\end{figure*}

The off-axis normalized modulation function for a Compton polarimeter, calculated dividing the
modulation function obtained with our treatment by the same function when $\P=0$, is reported in
Figure~\ref{fig:VsNorm_1_1} with a solid black line for a representative choice of the photon
parameters. We assume completely polarized photons of energy $\varepsilon=0.1$,
$E\approx50~\mathrm{keV}$, with an angle of polarization $\varphi_0=\pi/2$ and an incident direction
with inclination $\delta=0.7$ and azimuth $\eta=0$. Such a function is compared with a cosine
square $A+B\cos^2(\varphi-\varphi_0+\pi/2)$, which is reported in the figure as the light-gray
filled curve. As a matter of fact, the two peaks of the off-axis normalized modulation function are
not identical and then only one of them can be adequately represented with a cosine square with an
appropriate choice of the $A$ and $B$ parameters. Moreover, if we change the angle of polarization,
the off-axis modulation function does not simply shift according to the change of $\varphi_0$ as it
should do if the on-axis behavior was recovered. This is evident in Figure~\ref{fig:VsNorm_2_1},
where the off-axis normalized modulation function and the cosine square are reported for
$\varphi_0=\pi/6$. While the cosine square is just horizontally shifted of an angle $\pi/3$, the
change of the angle of polarization causes also a change in the maximum and minimum values of the
off-axis normalized modulation function. As a consequence, the measured modulation factor and then
the value of the polarization degree depend systematically on the polarization angle. This effect
was possibly already reported, although the authors attributed it to different instrumental effects,
see Figure~8 and Figure~9 of \citet{Xiong2009}.

A rough estimate of the error $\xi$ on the polarization degree that it is made by assuming that the
off-axis normalized modulation curve has a cosine square behavior can be derived by a simple
argument. The fact that the two peaks are not identical causes a ``systematic modulation'' whose
amplitude is of the order of their height difference. As the degree of polarization is obtained by
dividing the measured modulation amplitude by its value for completely polarized photons, we can
evaluate $\xi$ as the ratio between such a systematic modulation and the amplitude for 100\%
polarized radiation. The value of $\xi$ in the configuration discussed in
Figure~\ref{fig:VsNorm__1}, that is at $\varepsilon=0.1$, is a few percent in absolute value, but
our analysis suggests that this value increases to $\sim$10\% at a few hundreds keV. The sensitivity
promised by future instrumentations dedicated to GRBs polarimetry will be able to measure
polarization which are lower that this level of systematics and then we advocate a critical
discussion on the adequacy of the analysis performed by normalizing off-axis the modulation
curve.

\section{Conclusions} \label{sec:Conclusions}

We have presented a novel approach which allowed us to calculate analytically the response of a
photoelectric or Compton polarimeter when photons are incident off-axis starting from first
principles, that is, from the differential cross section of the involved photon interaction.
Our results show that the modulation curve is no more a cosine square and depends not only on the
state of polarization of the photons, but also on their direction of incidence and energy in a
complex way. The amplitude of the modulation, which on-axis is proportional to the degree of
polarization, is affected also by the inclination and its phase loses any direct relation with the
polarization angle, being related to the polarization degree also. We replaced the concept of
modulation factor with a new quantity, which we called f-factor, to fully take account of this.

There is a certain degeneracy among all of the parameters which the modulation curve depends on.
Although a complete analysis was out of the scope of this paper, we showed that it is easy to find
configurations for which unpolarized and 10\% polarized photons incident with a different
inclination give rise to essentially identical modulation curves. Therefore, an adequate knowledge
of the source position and a reasonable energy resolution is necessary to extract unambiguously the
state of polarization from the instrumental response. For a small instrument, a source positioning
at the level of 1 degree may be sufficient to make negligible the systematic error on the
polarization measurement with respect to the statistical one, whereas a stricter requirement has to
be put for larger instruments. As well, an energy resolution at the level of 10--20 percent may be
enough to avoid a significant impact on the polarization determination. We will make both of these
claims more quantitative in future papers.

In our view, the fact that the response to off-axis photons is different from that for on-axis
radiation is an unavoidable consequence of how current instruments work. As a matter of fact,
state-of-the-art photoelectric and Compton polarimeters are designed to be sensitive only to the
azimuthal distribution of either the emitted photoelectrons or the scattered photons, because it is
well known that the polar distribution does not carry any information on polarization. However, the 
instrument is sensitive to the ``intrinsic'' azimuthal event distribution, which is cosine square
modulated for polarized radiation, only if photons are incident on-axis. When the radiation is
impinging from an inclined direction, the distribution of the events as it is seen by the instrument
does not coincide with that intrinsic and this ultimately gives rise to the forest of effects
discussed in this paper. At this regard photoelectric and Compton polarimeters are substantially
equivalent and this explains the emergence of analogous effects in the two classes of instruments.
Such effects would not be removed by rotating the instrument around an axis orthogonal to the
detection plane, even assuming that it would be possible to perform a high number of rotations
during the observation. In fact this procedure is effective only to cancel nonuniformities in the
instrument response which, however, are not included in our discussion. The usual cosine square
modulation can be recovered only if the instrument can measure both the azimuthal and the polar
direction of the event, as it happens for Compton telescopes which by design have to be sensitive to
both to reconstruct the photon incidence direction. \citet{Lei1997} already outlined how to proceed
in this case.

The dependency of the modulation curve on the photon incident direction and energy that we put
forward can easily explain the systematic effects reported by other authors thanks to Monte Carlo
simulations but attributed to generic instrumental non-uniformities. This ascription has led to
the common practice of normalizing the measured off-axis modulation curve by the modulation curve
obtained for the same source position and unpolarized photons, in the assumption that such an
``off-axis normalized'' modulation curve recovers the usual behavior observed on-axis. We find that
actually the obtained curve is not a cosine square and therefore this procedure is inherently
inaccurate. Simple arguments suggest that the error on the determination of the polarization degree
can be significant, at the level of $\sim$10\% in absolute value, and then we encourage a critical
discussion on the adequacy of this method.

As a final note, it is worth stressing that the results presented in this paper were intentionally
based for better clarity on simplified assumptions which, although effective for obtaining a
qualitative picture, may turn out to be too inaccurate for any practical use. The most relevant
example is the choice of expanding the energy dependence of the event angular distribution only at
the first order. While this allowed us to write down explicitly the modulation function at least
in the simple scenario presented in Section~\ref{sec:SimpleScenario}, the error with respect to
the exact function is below a few percent only at very low energy, a few keV for photoelectric
polarimeters and a few tens of keV for Compton ones. The operative energy range of any realistic
instrument would largely exceed these limits and therefore some kind of \emph{quantitative}
disagreements has to be reckoned with. At this regard, we have also to remember that the analytical
expressions used to treat the photoelectric absorption and the Compton scattering, see
Equations~(\ref{eq:dsdO_Ph}) and (\ref{eq:dsdO_Cm}), are approximated. The former assume that the
energy of the incident photons is well above the ionization potential but well below the rest mass
energy of the electron since it neglects relativistic corrections; the latter assume that the
electron is free and at rest, which is a simplified treatment of the atomic electrons in the
scattering material. Another point which deserves a better investigation, especially for Compton
polarimeters, is the interval of polar angles in which the events has to be summed up.
Therefore, this work has to be intended as a first step to better understand the off-axis response
of photoelectric and Compton polarimeters and it does not pretend to supersede Monte Carlo
simulations in modeling the instrumental response. Notwithstanding these necessary notes of
caution which will be further discussed in subsequent works, we mention that a comparison of our
findings with real data in case of a photoelectric polarimeter and the simple geometry discussed in
Section~\ref{sec:SimpleScenario} was already presented elsewhere and it came out to be very
encouraging \citep{Muleri2008c,Muleri2010b}.

\begin{acknowledgements}
The author acknowledges Enrico Costa, Paolo Soffitta and Riccardo Campana, whose comments allowed
him to greatly increase the clarity of this paper, and the financial support on the INAF contract
PRIN-INAF-2009.
\end{acknowledgements}

\appendix

\section{Derivation of the distribution of event directions in the instrument frame of reference}
\label{app:CoordTransf}

In this Appendix we discuss how to calculate the angular distribution of the events in the
instrument frame of reference $xyz$ starting from that in the photon frame of reference
$x_\gamma y_\gamma z_\gamma$. Basically, it is necessary to calculate how the spherical coordinates
$(\phi;\theta)$ defined in the latter depend on those $(\varphi;\vartheta)$ defined in the former,
that is, we have to derive explicitly the functions $\phi=\phi(\varphi,\vartheta)$ and
$\theta=\theta(\varphi,\vartheta)$. After that, we can substitute them in Equations~(\ref{eq:DPh})
and (\ref{eq:DCm}) so that the angular distribution of the events is expressed as a function of the
variables $(\varphi;\vartheta)$. For comparison, in case of on-axis photons we have that
$\phi=\varphi+\mbox{constant}$ and $\theta\equiv\vartheta$ and therefore the change of variable to
switch to the instrument frame of reference was trivial and we did it implicitly.

The first step in our procedure is to find out the coordinate transformation from the $xyz$ to the
$x_\gamma y_\gamma z_\gamma$ frame of reference. The origins of the two frames of reference are not
relevant for our discussion because we are interested only in the angular direction of the events
and therefore one frame of reference can be unambiguously identified with respect to the other by
means of three angles. As discussed in Section~\ref{sec:OffAxisProcedure}, we will use the
inclination $\delta$ and the azimuth $\eta$ of the incident direction and the angle of polarization
$\varphi_0$ (see Figure~\ref{fig:Inclination}). As the position of the photon frame of reference is
completely characterized by these three angles, the coordinate transformation from $xyz$ to
$x_\gamma y_\gamma z_\gamma$ can be decomposed in three different rotations, a rotation
$\R_{\eta}^z$ of an angle $\eta$ around the $z$ axis, a rotation $\R_{\delta}^{y'}$ around $y'$ of
$\delta$ and a rotation $\R_{\varphi_0}^{z''}$ around $z''$ of $\varphi_0$. Here we indicated as
$y'$ the ordinate axis of a frame of reference which is rotated of $\eta$ around $z$ and as $z''$
the $z$-axis of a frame of reference which is also rotated of $\delta$ around $y'$ (see
Figure~\ref{fig:Inclination}). The explicit expression of the coordinate transformation is quite
complicated when expressed in spherical coordinates and therefore it is somehow convenient to switch
to Cartesian coordinates. In this case, we have that
\begin{equation}
\left(
\begin{array}{c}
x_\gamma \\
y_\gamma \\
z_\gamma 
\end{array}
\right) = \R_{\varphi_0}^{z''}\;\R_{\delta}^{y'}\;\R_{\eta}^z
\left(
\begin{array}{c}
x \\
y \\
z
\end{array}
\right) \; ,
\label{eq:CartRot}
\end{equation}
with
\begin{align*}
\R_{\eta}^z &= \left(
\begin{array}{ccc}
\cos\eta & \sin\eta & 0 \\
-\sin\eta & \cos\eta & 0 \\
0 & 0 & 1 \\
\end{array}
\right) \; ; \\ 
\R_{\delta}^{y'} &= \left(
\begin{array}{ccc}
\cos\delta & 0 & -\sin\delta \\
0 & 1 & 0 \\
\sin\delta & 0 & \cos\delta \\
\end{array}
\right) \; ; \\ 
\R_{\varphi_0}^{z''} &= \left(
\begin{array}{ccc}
\cos\varphi_0 & \sin\varphi_0 & 0 \\
-\sin\varphi_0 & \cos\varphi_0 & 0 \\
0 & 0 & 1 \\
\end{array}
\right) \; . 
\end{align*}

Equation~(\ref{eq:CartRot}) allows us to calculate the Cartesian coordinates of a point in the
photon frame of reference from the Cartesian coordinate of the same point in the instrument frame of
reference, that is it provides $x_\gamma=x_\gamma(x,y,z)$, $y_\gamma=y_\gamma(x,y,z)$ and
$z_\gamma=z_\gamma(x,y,z)$. We can fully exploit these relations by noting that simple combinations
of the Cartesian coordinates of a point on a sphere of radius 1 and center in the origin have the
same $(\phi;\theta)$ dependency as the event distribution $\D$. In fact, $\D$ depends on the
spherical coordinates by means of a combination of $\sin^2\theta\cos^2\phi$,
$\sin^2\theta$ and $\cos\theta$, cf. Equations~(\ref{eq:DPh}) and (\ref{eq:DCm}). Instead, the
Cartesian coordinates of a point on the sphere of radius 1 and center in the origin, indicated in
the following with a hat over them, are expressed in $x_\gamma y_\gamma z_\gamma$ as a function of
$(\phi;\theta)$ as 
\begin{equation}
\begin{cases}
\hat{x}_\gamma = \sin\theta\cos\phi \\
\hat{y}_\gamma = \sin\theta\sin\phi \\
\hat{z}_\gamma = -\cos\theta
\end{cases} \; . \notag
\end{equation}
The minus sign in the definition of $\hat{z}_\gamma$ derives from the fact that the angle $\theta$
is measured from the negative z-axis (see Figure~\ref{fig:Angles}) instead that from the positive
one as usual.

Then, we have that
\begin{subequations}
\label{eq:Transf1}
\begin{align}
&\sin^2\theta\cos^2\phi = \hat{x}_\gamma^2 = \hat{x}_\gamma^2(\hat{x},\hat{y},\hat{z})\; ; \\
&\sin^2\theta = 1-\hat{z}_\gamma^2 = 1-\hat{z}_\gamma^2(\hat{x},\hat{y},\hat{z}) \; ; \\
&\cos\theta = -\hat{z}_\gamma = -\hat{z}_\gamma(\hat{x},\hat{y},\hat{z}) \; ,
\end{align}
\end{subequations}
where we used Equation~(\ref{eq:CartRot}) to write the last equalities. Coordinates
$(\hat{x};\hat{y};\hat{z})$ defines in $xyz$ the point with coordinates
$(\hat{x}_\gamma;\hat{y}_\gamma;\hat{z}_\gamma)$ in $x_\gamma y_\gamma z_\gamma$. Since we applied
only rotations to switch from a frame of reference to the other, the point
$(\hat{x};\hat{y};\hat{z})$ is still on a sphere of radius 1 and then 
\begin{equation}
\label{eq:CoordRot}
\begin{cases}
\hat{x} = \sin\vartheta\cos\varphi \\
\hat{y} = \sin\vartheta\sin\varphi \\
\hat{z} = -\cos\vartheta 
\end{cases}
\end{equation}
Substituting Equation~(\ref{eq:CoordRot}) in Equation~(\ref{eq:Transf1}), we eventually obtain that 
\begin{subequations}
\label{eq:Transf}
\begin{align}
&\sin^2\theta\cos^2\phi =
\hat{x}_\gamma^2(\sin\vartheta\cos\varphi,\sin\vartheta\sin\varphi,-\cos\vartheta)\; ; \\
&\sin^2\theta =1-\hat{z}_\gamma^2(\sin\vartheta\cos\varphi,\sin\vartheta\sin\varphi,-\cos\vartheta)
\; ; \\
&\cos\theta = \hat{z}_\gamma(\sin\vartheta\cos\varphi,\sin\vartheta\sin\varphi,-\cos\vartheta) \; .
\end{align}
\end{subequations}
Equations~(\ref{eq:CartRot}) and (\ref{eq:Transf}) provide all of the ``ingredients'' to calculate,
with some algebra, the angular distribution of the events in the instrument frame of reference,
$\D(\varphi,\vartheta)$. However, the resulting expressions is cumbersome and rather
difficult to handle and, eventually, of not much relevance for our discussion. Therefore, we will
leave it implicit, limiting ourselves in an explicit derivation of $\D(\varphi,\vartheta)$ only
in the simple case discussed in Section~\ref{sec:SimpleScenario} and
Appendix~\ref{app:SimpleScenario}.

\section{Explicit derivation of the modulation function for photoelectric polarimeters in a
``simple'' scenario}
\label{app:SimpleScenario}

In the simple geometry assumed in Figure~\ref{fig:InclinationSimple}, the Cartesian coordinate
transformation to switch from the photon to the instrument frame of reference is (cf.
Equation~(\ref{eq:CartRot})):
\begin{align}
\left(
\begin{array}{c}
x_\gamma \\
y_\gamma \\
z_\gamma 
\end{array}
\right) &= \R_{\varphi_0=0}^{z''}\;\R_{\delta}^{y'}\;\R_{\eta=0}^z
\left(
\begin{array}{c}
x \\
y \\
z
\end{array}
\right) = \notag \\ &= \left(
\begin{array}{ccc}
\cos\delta & 0 & -\sin\delta \\
0 & 1 & 0 \\
\sin\delta & 0 & \cos\delta \\
\end{array}
\right) \left(
\begin{array}{c}
x \\
y \\
z
\end{array}
\right) \notag
\end{align}
because $\eta=0$ and $\varphi_0=0$ and therefore $\R_{\eta=0}^z$ and $\R_{\varphi_0=0}^{z''}$ are
identity matrices. Then,
\begin{equation}
\begin{cases}
x_\gamma = x\cos\delta - z\sin\delta \\
y_\gamma = y \\
z_\gamma = x\sin\delta + z\cos\delta \\
\end{cases} \; . \notag
\end{equation}
Using such a coordinate transformation in Equations~(\ref{eq:Transf}), we obtain
\begin{subequations}
\label{eq:Transf2}
\begin{align}
\sin^2\theta\cos^2\phi &= \hat{x}_\gamma^2 = [\hat{x}\cos\delta - \hat{z}\sin\delta]^2 = \notag \\ 
&= [(\sin\vartheta\cos\varphi)\cos\delta - (-\cos\vartheta)\sin\delta]^2  = \notag \\
&= [\sin\vartheta\cos\varphi\cos\delta + \cos\vartheta\sin\delta]^2  \; ; \label{eq:sin2cos2}\\
\sin^2\theta &= 1-\hat{z}_\gamma^2 = 1-[\hat{x}\sin\delta + \hat{z}\cos\delta]^2 = \notag \\
&= 1-[(\sin\vartheta\cos\varphi)\sin\delta + (-\cos\vartheta)\cos\delta]^2 = \notag \\
&= 1-[\sin\vartheta\cos\varphi\sin\delta - \cos\vartheta\cos\delta]^2 \; ; \label{eq:sin2}\\
\cos\theta &= -\hat{z}_\gamma = -[\hat{x}\sin\delta + \hat{z}\cos\delta] = \notag \\
& = -[(\sin\vartheta\cos\varphi)\sin\delta + (-\cos\vartheta)\cos\delta] = \notag \\
& = -\sin\vartheta\cos\varphi\sin\delta + \cos\vartheta\cos\delta \; . \label{eq:cos}
\end{align}
\end{subequations}

The distribution of the events in the instrument frame of reference can be calculated by
substituting Equations~(\ref{eq:Transf2}) in Equations~(\ref{eq:DPh}). We have, for polarized
radiation, that:
\begin{align}
\vDPhPol &=\frac{\sin^2\theta\,\cos^2\phi}{\left\{1-\beta\cos\theta\right\}^4} = \notag \\
&=\frac{[\sin\vartheta\cos\varphi\cos\delta +
\cos\vartheta\sin\delta]^2}{\left\{1-\beta [-\sin\vartheta\cos\varphi\sin\delta +
\cos\vartheta\cos\delta] \right\}^4} \; ,
\label{eq:vDPhPol}
\end{align}
where we made use of Equations~(\ref{eq:sin2cos2}) and (\ref{eq:cos}). Analogously, for unpolarized
radiation
\begin{align}
\vDPhUnP &=\frac{1}{2}\frac{\sin^2\theta}{\left\{1-\beta\cos\theta\right\}^4} = \notag \\
&=\frac{1}{2}\frac{1-[\sin\vartheta\cos\varphi\sin\delta -
\cos\vartheta\cos\delta]^2}{\left\{1-\beta [-\sin\vartheta\cos\varphi\sin\delta +
\cos\vartheta\cos\delta] \right\}^4} \; .
\label{eq:vDPhUnP}
\end{align}

We are now in the position of applying Equation~(\ref{eq:M_Phi2}). Unfortunately, the angular
distribution of the events in the instrument frame of reference, expressed with
Equations~(\ref{eq:vDPhPol}) and (\ref{eq:vDPhUnP}), is not easily integrable over $\vartheta$
because of the polynomial at the denominator. To simplify our discussion, we can expand $\D_\Ph$ in
Maclaurin series with respect to the variable $\beta$, so that the integrand becomes a polynomial of
trigonometric functions which is integrable with standard techniques for any order of approximation.
For example, at the first order in $\beta$ we have that 
\begin{align}
\DPhPol &\approx \sin^2\theta\,\cos^2\phi\left(1+4\beta\cos\theta\right) \; ; \notag \\
\DPhUnP &\approx \frac{1}{2}\sin^2\theta\left(1+4\beta\cos\theta\right) \; . \notag
\end{align}
Any other better approximation of the modulation function can be easily obtained adding a sufficient
number of higher order terms to the series, at the cost of some additional algebra.

The calculus of the normalized azimuthal distribution of the emitted events is tedious but trivial
in the case of the first order approximation. The result, which we obtained with the help of the
\textsc{Maxima} software, is
\begin{align}
\Phi_\Ph^\Pol(\beta,\varphi,\delta) &=
-\frac{9\beta\cos^{2}\delta\sin\delta}{8} {\cos^{3}\varphi} + \frac{{\cos^{2}\delta }}{\pi
}{\cos^{2}\varphi} + 
\notag \\ &
+\frac{3\beta\left(3{\cos^{2}\delta }-1\right)\sin\delta }{8}\cos\varphi +\frac{\sin^{2}\delta}
{2\pi} 
\notag \\
\Phi_\Ph^\UnP(\beta,\varphi,\delta) &=
\frac{9\beta{\sin^3\delta}}{16}{\cos^3\varphi}-\frac{{\sin^2\delta}}{2\pi}{\cos^2\varphi}+ 
\notag \\ & 
+\frac{3\beta(3{\cos^2\delta}-4)\sin\delta}{16}\cos\varphi+\frac{3-{\cos^2\delta}}{4\pi} 
\notag
\end{align}
from which the modulation function is
\begin{align}
\M_\Ph(\beta,\varphi,\delta) &= f \N_{\tot} \left\{\P\left[
-\frac{9\beta\cos^{2}\delta\sin\delta}{8} {\cos^{3}\varphi} + \frac{{\cos^{2}\delta }}{\pi
}{\cos^{2}\varphi} + 
\right. \right. \notag \\ & \left.\left. 
+\frac{3\beta\left(3{\cos^{2}\delta }-1\right)\sin\delta }{8}\cos\varphi +\frac{\sin^{2}\delta}
{2\pi}
\right]+\right. \notag \\ &\left.+(1-\P)\left[ 
\frac{9\beta{\sin^3\delta}}{16}{\cos^3\varphi}-\frac{{\sin^2\delta}}{2\pi}{\cos^2\varphi}+ 
\right. \right. \notag \\ & \left. \left.
+\frac{3\beta(3{\cos^2\delta}-4)\sin\delta}{16}\cos\varphi+\frac{3-{\cos^2\delta}}{4\pi} 
\right]\right\} + \notag \\
& + \N_\tot\frac{1-f}{2\pi} \; .
\notag
\end{align}

\begin{figure}[htbp]
\begin{center}
\includegraphics[angle=0,totalheight=5.6cm]{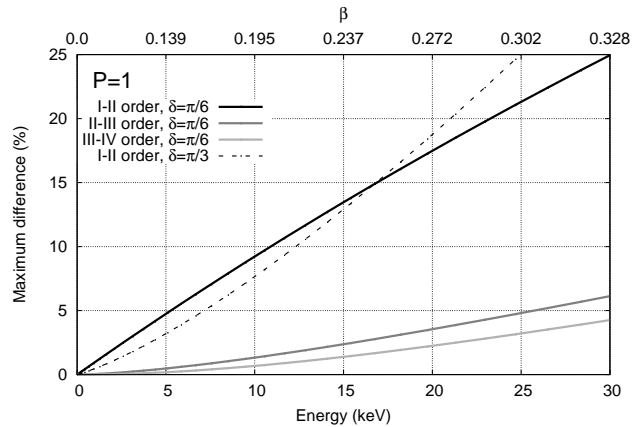}
\end{center}
\caption{Maximum percent difference between the modulation functions obtained for photoelectric
polarimeters by expanding the energy dependence of the event angular distribution at different
orders of approximation. The inclination is $\pi/6$ and $\pi/3$ for solid and dashed lines,
respectively. It is assumed that $\P=1$ and $f=1$.}
\label{fig:PhCompareOrders}
\end{figure}

We conclude this appendix by arguing on how appropriate is the modulation function obtained by
developing at the first order the energy dependence of the event angular distribution. An indication
of the deviation with respect to the actual behavior is the difference to the modulation function
derived by expanding $\D_\Ph(\beta,\phi,\theta)$ at the second order in $\beta$ because, by
definition, higher the order lower the magnitude of the correction. Therefore, we report in
Figure~\ref{fig:PhCompareOrders} as solid lines the maximum percentage difference between the
modulation functions obtained with subsequent higher approximations, assuming completely polarized
photons which are incident at 30$^\circ$ off-axis. As expected, higher orders differ less as the
approximated modulation function approaches the actual behavior, although the effective deviation
depends on the inclination and on the polarization degree, see for example the dashed line in
Figure~\ref{fig:PhCompareOrders} which refers to $\delta=\pi/3$. The relevant result is however that
the first order approximation significantly deviates from the actual behavior also at relatively low
energy. Whereas the qualitative description carried out in this paper remains valid for any degree
of approximation, it is evident that higher orders are necessary to model the response of real
instruments at few percent level, that is the accuracy relevant for astrophysical observations, as
soon as the energy range is above a few keV.

\bibliography{/home/fabio/Bibliografia/Latex/References.bib}
\bibliographystyle{apj}

\end{document}